\documentclass[aps,prb,twocolumn,superscriptaddress,longbibliography]{revtex4-2}

\usepackage{graphicx}
\usepackage{hyperref}
\usepackage{url}
\usepackage{amsmath}
\usepackage{amssymb}
\usepackage{verbatim}
\usepackage{xcolor}
\usepackage{enumerate}
\usepackage[normalem]{ulem}
\usepackage{float}
\usepackage{hyperref}

\def\beq {\begin{equation}}
\def\eeq {\end{equation}}

\def\bfr {\mathbf{r}}

\date{\today}

\begin{document}

\newcommand{\refeq}[1]{Eq.~\eqref{#1}}
\newcommand{\refEq}[1]{Eq.~\eqref{#1}}
\newcommand{\refeqs}[1]{Eqs.~\eqref{#1}}
\newcommand{\refsec}[1]{Sec.~\ref{#1}}
\newcommand{\qP} {quasi-particle}
\newcommand{\ie}{{i.e.,\ }}
\newcommand{\eg}{{e.g.,\ }}
\newcommand{\nrho}{n}
\newcommand{\reffig}[1]{Fig.~\ref{#1}}
\newcommand{\refFig}[1]{Fig.~\ref{#1}}
\newcommand{\GF}{Green's function}
\newcommand{\ham}{hamiltonian}
\newcommand{\xc}{exchange-correlation}
\newcommand{\rij}{\mbox{${r_{ij}}$}}
\newcommand{\modrij}{\mbox{${|{\bf r}_{i} - {\bf r}_{j}|}$}}
\newcommand{\modrijt}{\mbox{${|\tilde{{\bf r}}_{i} - \tilde{{\bf r}}_{j}|}$}}
\newcommand{\modrrp}{\mbox{${|{\bf r} - {\bf r}^{\prime}|}$}}
\newcommand{\modriI}{\mbox{${|{\bf r}_{i} - {\bf R}_{I}|}$}}
\newcommand{\modRIJ}{\mbox{${|{\bf R}_{I} - {\bf R}_{J}|}$}}

\newcommand{\rmb}{\mbox{$\{ \vecr_{i} \} $}}
\newcommand{\Rmb}{\mbox{$\{ \vecR_{I} \} $}}

\newcommand{\rp}{\mbox{${r^{\prime}}$}}
\newcommand{\vecrp}{\mbox{${{\bf r}^{\prime}}$}}
\newcommand{\qp}{\mbox{${q^{\prime}}$}}
\newcommand{\vecqp}{\mbox{${{\bf q}^{\prime}}$}}
\newcommand{\kp}{{k^{\prime}}}
\newcommand{\veckp}{\mbox{${{\bf k}^{\prime}}$}}
\newcommand{\bDelta}{\mbox{$\bf \Delta$}}
\newcommand{\bG}{{\bf G}}
\newcommand{\bGp}{\mbox{$\bG^{\prime}$}}
\newcommand{\bk}{{\bf k}}
\newcommand{\bnabla}{\mbox{\boldmath $\nabla$}}
\newcommand{\bq}{{\bf q}}
\newcommand{\bR}{{\bf R}}
\newcommand{\br}{{\bf r}}
\newcommand{\brp}{\mbox{$\br^{\prime}$}}
\newcommand{\bx}{{\bf x}}
\newcommand{\half}{{1 \over 2}}
\newcommand{\vxc}{\mbox{{$v_{\rm xc}$}}}
\newcommand{\vxcr}{\mbox{{$v_{\rm xc}({\bf r})$}}}

\title{Accurate Kohn-Sham auxiliary system from the ground state density of solids}

\newcommand{\lsi}{LSI, CNRS, CEA/DRF/IRAMIS, \'Ecole Polytechnique, Institut Polytechnique de Paris, F-91120 Palaiseau, France}
\newcommand{\etsf}{European Theoretical Spectroscopy Facility (ETSF)}
\newcommand{\soleil}{Synchrotron SOLEIL, L'Orme des Merisiers, Saint-Aubin, BP 48, F-91192 Gif-sur-Yvette, France}

\newcommand{\wm}{Department of Physics, College of William \& Mary, Williamsburg, Virginia 23185, USA}
\newcommand{\flatiron}{Center for Computational Quantum Physics, Flatiron Institute, New York, New York 10010, USA}

\author{Ayoub Aouina}
\affiliation{\lsi}
\affiliation{\etsf}

\author{Matteo Gatti}
\affiliation{\lsi}
\affiliation{\etsf}
\affiliation{\soleil}

\author{Siyuan Chen}
\affiliation{\wm}

\author{Shiwei Zhang}
\affiliation{\flatiron}
\affiliation{\wm}

\author{Lucia Reining}
\affiliation{\lsi}
\affiliation{\etsf}
\email{lucia.reining@polytechnique.fr}

\begin{abstract}
The Kohn-Sham (KS) system is an auxiliary system whose effective potential is unknown in most cases. It is in principle determined by the ground state density, and it has been found numerically for some low-dimensional systems by inverting the KS equations starting from a given accurate density. For solids, only approximate results are available. In this work, we determine accurate exchange correlation (xc) potentials for Si and NaCl using the ground state densities obtained from Auxiliary Field Quantum Monte Carlo calculations. 
We show that these xc potentials can be rationalized as an ensemble of a few local functions of the density, whose form depends on the specific environment and can be well characterised by the gradient of the density and the local kinetic energy density. The KS band structure can be obtained with high accuracy. 
The true KS band gap turns out to be larger than the prediction of the local density approximation, but significantly smaller than the measurable photoemission gap, which confirms previous estimates.
Finally, our findings show that the conjecture that very different xc potentials can lead to very similar densities and other KS observables is true also in solids, which questions the meaning of details of the potentials and, at the same time, confirms the stability of the KS system. 
\end{abstract} 


\maketitle

\section{Introduction}
Density Functional Theory \cite{DFT0,DFT1}  (DFT) is undoubtedly one of the biggest success stories of condensed matter theory, since it has made realistic electronic structure calculations possible for a huge range of materials, and since it has led to numerous insights \cite{Kohn1999,martin2004}. Two main problems had to be overcome in order to make DFT applicable in practice: first, it was necessary to find reliable approximations for the total energy as functional of the ground state density; second, an efficient way to determine the ground state density itself was needed. The solution to both problems relies on the approach of Kohn and Sham \cite{DFT1}, where the interacting system is mapped onto an auxiliary system of non-interacting electrons with an effective Kohn-Sham (KS) potential that is designed to yield the ground state density. The exchange-correlation (xc) contribution to this potential, \vxcr , and to the xc energy density per particle, $\epsilon_{\rm xc}({\bf r})$, is unknown in most systems. The initial breakthrough came with the Local Density Approximation  \cite{DFT1} (LDA). This approximation takes the energy density, and hence \vxcr , locally from the homogeneous electron gas (HEG), where it was calculated using Quantum Monte Carlo  \cite{Cepe1980} (QMC). However, to find approximations that are systematically better than the LDA has turned out to be exceedingly difficult \cite{Medvedev2017,Becke2022}. Today, in spite of the developments of successful gradient corrections and sophisticated approximations tailored by exact constraints \cite{PerdewWang1986,PBE1996,Perdew2008,Sun2015}, one may say that there is still no generally established multi-purpose approximation beyond the LDA. One of the difficulties is that it is not easy to benchmark \vxc . First, the Kohn-Sham potential is not an observable by itself, which means, there are no experimental data to compare with. Second, since \vxc\ is  the potential of an \textit{auxiliary} system, besides the density, any other observables calculated in the KS system can in principle be arbitrarily far from measured values. The prototype example for this dilemma is the KS eigenvalue band gap\cite{Perdew1982,ShamSchlueter1983,Perdew1983}. For example in the LDA, this KS gap is in general much smaller than the fundamental electron addition-removal gap  that can be extracted from direct and inverse photoemission \cite{Perdew1981,Schilfgaarde2006}. 
It would, however, be too simple to just blame the LDA. While the exact direct gap of the auxiliary KS system equals the optical excitation gap in the limit of a single electron, this is only approximately true in real molecules or materials. Moreover, the fundamental gap is in general larger than the optical gap. The exact fundamental gap could in principle be determined as a difference of total energies with varying particle number. The exact KS eigenvalue gap should be smaller than the exact fundamental gap by a constant called derivative discontinuity. Therefore, there is no reason for the eigenvalue gap of the auxiliary KS system to equal the photoemission gap of the true material\cite{DFT1,Sham1966}. However, the difference, i.e., the derivative discontinuity, is in general unknown.  Indeed,
the respective errors of the approximate functionals and of the KS system itself have been a matter of debate for many years. Results derived from many-body perturbation theory to first order in the screened Coulomb interaction \cite{Godby1986,Godby87,Godby88,Niquet2004,Gruning2006,GruningJCP2006,Kotani1998,Klimes2014,Riemelmoser2021} gave evidence  that the error in simple semiconductors is mainly due to the auxiliary nature of the KS system, i.e., due to the missing derivative discontinuity\cite{Perdew1982,ShamSchlueter1983,Perdew1983}, rather than due to the LDA for $v_{\rm xc}$. However, these are merely estimates based on perturbation theory, and the numerically exact KS potential and KS band gap of solids remain to date unknown.

More information is available in low-dimensional, often finite, systems, where ways have been proposed to invert the KS equations and find the KS potential starting from a given density \cite{PhysRevA.49.2421,Zhao1994,Wu2003,Peirs2003}. This density could be determined by analytical or numerical methods. This has given precious insight about the potential and observables in the KS system \cite{Almbladh1984,Aryasetiawan1988,Goerling1992,Knorr1992,Knorr1994,Umrigar1994,Gritsenko1995,Tozer1996,Helbig2009,Burke1D2012,Varsano2014,Hollins2016,Hodgson2016,Hodgson2017,Wetherell2019,Nam2020}. For example, in the helium atom the exact highest occupied molecular orbital (HOMO) lies about 25 eV below the vacuum level, and an additional electron is unbound. The exact KS eigenvalue gap, instead,  turns out to be only 20.3 eV, since KS binds the lowest unoccupied molecular orbital (LUMO)\cite{Umrigar1994,Savin1998,Li2019}. On top of this underestimate, the LDA reduces the HOMO-LUMO gap further, yielding 15.85 eV. The inversion of the KS system is not an easy task, though, and in particular a finite basis set may lead to drastically modified results \cite{Schipper1997,Mura1997,Heaton-Burgess2007,Jacob2011,Gaiduk2013}. Moreover, small changes in the density can yield large differences in the potential \cite{Jensen2018,Shi2021}. Altogether, a reliable inversion of the KS equations remains a difficult task even for finite systems, and while various methods have been proposed to overcome the problems, research in this direction is still ongoing \cite{Ryabinkin2012,Jensen2018,Ou2018,Kanungo2019,Kumar2019,Kumar2020,Kumar2020a,Callow2020,Nam2021,Shi2021,erhard2022,Shi2022}.

In realistic three-dimensional periodic systems, the density of beryllium obtained from x-ray diffraction experiments has been used to determine an auxiliary non-interacting system\cite{Jayatilaka1998}. However, to the best of our knowledge, no results for \vxc\ obtained directly from a numerically exact density are available. This has several reasons, including the fact that 
data for numerically exact densities of solids were not available in the literature, and that the inversion in extended three-dimensional systems may bear new technical difficulties.
Therefore,
to date a series of important fundamental questions remain to be answered, in particular: \textit{How can we adapt an inversion approach designed for finite systems to the case of solids, and which kind of precision can be obtained in solids?
How different is the resulting \vxc\ from standard approximations for solids, such as the LDA or Perdew-Burke-Ernzerhof (PBE) generalised gradient approximation\cite{PBE1996} (GGA)? What about observables in this numerically exact KS system, and in particular, the band gap?} \textit{How much does \vxc\ depend on details of the density? And if it depends significantly, do the resulting changes have an impact on other KS observables?} Starting from {nearly} numerical exact densities\footnote{Given the pseudopotential. In other words, all discussions are valid because we have used the same hamiltonian, with a fixed LDA pseudopotential, for the valence electron problem throughout, including for the QMC calculations. A full all-electron $v_{\rm xc}$ would of course look different.} for the simple semiconductor bulk silicon and insulating sodium chloride obtained  {by} the Auxiliary Field (AF) QMC  {method \cite{ Zhang2003, Motta2018}} in Ref. \cite{SiyuanChen}, in the present work we answer these long-standing questions.

\section{How to invert the KS problem in infinite systems}
The probably simplest algorithm to obtain the KS potential from a given density $n_{\rm ref}$ has been proposed for finite systems by van Leeuwen and Baerends\cite{PhysRevA.49.2421}. In its original form it was derived by solving the KS equations for the KS potential $v_{\rm KS}$. The result was then translated into an iteration procedure which relates a potential $v^{\rm i+1}$ at step $i+1$ to the potential $v^i$ at step $i$ by the ratio of the target density $n_{\rm ref}$ and the density $n^i$ at step $i$. As pointed out in \cite{Peirs2003}, the best use of this ratio depends on the sign of the potential that is updated: for example, $v$ may be either the usually negative total $v_{\rm KS}$, or its rather positive interaction part $v_H+v_{\rm xc}$ with $v_H$ the Hartree potential. In the present work we use
\begin{equation}
    \label{invertvxc}
    v_{\rm xc}^{i+1}({\bf r}) = \frac{n_{\rm ref}({\bf r})+a}{\tilde n^{i}({\bf r})+a} v_{\rm xc}^i({\bf r})\,,
\end{equation}
where $a$ is a parameter that avoids instabilities in regions of very low density as suggested in \cite{PhysRevA.49.2421},
and the mixing $\tilde n^i=\alpha n^{\rm i-1}+(1-\alpha)n^i$, with $0<\alpha<1$, is introduced to smooth the convergence. This density $\tilde n^i$ is also used to update the Hartree potential at each iteration. \refEq{invertvxc} is clearly a good strategy if $v_{\rm xc}$ is negative, and if the density { at} a point $\bfr$ is determined only by the KS potential {at} that same point. Suppose that at a given iteration $\tilde n^i(\bfr)$ is larger than $n_{\rm ref}(\bfr)$. The algorithm then decreases the absolute value of $v_{\rm xc}(\bfr)$. If the \xc\ potential is negative, this step makes the potential more shallow, and less density will be attracted to the point ${\bfr}$ in the next iteration, which pushes the solution in the good direction. Of course, it is not true that $n({\bf r})$ depends only on the KS potential in the same point ${\bf r}$, and it has to be seen to which extent the relation is nearsighted enough to make the algorithm work in a solid.

The negative sign of the  potential that is updated in \eqref{invertvxc} is crucial for the algorithm to work, because a positive sign would drive the result in the wrong direction. 
However, contrary to the HEG, 
a real system can also exhibit regions of positive $v_{\rm xc}$. Moreover, while in a low-dimensional system one can impose that the potential tends to zero at large distances, in a three-dimensional solid  the zero of the potential is not defined.
One cannot even use the ionization potential theorem, which would force the eigenvalue of the highest occupied state to be minus the ionization energy, since this theorem does not hold in extended systems\cite{Nazarov2021}. 
The arbitrary energy scale
represents both an advantage and a drawback. On the upside, it allows us to introduce a rigid negative shift such that the potential remains negative throughout the iteration. This shift is arbitrary within reasonable limits: if it is too small, positive regions may appear and become an obstacle for convergence. If it is too large, the algorithm becomes unstable, as the shift is multiplied at every step by the density ratio. Reasonable values lie within the maximum amplitude of the potential. On the downside, iteration of \eqref{invertvxc} yields $v_{\rm xc}$  only up to a constant. This is not due to our introduction of a shift, but to the fact that the density does not contain information about the absolute value of the potential\cite{DFT0}. Therefore, this limitation cannot be avoided. The resulting potential can, however, be used to calculate a well defined density and KS observables such as {the KS band structure} (besides the meaningless constant shift).  

We have tested the algorithm using known functionals. As documented in Appendix \ref{subsec:algorithm}, the inversion works straightforwardly for the LDA, but two further aspects have to be verified. On one side, the true functional, unlike the LDA, is non-local, which might influence the behavior of the algorithm. This is sorted out in Appendix \ref{subsec:algo-nonloc}. Second, the QMC data contains statistical noise; this aspect is deepened in Appendix \ref{subsec:noisy-densities}. Finally, Appendix \ref{sec:QMC-invert} shows that the inversion starting from the QMC density behaves as expected. It also demonstrates that the final results do not depend on the starting point of the iterative procedure, including starting points as far from the final result as, e.g., $0.1\times v_{\rm xc}^{\rm LDA}$. For the results shown in the following, we use as starting point $0.3\times v_{\rm xc}^{\rm LDA}$ with a rigid downwards shift of 0.2 Hartree for silicon and 0.4 Hartree for NaCl. 
In the following, we report results in atomic units for both densities (expressed in bohr$^{-3}$) and xc potentials (in Hartree).

 \section{Results}
 \subsection{ Kohn-Sham potential of silicon and sodium chloride}

\begin{figure}
     \centering
       \includegraphics[width=\linewidth]{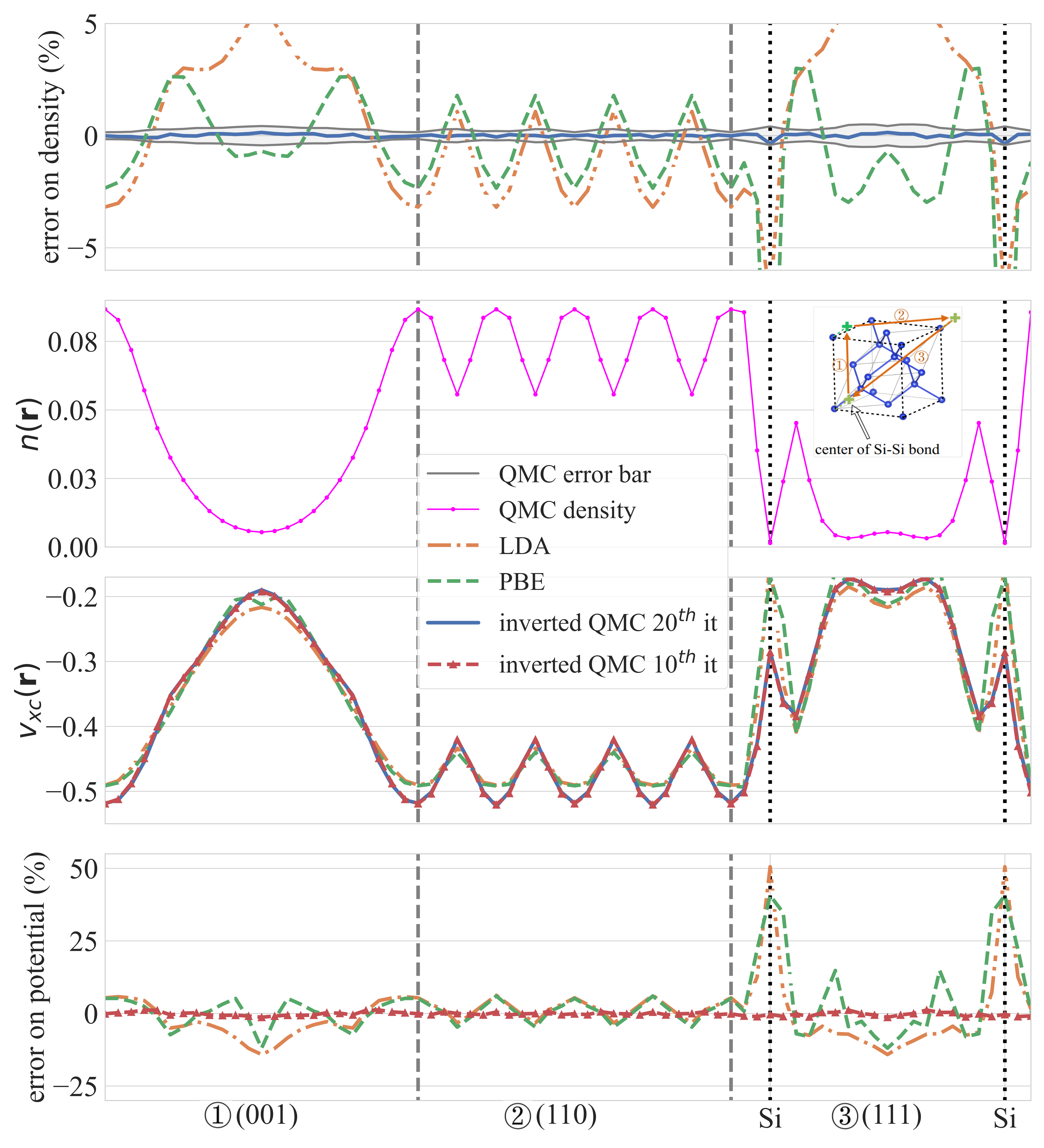}
     \caption{      {Density and xc potential of bulk silicon along the same path across the unit cell as in Ref. \cite{SiyuanChen}. The positions of atoms are indicated by dotted vertical lines.
     The iterative inversion follows} 
     \refEq{invertvxc} with the QMC  density $n_{\rm QMC}$ of silicon as reference density. 
     The potential $v_{\rm xc}^{\rm QMC,20}$ is obtained after $i=20$ iterations. The density $n^{\rm QMC,20}$ is calculated using $v_{\rm xc}^{\rm QMC,20}$ in the KS equation. The MaAPE  at $i=20$ compared to $n_{\rm QMC}$ is 0.38 \%, and the MeAPE is 0.04\%. 
     Top  panel: LPD of $n^{\rm QMC,20}$  (blue), self-consistent LDA $n_{\rm LDA}$ (orange), and PBE $n_{\rm PBE}$ (green) densities  with respect to $n_{\rm QMC}$.
     The grey area is the stochastic error bar of the QMC density.
   {Second panel: The  QMC density $n_{\rm QMC}$ (magenta line)}. {The inset shows the chosen path across the crystal from Ref. \cite{SiyuanChen}.}
   {Third} panel: 
    $v_{\rm xc}^{\rm QMC,20}$ (blue), $v_{\rm xc}^{\rm LDA}$ (orange), $v_{\rm xc}^{\rm PBE}$ (green), and $v_{\rm xc}^{\rm QMC,10}$ (red). {Note that the two QMC potentials (blue and red lines) are almost indistinguishable.}
    The average potentials are aligned. 
    Bottom panel: LPD of xc potentials with respect to $v_{\rm xc}^{\rm QMC,20}$ for LDA (orange), PBE (green), and QMC at $i=10$ {(red)}. 
}
     \label{fig:dens_xcpot_SILICON-QMC}
 \end{figure}

We have applied the algorithm to obtain the \xc\ potential from the charge density obtained by  AFQMC calculations. For silicon, we have used the results  of Ref. \cite{SiyuanChen}. For NaCl, we have applied additional symmetry operations to the density from the same Ref. \cite{SiyuanChen}. 
Ideally, from the inversion procedure for iteration $i\to \infty$ one should find $n^{\rm QMC,i}(\bfr)\to n_{\rm QMC}(\bfr)$. However, since the QMC data contains statistical noise, the inversion has a more limited precision than in the case of, e.g., clean LDA  data (see App. \ref{subsec:noisy-densities}). 
The Mean Absolute value of the Percentage Error (MeAPE) of the density of silicon $n^{\rm QMC,i}$ at iteration $i$ compared to $n_{\rm QMC}$,
$ 100\times {\rm mean}_{\bf r} \left| 1-n^{\rm QMC,i}({\bf r})/n_{\rm QMC}({\bf r})\right|
$,
does not fall below 0.02\%, while
the Maximum (over the unit cell) of the Absolute value of the Percentage Error (MaAPE) of the density,
$ 100\times {\rm max}_{\bf r} \left| 1-n^{\rm QMC,i}({\bf r})/n_{\rm QMC}({\bf r})\right|
$,  decreases to 0.38\%  at $i=20$ iterations{. This} is in any case sufficient to make significant distinctions between different densities and potentials.
The upper panel of Fig. \ref{fig:dens_xcpot_SILICON-QMC} shows the Local Percentage Difference (LPD)  of the iterative density with respect to the QMC  density after 20 iterations {(blue line)}, $100\times(n^{\rm QMC,20}({\bf r})/n_{\rm QMC}({\bf r})-1)$, along a path through the unit cell (the same as in Ref. \cite{SiyuanChen}, see the inset to the second panel of Fig. \ref{fig:dens_xcpot_SILICON-QMC}). 
The result stays well within the stochastic error bar of the QMC calculation (grey area). For comparison, we also show the LPD of the  LDA and PBE densities (dot-dashed orange and dashed green lines, respectively), with respect to the QMC.
As also shown in Ref. \cite{SiyuanChen}, differences between LDA, PBE and the QMC densities are largest on the atoms and also in other regions of low density\footnote{Note that we have defined the error with opposite sign with respect to Ref. \cite{SiyuanChen}} {(see the magenta line in the second panel of Fig. \ref{fig:dens_xcpot_SILICON-QMC})}, but they are still  significant in regions of higher density, along the (110) direction, where LDA and PBE are very similar, but differ from the QMC result. Most importantly, the differences between different densities are much larger than  the error due to the inversion of the QMC density: while the MeAPE at  $i=20$ is 0.04\%, the mean absolute relative difference between the LDA and QMC densities is  1.93 \%, and it is 1.07 \% between the PBE and QMC densities. 

The xc potentials are compared in the {third} panel of Fig. \ref{fig:dens_xcpot_SILICON-QMC}. {For this comparison, the potentials are aligned at their average value.} Our numerically determined and supposedly most accurate KS xc potential{, obtained from the QMC density,} is similar to the local and semi-local approximations. This result is stable: the QMC result obtained at $i=10$, where the MaAPE and MeAPE on the density are 0.90 \% and 0.09 \%, respectively, is almost indistinguishable from the one at $i=20$.  The differences between QMC on one side, and LDA and PBE on the other side, can be appreciated in the bottom panel, which shows the LPD of LDA and PBE with respect to the QMC xc potential obtained at $i=20$.  These differences are  similar for LDA and PBE along most of the path. 
The MeAPE with respect to the QMC result for potentials is 3.90~\% for the LDA and 3.88~\% for the PBE: of similar order, though larger, than the MeAPE of the densities.  Instead, the LPD of the QMC potential at $i=10$ with respect to the potential obtained at $i=20$ can hardly be seen. 
 We have hence reached sufficient precision on the density, which lies within the QMC error bar, and the xc potential, which shows some differences with respect to common functionals. The effect of iterating further using the noisy QMC  data is discussed in the App. \ref{subsec:noisy-densities}.

\begin{figure}
    \centering
    \includegraphics[width=\linewidth]{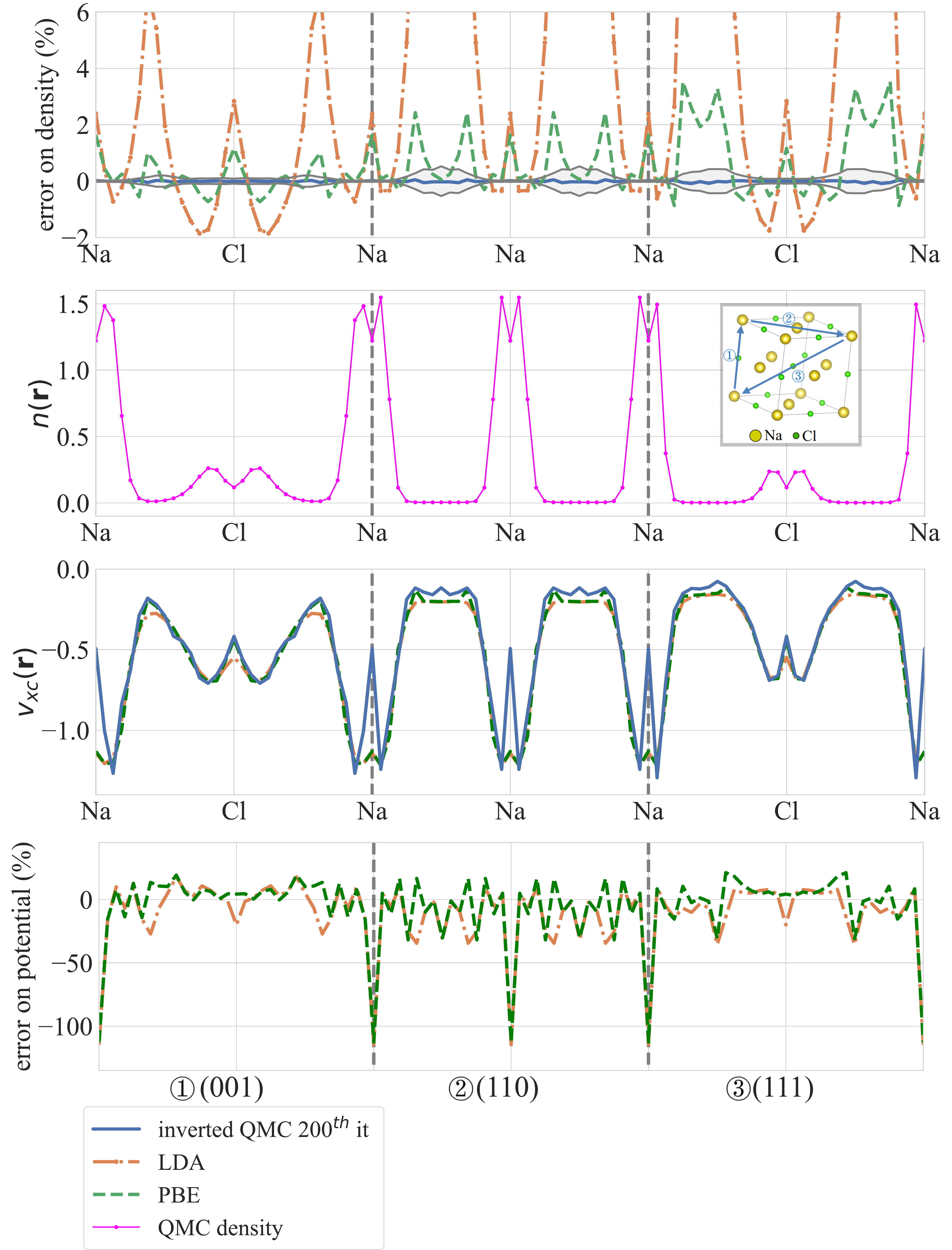}
    \caption{Density and xc potential of NaCl along the same path across the unit cell as in Ref. \cite{SiyuanChen}. The positions of atoms are indicated by dotted vertical lines. The iterative inversion follows \refEq{invertvxc} with the QMC  density $n_{\rm QMC}$ of NaCl as reference density.
        The result of the QMC inversion is shown at $i=200$. The MaAPE on the density at $i=200$ compared to $n_{\rm QMC}$ is 0.29 \%, and the MeAPE is 0.03\%. 
    Top  panel:
     LPD of $n^{\rm QMC,200}$  (blue), self-consistent LDA $n_{\rm LDA}$ (orange), and PBE $n_{\rm PBE}$ (green) densities  with respect to $n_{\rm QMC}$.
     The grey area { in the top panel indicates} the stochastic error bar of the QMC density.   {Second} panel: {The   QMC density $n_{\rm QMC}$ (thin magenta line)}.
  {The inset shows the chosen path across the crystal from Ref. \cite{SiyuanChen}.} {Third panel:   $v_{\rm xc}^{\rm QMC,200}$ (blue), $v_{\rm xc}^{\rm LDA}$ (orange) and  $v_{\rm xc}^{\rm PBE}$ (green). The averages of the potentials are aligned.}
    Bottom panel: LPD of xc potentials with respect to $v_{\rm xc}^{\rm QMC,200}$ for LDA (orange) and PBE (green).
        }
    \label{fig:invqmc_nacl}
\end{figure}

Results for sodium chloride show similar trends, with even better convergence properties of the potential due to the fact that 
our QMC density for NaCl is less noisy than the one of silicon in the important regions of high density (see also App. \ref{subsec:noisy-densities}). For the density, we obtain a MeAPE of 0.03\%   and a MaAPE of 0.29\% at $i=200$. Here, analogous to Fig. \ref{fig:dens_xcpot_SILICON-QMC} for silicon, in Fig. \ref{fig:invqmc_nacl} we show the LPD of the density and of the xc potential along a path {(see the inset to the second panel)}.  
The QMC-derived xc potential differs from the LDA and PBE especially on the sodium atoms, where the density shows rapid changes. 
At first sight, however, and as in the case of silicon, it is difficult to rationalize the differences between the three potentials. While it is exciting to see the numerically exact xc potential for real semiconductors and insulators,  it is useful to switch to a representation that highlights the essence of the difference, {in order to gain more insight.}

 \subsection{Non-local dependence of the KS \xc\ functional on the density}

\begin{figure}
    \centering
    \includegraphics[width=\linewidth]{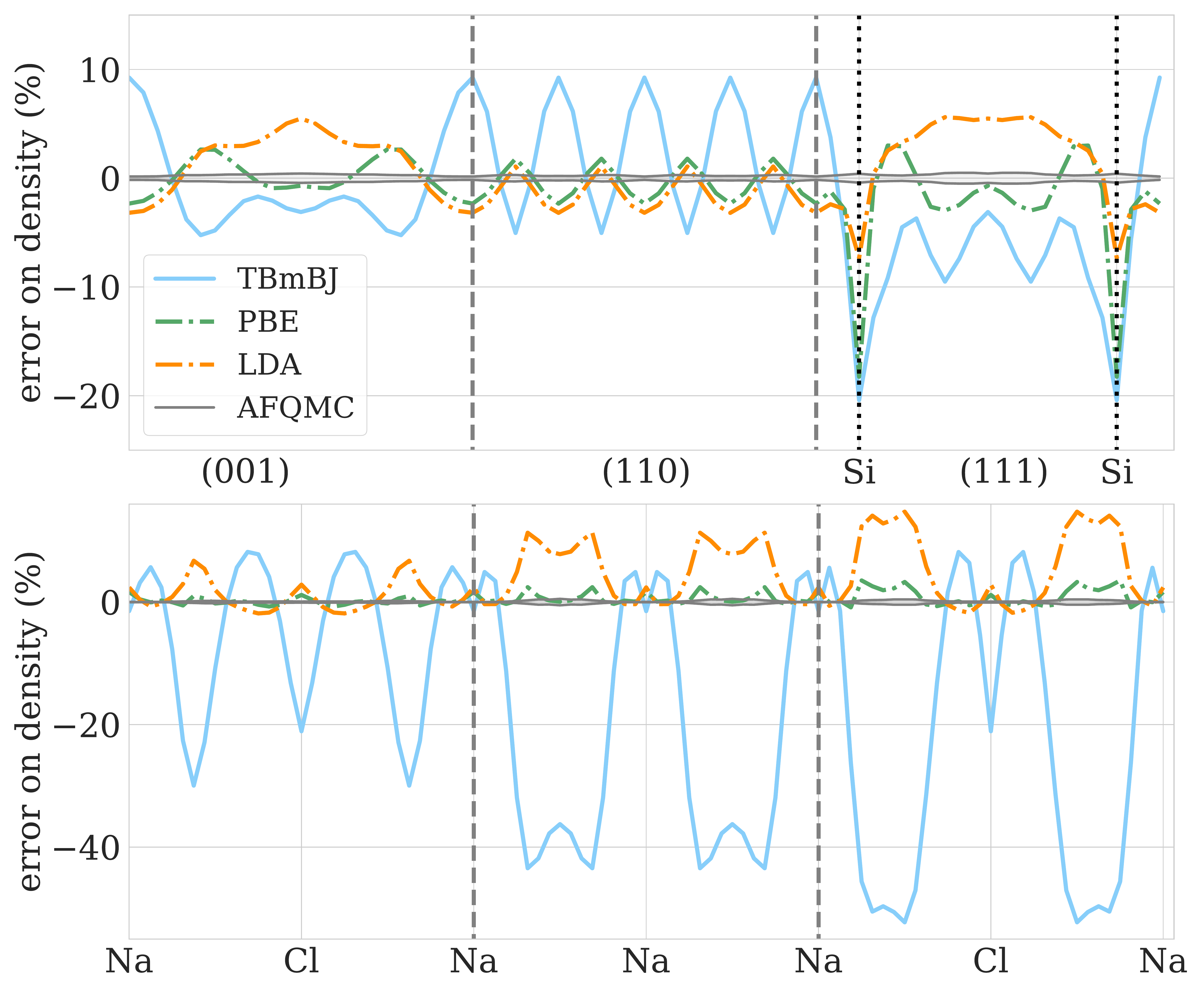}
    \caption{LPD of  self-consistent LDA $n_{\rm LDA}$ (orange), PBE $n_{\rm PBE}$ (green) and TBmBJ (light blue) densities  with respect to $n_{\rm QMC}$.
     The grey area is the stochastic error bar of the QMC density. (Top panel) Silicon. (Bottom panel) NaCl.} 
    \label{fig:dens_tbmbj}
\end{figure}

In order to appreciate the non-locality of the QMC derived potential, we will again compare it to known functionals.
It is interesting to add the Tran–Blaha modified Becke-Johnson potential\cite{TranBlaha2009} to the more common functionals LDA and PBE, since this functional is designed with a different purpose, namely, to yield an eigenvalue gap closer to the measurable fundamental gap than the exact KS potential. This requires a compromise concerning the resulting density. The deviation of the latter from the QMC one is shown in the top and bottom panels of Fig. \ref{fig:dens_tbmbj} and  for silicon and NaCl, respectively. The density errors have opposite sign with respect to those of the LDA throughout, and they are of similar order of magnitude in silicon, and about 4-5 times larger in NaCl. This raises the question how the potential and KS observables will compare. 
 
\begin{figure*}
    \centering
    \includegraphics[width=\linewidth]{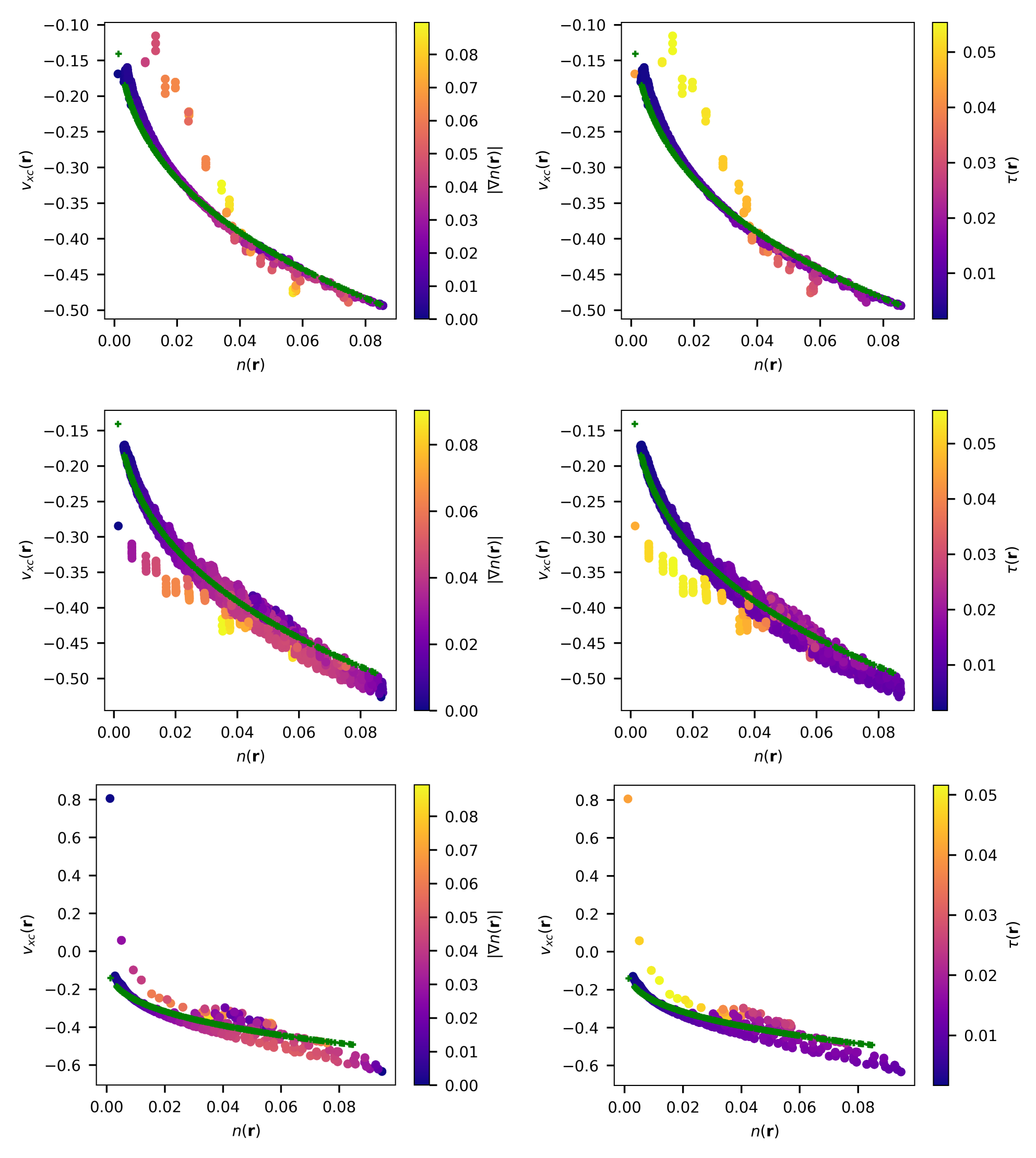}
    \caption{ Map of the xc potential of silicon with respect to the local density at all points in the unit cell. Color codes reflect the modulus of the local gradient of the density (left column) or the local kinetic energy density (right column). Upper figures are for PBE, middle figures for QMC, and bottom figures for TBmBJ. The LDA is shown in green. 
    }
    \label{fig:si_grad_tau}
\end{figure*}

\begin{figure*}
    \centering
    \includegraphics[width=\linewidth]{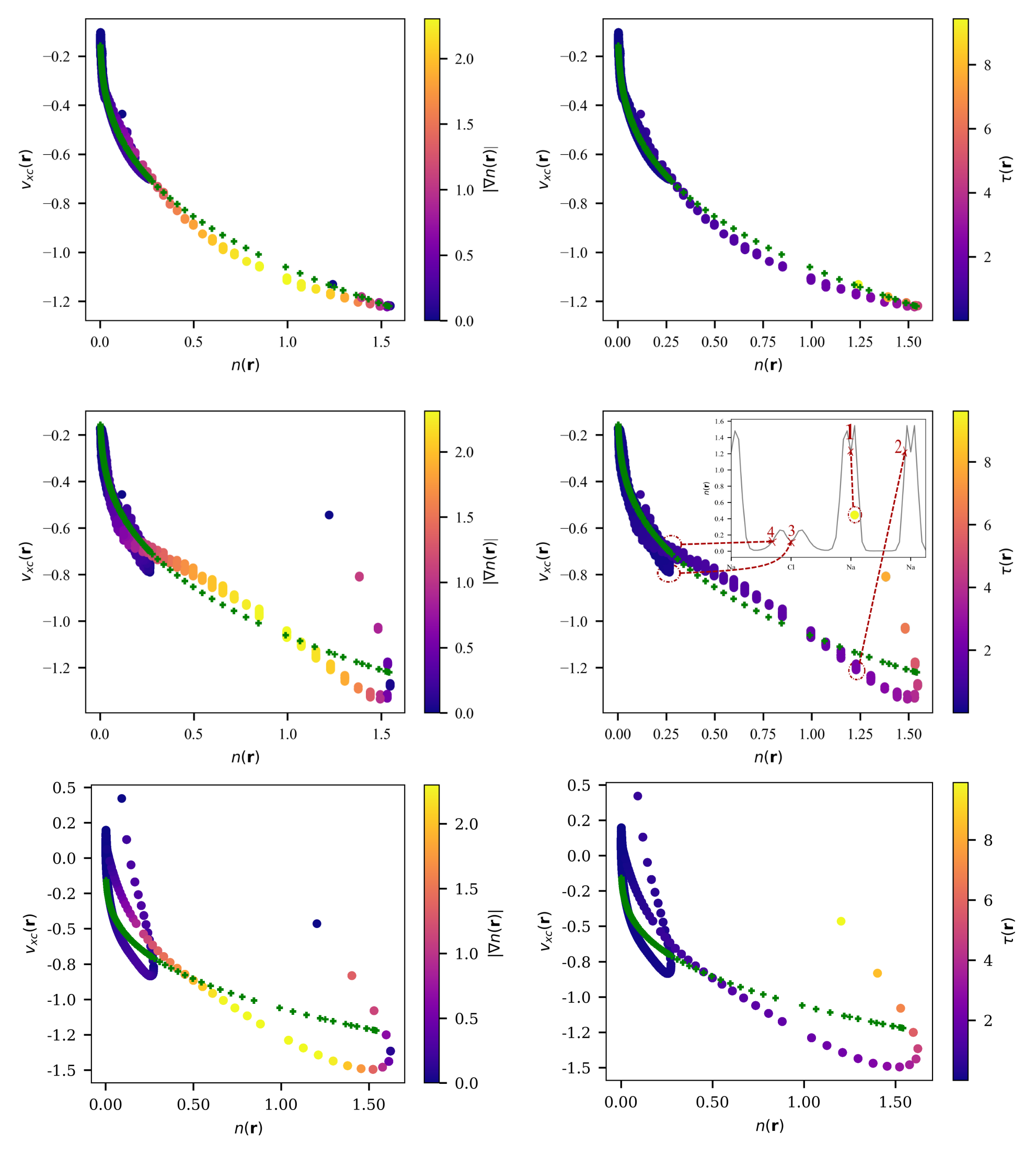}
    \caption{Same as Fig. \ref{fig:si_grad_tau}, for NaCl.  Note the much larger range for $n(\bf r)$ here.
    Upper panels are for PBE, middle panels for QMC, and bottom panels for  TBmBJ. The LDA is shown in green. The inset in the middle right panel shows the QMC density along a part of the path across the unit cell. Note that a data point in the main graph (marked by 1) falls into the inset area.
       }
    \label{fig:nacl_grad_tau}
\end{figure*}

In the LDA, \vxcr\ is a monotonic function of $n({\bfr})$.  The exact KS potential is a functional of the density everywhere, which means that { it} can take different values in different points $\bfr$ where the density, instead, is the  same. This  expresses  the  fact  that \vxcr\  depends not  just  on  the  local  density,  but  also  on  the  environment. In order to highlight {how} the true $v_{\rm xc}$ differs from a function of the local density, we create a map of $v_{\rm xc}({\bf r})$ with respect to $n({\bf r})$: for each point ${\bf r}$ in real space, we add {a point} [$v_{\rm xc}({\bf r}) \leftrightarrow n({\bf r})$] to Fig. \ref{fig:si_grad_tau} and Fig. \ref{fig:nacl_grad_tau} for silicon and NaCl, respectively. 
In the case of the LDA, this plot shows the universal function $v_{\rm xc}(n)$, the same for silicon and NaCl, which is identical to the function in the HEG. Beyond the LDA, different environments may change this function, such that it is different for different materials. Moreover, in one and the same material the presence of different environments may lead to the presence of more than one function, and finally, if the result is very sensitive to details, one might find it  difficult to detect anything like a limited number of functions. All these effects are possible signatures of the non-local dependence of $v_{\rm xc}({\bf r})$ on the density, and being able to discern them, and to characterize different environments, may give precious input for further modeling of $v_{\rm xc}$.  

The maps in Fig. \ref{fig:si_grad_tau} and \ref{fig:nacl_grad_tau} contain the results of LDA, PBE, TBmBJ and QMC. 
The universal LDA result is given for reference in all panels of Fig. \ref{fig:si_grad_tau} and \ref{fig:nacl_grad_tau}. LDA, PBE and TBmBJ potentials are directly obtained from the KS calculation; only the QMC result stems from the inversion.
In silicon, the PBE result {(upper panels of Fig. \ref{fig:si_grad_tau})} is dominated by a simple monotonic function, but it is slightly steeper than the  LDA function. Moreover, the result appears to be a little more scattered. Finally, a new branch appears at low densities. For more insight, the colors reflect, respectively, the modulus of the local gradient of the density (left panels) and the KS kinetic energy density defined as
$
    \tau(\mathbf{r}) = \tfrac{1}{2}\sum_i^{\rm occ} \rvert\nabla\phi_i(\mathbf{r})\lvert^2
$, where $\phi_i$ are KS orbitals
(right panels). The TBmBJ potential in the bottom panels is similar to the PBE, with an extra upwards branch at low densities, some blurring at higher densities and a tendency to be lower than the LDA result at high density, but the latter two features appear to be much stronger than in the PBE.  

The QMC xc potential in the middle panels is also blurred. However, this cannot be interpreted as a reliable feature of the true KS potential, since the QMC data is noisy, which may also cause blurring, as we have demonstrated in App. \ref{subsec:noisy-densities}.  
The overall shape and branches of the QMC xc potential, instead, are significant. Similarly to the PBE and TBmBJ, one can identify a dominant curve, and, with respect to the LDA, two main changes are seen: the curve is steeper than that of the LDA, and at low densities an additional branch appears. The change in slope of the main branch with respect to the LDA goes in the same direction as in the PBE and TBmBJ results, and it is  more pronounced than in the case of PBE, similarly to TBmBJ. Also the branching happens in a similar region as in the case of PBE and TBmBJ. However, the branch departs in the opposite direction. 
 
 In all cases, the extra branch is characterized by a very different gradient and kinetic energy density with respect to the main branch at the same local density. Indeed, the region in space where the potential lies on the extra branch is close to the center of the atoms, where the density varies rapidly. It should therefore be noted that it will be particularly sensitive to details of the pseudopotential. This, together with the fact that the inversion error on the density is largest on the atoms (see Fig. \ref{fig:dens_xcpot_SILICON-QMC-noise} in the Appendix), which then also influences the large density gradients in the vicinity, suggests that 
the different directions of the branches observed here would deserve 
more studies including many more QMC datasets using different pseudopotentials and including different materials,
which is beyond the scope of the present work.  The changed slope of the main branch, instead, happens over the complete range of densities and should be a feature of silicon independent of the pseudopotential and other ingredients of the calculation. 

The modifications of the different branches with respect to the LDA $v_{\rm xc}^{\rm LDA}$ may be translated in different ways, for example: $v_{\rm xc}^{\rm e}({\bf r})=F^e(n({\bf r}))v_{\rm xc}^{\rm LDA}(n({\bf r}))$ with a correction factor $F^e$ that depends on the local density and on an environment $e$, which must be characterized. Another possibility would be $v_{\rm xc}^{\rm e}({\bf r})=v_{\rm xc}^{\rm LDA}(\mathcal{F}^e(n({\bf r}))\times n({\bf r}))$. The GGAs, for example, are an attempt to characterize the environment by the local gradient of the density (see, e.g., \cite{PerdewChevary1992}). Our results motivate further search for improved approximations of the true \textit{functional} that can be expressed as \textit{functions} of a limited number of parameters, such as the local density and its gradients. 

Consistently with the fact that the QMC density for NaCl is less noisy than in the case of silicon, the map for NaCl in Fig. \ref{fig:nacl_grad_tau}  (middle panels) shows less scattering. As for silicon,
we find a main branch that corresponds to a modified LDA. Moreover, there is an additional branch at low densities and another  branch at high density, both characterized by differences in the gradient or kinetic energy density. 
{The analogous secondary branches for PBE (see upper panels) are less pronounced whereas these features are much stronger in TBmBJ}.
 The inset in the middle right panel also shows the QMC density along {part of} the path. Numbers indicate to which {locations} selected data points correspond. For example, data point 1 on the additional high-energy branch corresponds to the potential on the sodium atom, with an environment where the density is very quickly varying, which explains why the LDA completely fails. Data point 2, instead, corresponds to a place with similar density but located in a more gentle environment, although the gradient of the density is  significant. As expected, in this point we are on the main branch, which is, however, modified with respect to the LDA.

Similarly, points 3 and 4 on the chlorine atom explain the extra branch at lower density. These results show that the potential-density relation presented as maps such as the one in Figs. \ref{fig:si_grad_tau} and \ref{fig:nacl_grad_tau} may give { further} insight about the most efficient way to introduce correction factors, and about the most important features distinguishing different environments, which could eventually be combined with machine learning approaches \cite{Kalita2021}.

 \subsection{Kohn-Sham approximation to the band gap}

 It is most interesting to look at KS observables other than the density, in particular, KS eigenvalues: even though, as discussed above, these do not by themselves have direct physical meaning in an exact sense, they can still be seen as an approximation to the physical quantities \cite{Chong2002,Filippi1997}, and they are frequently used as starting points for calculations in a more appropriate framework, such as many-body perturbation theory\cite{Martin2016}.  
 In particular, the study of the KS band gap is interesting by itself, since it is a matter of a 
long-standing debate. In absence of knowledge of the exact Kohn-Sham potential, it was not possible to distinguish between the discrepancies due to approximations of the functional, and those due to the difference between the (even exact) Kohn-Sham system itself and the real material. 
 Precious hints were already given by work on model systems; for example, Knorr and Godby \cite{Knorr1992,Knorr1994} determined the xc potential by inversion from the density of a finite one-dimensional model semiconducting wire that was then extrapolated to infinite length. Most of the band gap error was shown to be due to the fact that the exact KS eigenvalue gap differs from the fundamental electron addition and removal gap, and not due to approximations. Indeed, the KS  eigenvalue gap calculated at fixed particle number disregards the derivative discontinuity of the exact xc potential upon change of particle number \cite{Perdew1982,ShamSchlueter1983,Perdew1983}. Since the numerically exact density and/or xc potential could be obtained only for very few, low-dimensional, systems, several studies used the link between the xc potential and the self-energy given by the Sham-Schl\"uter equation \cite{ShamSchlueter1983} in order to extract $v_{\rm xc}$ from the self-energy. These include work on a two-plane wave model \cite{ShamSchlueter1983,Lannoo1985},  the surface barrier for semi-infinite jellium \cite{Eguiluz1992}, finite systems \cite{Niquet2004,Hellgren2007,Hellgren2010,Bleiziffer2013}, and the study of several real semiconductors and insulators \cite{Godby1986,Godby88,Niquet2004,Gruning2006,GruningJCP2006,Kotani1998,Klimes2014,Riemelmoser2021}. These studies confirmed that the error inherent in using Kohn-Sham eigenvalues instead of true electron addition and removal energies is significant. However, the approaches used to determine the potential involved themselves approximations whose quantitative impact on the findings are not known: first, the Sham-Schl\"uter equation was linearized in all studies; second, the self-energy itself was approximated in many-body perturbation theory, mostly on the GW\cite{Hedin-GW1965} level.
With the present work, we finally do have an almost numerically exact Kohn-Sham potential at hand for real materials, and we can therefore draw definite conclusions concerning the band structure, and in particular the band gap, of standard semiconductors and insulators. 

\begin{figure}
    \centering
   \includegraphics[width=\linewidth]{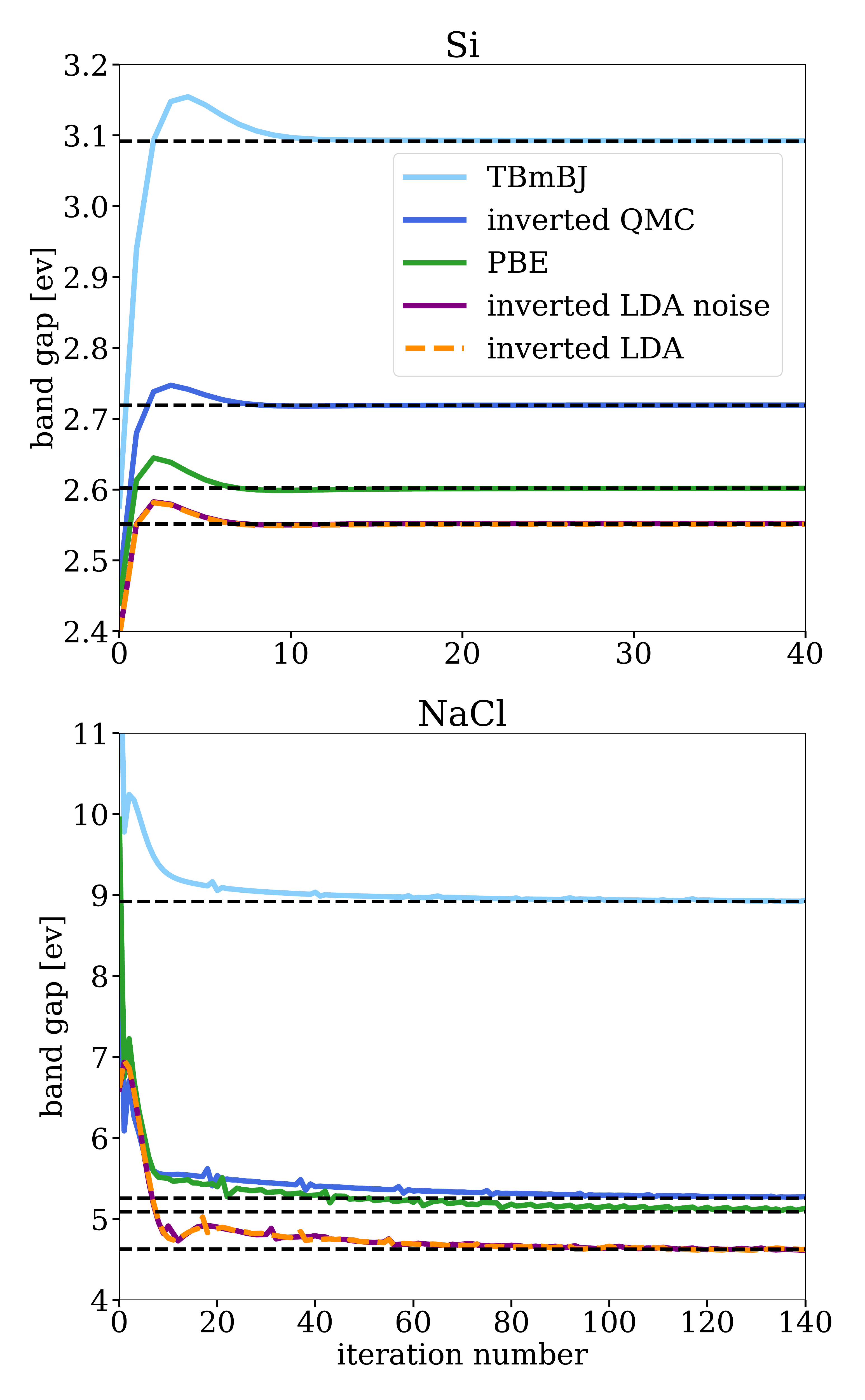}
       \caption{Convergence with the number of iterations of the direct band gap at $\Gamma$ obtained from the inversion, for silicon (upper panel) and NaCl (lower panel). Shown are the results of LDA (orange), noisy LDA (purple), PBE (green), TBmBJ (light blue) and QMC (blue). The horizontal continuous lines are the reference results, which are obtained directly from the KS calculation except for the noisy LDA where the reference is the result at $i_0=24$, and for QMC, where the reference is the result at $i_0=20$.  The clean and noisy LDA results are almost entirely overlapping and converge to the reference result of clean LDA, 2.55 eV. The QMC result converges to 2.72 eV.
       }
    \label{fig:bandgap-noisy}
\end{figure}

With the fact in mind that errors of the potential can be much larger than errors of the density, the quality of the gap resulting from inversion has to be checked separately. To this end, we show in 
 Fig. \ref{fig:bandgap-noisy} the  direct KS band gap of silicon (upper panel) and sodium chloride (lower panel) at $\Gamma$
 as a function of the number of iterations at which the KS potential and corresponding density were calculated. For all functionals, also when noise is included, the result converges very rapidly and remains stable, within 1 meV, over a wide range of iterations even after the potential has developed huge spikes (in the case of noisy density for silicon). This means that the results for the band gaps are reliable with high precision. 
 
Results for the converged band gaps of silicon and NaCl
are shown in Table \ref{tab:bandgap_si}. For silicon, our numerically exact minimum indirect KS band gap is 0.69 eV, about 40 \% larger than the KS gap of 0.49 eV calculated in LDA, and significantly smaller than the experimental gap \cite{silicon-gap} of 1.17 eV. The PBE gap of 0.66 eV is close to the QMC-derived value. The direct band gap opening of QMC with respect to LDA is analogously 0.17 eV at  $\Gamma$ and 0.12 eV at X. TBmBJ yields a direct gap of 3.09 eV, close to experiment.
In NaCl, the situation is similar, with the QMC-derived gap about 14\% larger than the LDA one, and only 3\% larger than the PBE gap.
The 5.25 eV QMC-derived KS gap is again much smaller than the 8.5 eV experimental gap \cite{nacl-gap}, while TBmBJ is again close to experiment, with a direct gap of 8.93 eV.
The QMC bandwidths do not change in a noteworthy way with respect to LDA (or PBE):
in silicon the QMC valence bandwidth is reduced by 0.1 eV compared to LDA (and 0.05 eV compared to PBE), while in NaCl the QMC bandwidth is 0.15 eV smaller than LDA and 0.04 eV larger than PBE.\footnote{Note that the difference between the exact KS band structure and the true quasiparticle one is not limited to a rigid shift of conduction states, since the KS potential is only supposed to correctly describe the highest occupied state of the N-electron and (when the derivative discontinuity is taken into account) N+1-electron systems.}

Our QMC derived KS gaps confirm the conclusion of Ref.\cite{Godby1986,Godby88} and thus definitely highlight  the fact that the true multiplicative  KS potential does  not yield a ``good'' eigenvalue band gap in solids.
Overall, the band gap is an excellent illustration for the fact that the exact Kohn-Sham system is an auxiliary system designed to yield the density in principle exactly, but for other observables, it can only give an approximation. At the same time, the TBmBJ results illustrate that by enhancing the deviations of the true potential from the LDA one can design a multiplicative potential which overcorrects the LDA errors on the density and on the KS eigenvalue gap, bringing the latter close to the experimental fundamental gap. This may yield a compromise between the accuracy of the density and the search for a simple estimate of the fundamental gap.

 \begin{table}
        \begin{tabular}{c|cc|c}
        \hline\hline
       &   \multicolumn{2}{c|}{Si} &  NaCl \\ 
       & indirect & \ \ \ direct at $\Gamma$  &  direct at $\Gamma$\\
      \hline 
      QMC derived& 0.69 & 2.72&5.25\\ 
      PBE & 0.66 & 2.60 &5.08\\
      LDA & 0.49 &2.55 &4.59\\
      TBmBJ & 1.19& 3.09 & 8.93\\
            \hline
      Exp. & 1.17 \cite{silicon-gap} & 3.05\cite{Ortega1993} & 8.5\cite{nacl-gap} \\
       & & 3.40\cite{silicon-gap} & \\
      \hline\hline
 \end{tabular}
       \caption{KS minimum band gaps and direct band gaps at $\Gamma$ (eV) in comparison with experimental photoemission gaps from Refs. \cite{silicon-gap,Ortega1993,nacl-gap}. 
       }    \label{tab:bandgap_si}
 \end{table}

There is another interesting aspect to this study of the band gap: while in certain cases (in particular, the noisy LDA) the potential can develop huge spikes during the iterations, the gap, similarly to the density, remains close to the reference value. This means that very different xc potentials can yield not only very similar densities, but also very similar KS observables more in general. Fig. \ref{fig:bandgap-noisy} also shows that the gaps corresponding to clean and noisy LDA densities are almost indistinguishable, i.e., the noise does not affect KS observables. The band gap results confirm the statement, mostly based on findings from low-dimensional systems, that examining the xc potential alone is not sufficiently meaningful\cite{Savin2003,Kim2013,Wasserman2017}. 
Our study also suggests that an effort is needed to distinguish in the KS potential crucial features, which must be contained in good functionals, from others that may be quantitatively strong in the potential, but insignificant for its effects.
 
\section{Conclusion}
In conclusion, we have shown that a simple algorithm allows one to obtain the Kohn-Sham xc potential for periodic semiconductors and insulators, given their ground state density. The precision that can be obtained is limited by the quality of the input data. Here, we use densities taken from AFQMC calculations, and the limiting factor is the stochastic noise. Nevertheless, meaningful results are obtained, with an error bar smaller than the difference between the resulting potentials and their LDA  PBE, or TBmBJ counterparts, which allows us to safely draw conclusions. In particular, for the materials studied here, namely bulk silicon and NaCl, the xc potential \textit{functional of the density everywhere} can be represented in terms of two or three \textit{functions of the local density}, each of which is determined by a specific environment. These environments appear to be characterized by the local gradient of the density or, even more clearly, by the local kinetic energy density.
The function that represents most of the data points is close to the LDA, but with slight material-dependent deviations. PBE also predicts deviations and the existence of the additional functions, although it does not always describe them well. On the other hand, our results clearly illustrate that very different potentials may lead to very similar densities, and more generally, to very similar KS observables. In particular, the KS band gap converges rapidly with the number of iterations of the inversion process, while the xc potential still undergoes violent modifications. More work is needed to discern important features of the xc potential from those that do not influence KS observables; sum rules and other exact constraints may be helpful for this \cite{PerdewWang1986,PBE1996,Perdew2008,Sun2015}. Our results for the KS band gap confirm previous conjectures based on model systems and/or many-body perturbation theory, which predict that the exact KS band gap is closer to the LDA one than to the measurable electron addition and removal gap, in other words, that the derivative discontinuity of the true xc potential is sizable. Still, the LDA error is non negligible, whereas PBE predicts the exact KS gap with an error of less than 5\% for the materials studied here. Our work highlights directions for the improvement of density functionals, stressing the need for, and usefulness of, QMC calculations of the density in many more materials.

\section{Acknowledgements}
{Fruitful discussions with Kieron Burke and Rex Godby are acknowledged. S.C. was supported by the U.S. Department of Energy (DOE) under Grant No. DE-SC0001303, and by the Center for Computational Quantum Physics, Flatiron Institute. The Flatiron Institute is a division of the Simons Foundation.}

\appendix
\section{Computational details}

We have adopted the same computational parameters (lattice constants, cutoff energies and k-point grids) and pseudopotentials as in Ref. \cite{SiyuanChen}.
Following Ref. \cite{SiyuanChen}, all the LDA, PBE, TBmBJ and QMC results have been obtained with the same optimized norm-conserving LDA pseudopotentials\cite{Hamann2013}.
We have employed the Abinit code\cite{abinit2020} and verified that it gives the same numerical results as the Quantum Espresso code\cite{quantum-espresso} used in Ref. \cite{SiyuanChen} for LDA and PBE. For the calculations with the TBmBJ functional, composed of a modified version of the Becke-Johnson
exchange potential\cite{Becke2006} and a LDA correlation part, we have used the Abinit implementation\cite{Waroquiers2013} with the libxc library\cite{Marques2012}.
The KS inversion algorithm has been implemented in our own KS python code, which is interfaced with the Abinit code.

\section{Accuracy of the algorithm: inverting the LDA}
\label{subsec:algorithm}

\begin{figure}
  \centering
    \includegraphics[width=\linewidth]{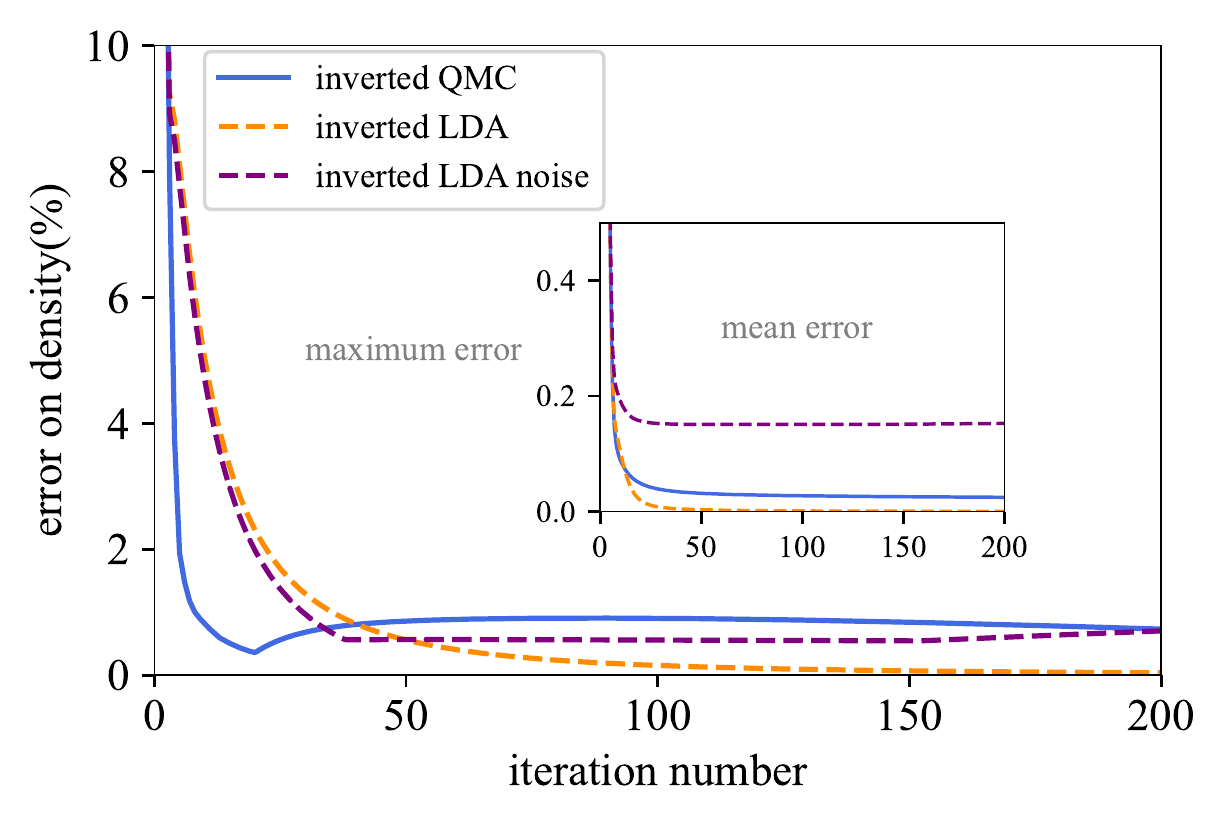}
    \includegraphics[width=\linewidth]{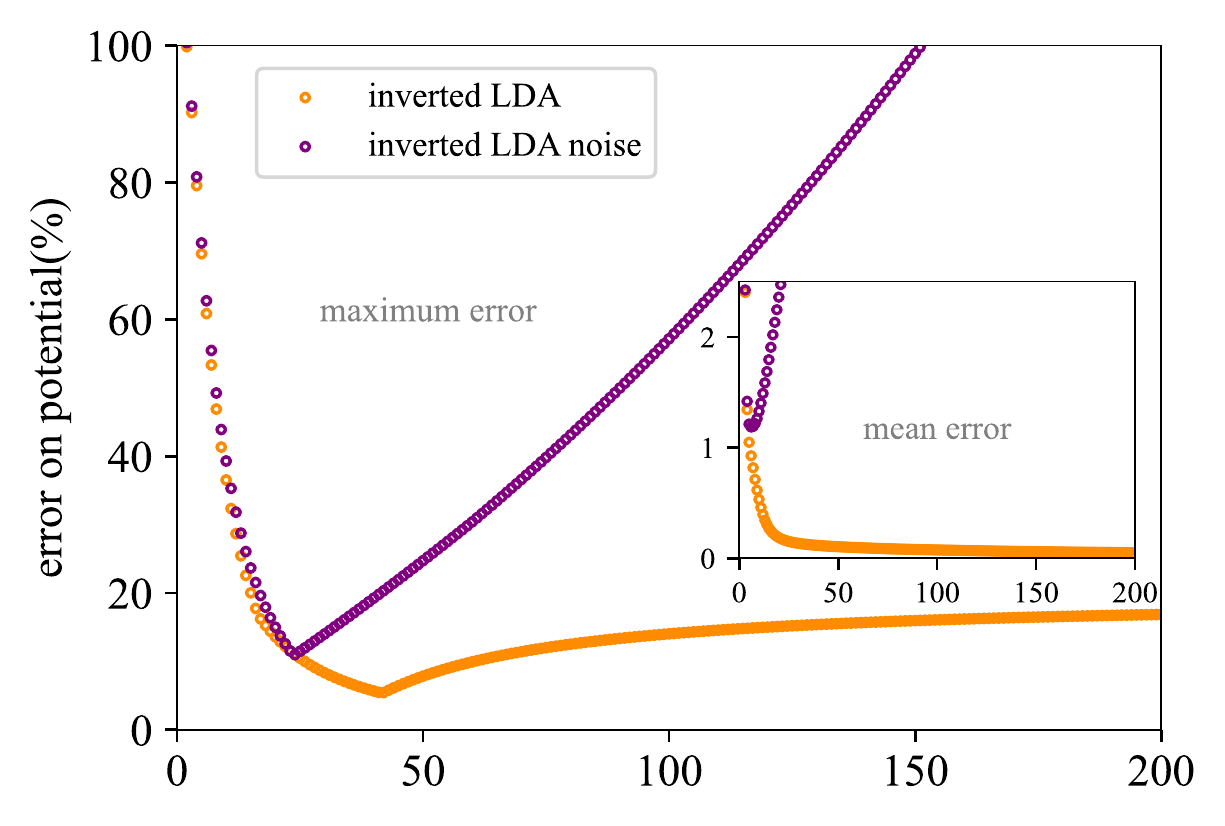}
    \caption{
    { {Errors} of the iteration procedure as a function of number of iterations $i$ {in silicon}. Top panel:  MaAPE of the density  for the inverted LDA {(orange)}, inverted noisy LDA {(purple)} and QMC {(blue)} xc potentials. Each inverted density is compared to its corresponding reference result. In the inset: MeAPE of the density  for the inverted LDA, noisy LDA and QMC xc potentials. Bottom panel:  MaAPE of the xc potential for the inverted LDA {(orange)} and inverted noisy LDA {(purple)}. In these cases the error is defined with respect to  $v_{\rm xc}$ of KS LDA, since the xc potential that yields the noisy density is unknown. In the inset: MeAPE of the xc potential for the inverted LDA and inverted noisy LDA potentials.}
    }
    \label{fig:error_lda_pbd_qmc}
\end{figure}

\begin{figure}
     \centering
    \includegraphics[width=\linewidth]{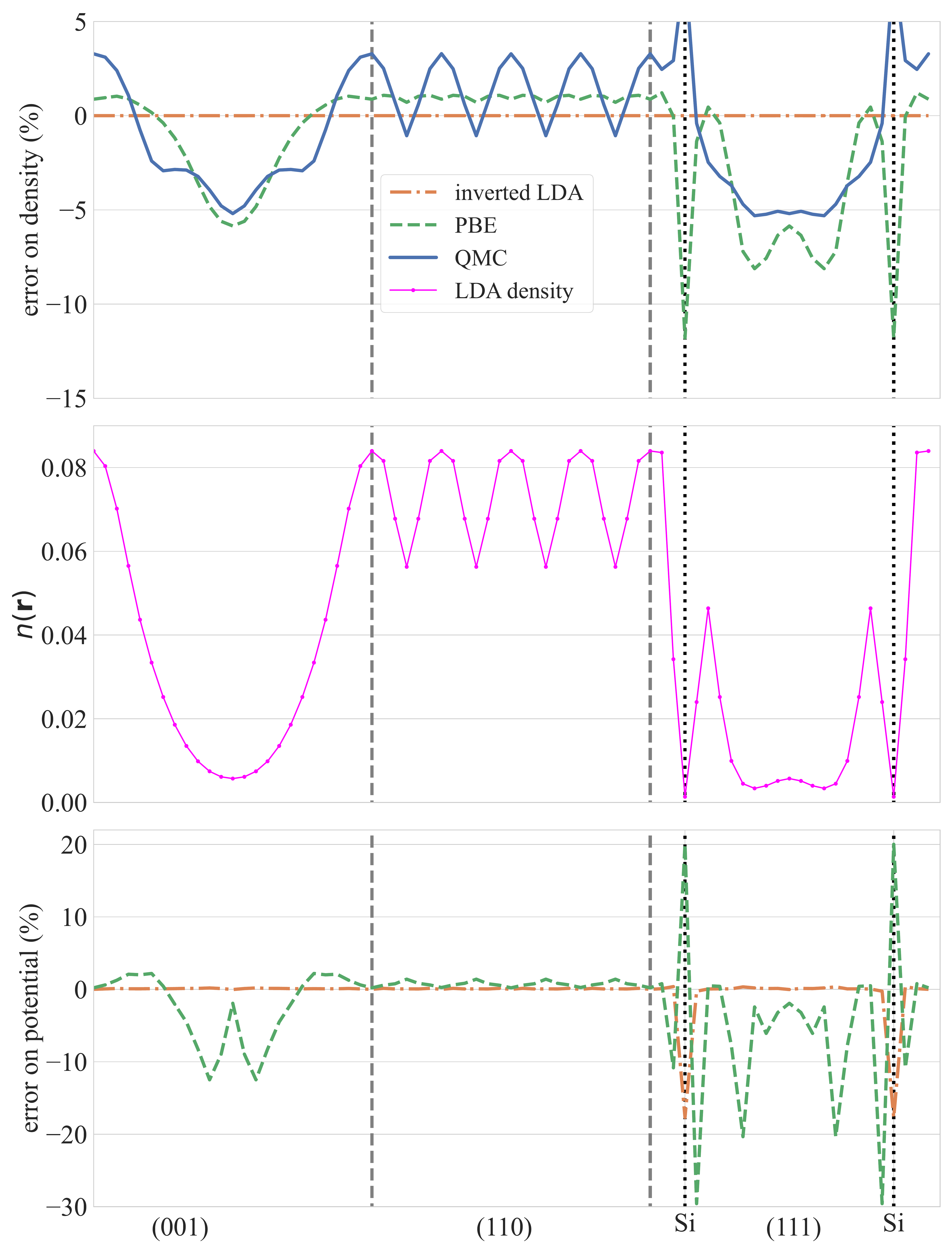}
     \caption{      Upper  panel:  The LPD of the density: in orange,
     LPD of $n^{\rm LDA,500}$  with respect to $n_{\rm LDA}$ (where the MaAPE is $6.55 \times 10^{-4} \%$); 
     in green LPD of the self-consistent PBE  and in blue QMC densities with respect to $n_{\rm LDA}$. {Middle panel: Self-consistent LDA density $n_{\rm LDA}$ (thin magenta line).}
     Lower panel:  LPD of $v_{\rm xc}^{\rm LDA,500}$ (orange) and  self-consistent PBE potential $v_{\rm xc}^{\rm PBE}$ (green) with respect to $v_{\rm xc}^{\rm LDA}$. 
     } 
     \label{fig:check-LDA}
 \end{figure}

To illustrate the reliability of the inversion algorithm, it is instructive to examine a case where the density-potential relation is well known; as a start, we choose the LDA \footnote{One might think that inverting the LDA is trivial, since the LDA xc potential is a monotonic  function of the density, and the density is therefore a function (and not a non-local functional) of the LDA xc potential. This might suggest that our algorithm should obviously work well for the LDA. However, one has to consider that at the start of the iteration procedure the xc potential is not the LDA one, so the density-potential relation is not local. Therefore, the algorithm could in principle be non-convergent or lead to a wrong solution, even for the LDA.}. This means that in Eq. \eqref{invertvxc}, $n_{\rm ref}=n_{\rm LDA}$ is the density obtained in a standard LDA Kohn-Sham self-consistent calculation with \xc\ potential $v_{\rm xc}^{\rm LDA}$ at convergence. Ideally, from the iteration procedure for $i\to \infty$ we should find $v_{\rm xc}^{\rm LDA,i}\to v_{\rm xc}^{\rm LDA}$ and $n^{\rm LDA,i}\to n_{\rm LDA}$. 
Fig. \ref{fig:error_lda_pbd_qmc} and \ref{fig:check-LDA} show results for silicon. 

In Fig. \ref{fig:error_lda_pbd_qmc} (upper panel) the  Maximum (over the unit cell) of the Absolute value of the Percentage Error (MaAPE) of the density $n^{\rm LDA,i}$ compared to the  LDA one, 
$ 100\times {\rm max}_{\bf r} \left| 1-n^{\rm LDA,i}({\bf r})/n_{\rm LDA}({\bf r})\right|
$, is shown as a function of the iteration number $i$.
It decreases smoothly and very fast. The same is true for the Mean Absolute (over the unit cell) Percentage Error (MeAPE), given in the inset.

In Fig. \ref{fig:check-LDA} snapshots for the errors on density and $v_{\rm xc}$ are presented. 
The upper panel gives $100\times \left(n^{\rm LDA,i}({\bf r})/n_{\rm LDA}({\bf r}) -1 \right)
$, the Local Percentage Difference (LPD) along a path through the unit cell (the same as in Ref. \cite{SiyuanChen}) 
of the density at $i=500$ iterations {(orange dashed line)} with respect to the LDA one.   In the scale of the figure, it is close to zero everywhere: it is largest, with a maximum of 6.55 $\times 10^{-4}$  \%, in places  of low LDA density, shown by the thin magenta line in the middle panel. Note that the density is close to zero on the silicon atoms.
The LPD of the potential is shown in the bottom panel {of Fig. \ref{fig:check-LDA}}.
Because of the arbitrary shift, only a comparison of the variations of the potential is pertinent. Indeed, during the iterations the average potential continuously moves upwards. The figure has been obtained by re-aligning at the end of the iterations the average potentials. This requires a final downwards shift of the iterative potential of 0.06 Hartree.  Again, the maximum error is found in places of lower density. The only significant error is found in the point exactly on the silicon atoms, where the density is almost vanishing. As we will also see below for other functionals, the result in this specific point cannot be considered to be reliable. Also for other regions, however, it is true that the LPD  is significantly larger for the potential than for the density. This can also be appreciated in Fig. \ref{fig:error_lda_pbd_qmc} {(bottom panel)}, where the open circles in the main panel give the MaAPE and MeAPE on the xc potential as a function of iterations.  
In order to illustrate that the remaining errors are small enough to make discussions meaningful, the errors in Fig. \ref{fig:check-LDA} are compared to the difference between two different functionals obtained directly from the KS calculation, here, between LDA and PBE (see green lines). The top panel contains the LPD of the PBE density with respect to the LDA self-consistent one, $100\times (n_{\rm PBE}({\bf r})/n_{\rm LDA}({\bf r})-1)$. This difference can be seen very clearly, since it is more than $10^4$ times larger than the inversion error.  The LPD  of the PBE with respect to the LDA xc potential, $100\times(v_{\rm xc}^{\rm PBE}({\bf r})/v_{\rm xc}^{\rm LDA}({\bf r})-1)$, can be found in the bottom panel. Differences can be seen throughout the path, although regions of lower density show larger differences. These differences are, though slightly larger, of the same order as the differences in the density. Except for immediate vicinity of
the atoms, they are  much larger than the error of the inversion{, like in the case of the density}. This demonstrates that the inversion yields meaningful results, with an error bar that is much smaller than the differences of interest, except for few points of very small density.  

\section{Inversion of non-local functionals}
\label{subsec:algo-nonloc}

\begin{figure}
    \centering
\includegraphics[width=\linewidth]{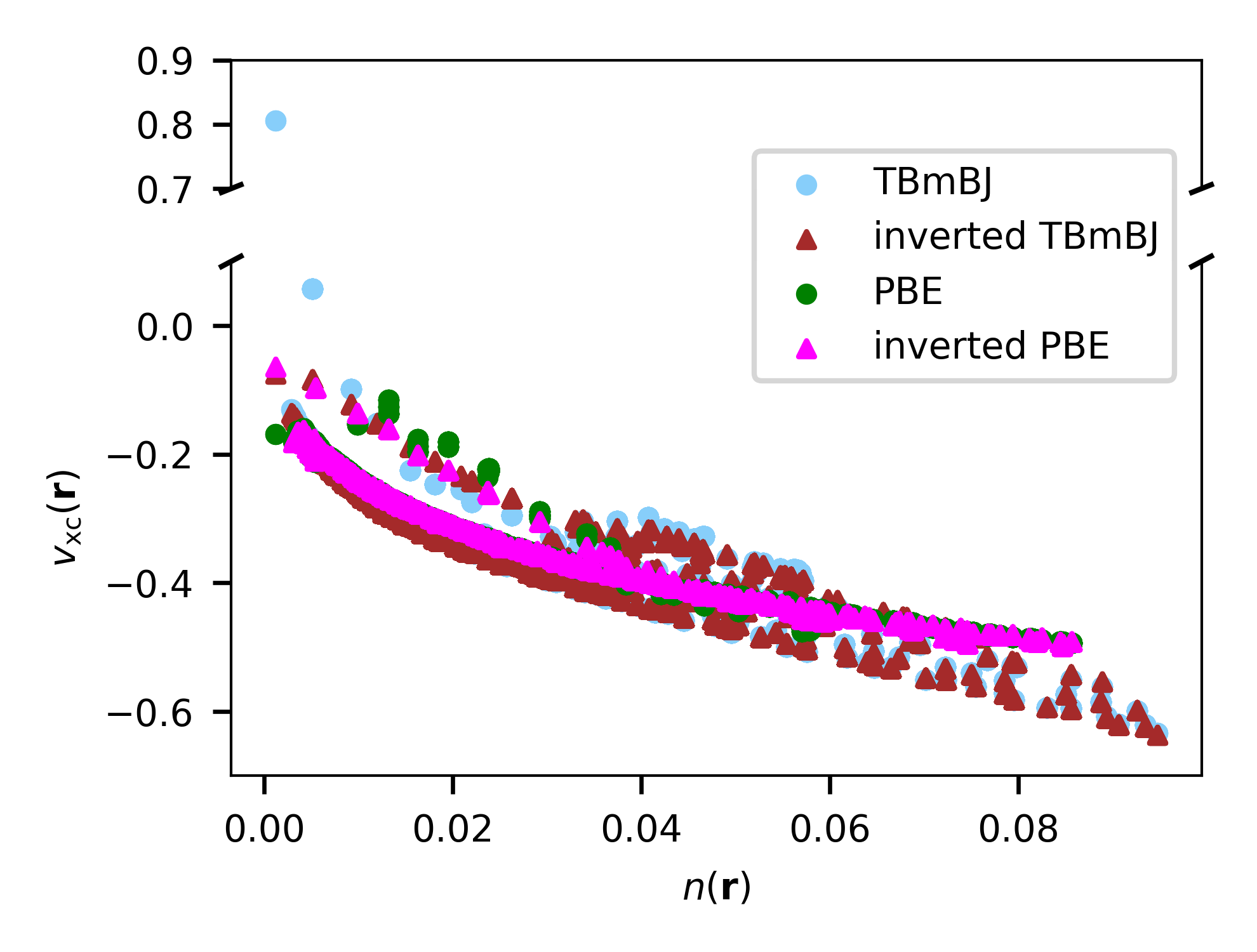}
    \caption{Map of the xc potential of silicon with respect to the local density at all points in the unit cell. The results of the KS cycle (circles) are compared to the result of inversion (triangles) for PBE (green) and TBmBJ (light blue).}
    \label{fig:inversion-general-silicon}
\end{figure}

\begin{figure}
    \centering
    \includegraphics[width=\linewidth]{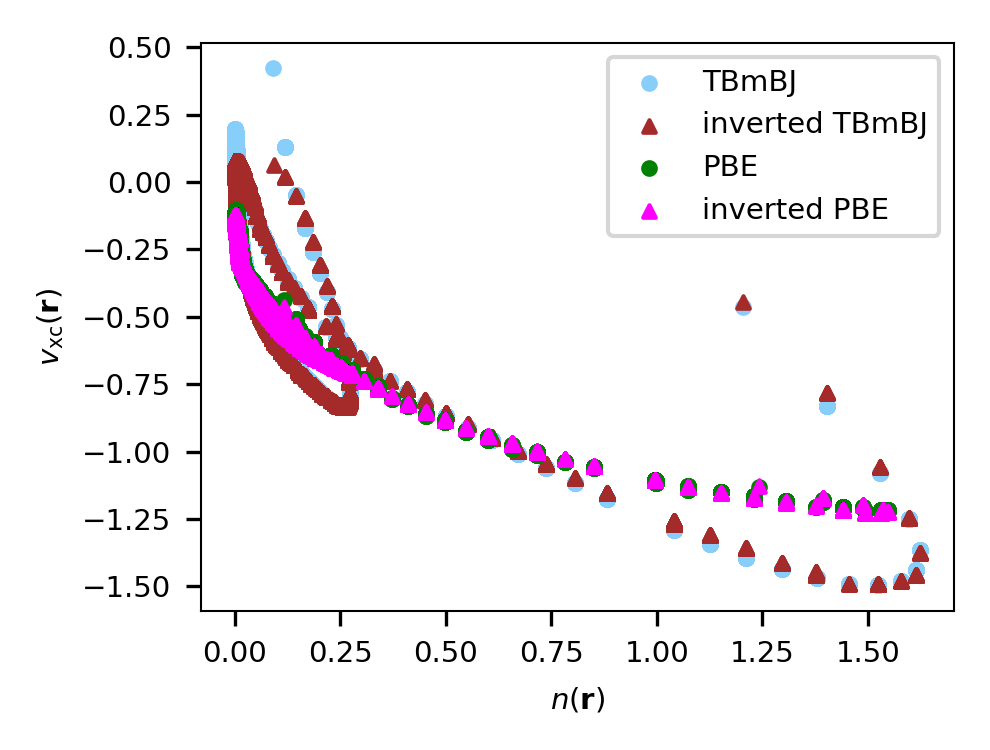}
    \caption{Map of the xc potential of NaCl with respect to the local density at all points in the unit cell. The results of the KS cycle (circles) are compared to the result of inversion (triangles) for PBE (green) and TBmBJ (light blue). }
    \label{fig:inversion-general-nacl}
\end{figure}

The true xc functionl is expected to be non-local, and testing the LDA alone is therefore not sufficient. For a conclusive test, we should invert functionals with a similar degree of non-locality as the QMC results. Figs. \ref{fig:si_grad_tau} and \ref{fig:nacl_grad_tau} show that the deviation from a local potential-density relation of the QMC-derived potential is larger than for the PBE and smaller than for the TBmBJ functional. This means that PBE and TBmBJ cover the range of non-locality of the true KS potential. We have therefore also performed the inversion tests with these two functionals, with the results given in Figs. \ref{fig:inversion-general-silicon} and \ref{fig:inversion-general-nacl}. 

As one can see, except for the points of lowest density the agreement between the reference result and the result of the inversion is again very good, and definitely accurate enough to distinguish the main features of the potentials. We conclude that non-locality in the range of that of the true KS potential does not hamper the use of our inversion algorithm.

\section{Inversion starting from noisy densities}
\label{subsec:noisy-densities}

 \begin{figure}
     \centering
       \includegraphics[width=\linewidth]{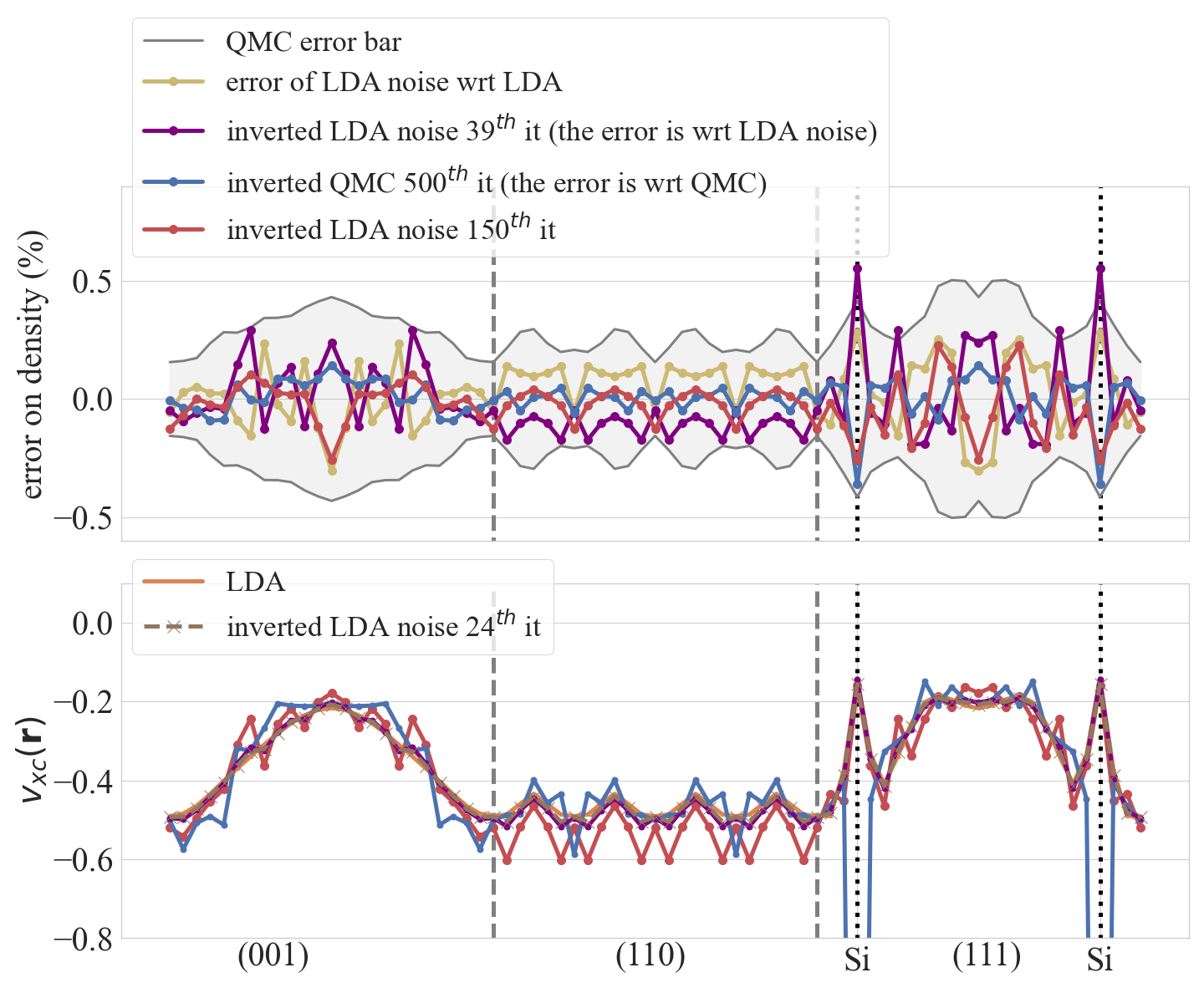}
     \caption{    Relation between noise and errors of the iterative procedures that use  $n_{\rm QMC}$ or  $n_{\rm LDA+{\rm noise}}$ as reference densities.
     Top panel:  
     In yellow, LPD of the LDA density from the KS calculation decorated with a point-wise Gaussian noise $n_{\rm LDA+{\rm noise}}$ with respect to the clean LDA density $n_{\rm LDA}$.
     The noise lies within the stochastic error bar of the QMC calculation (grey area).
      In purple, LPD of $n^{\rm LDA+{\rm noise},39}$ at the iteration  $i=39$, where the 
      MaAPE on the density with respect to $n_{\rm LDA+{\rm noise}}$ has its first minimum. Here, the MaAPE is 0.56\%, and the MeAPE is 0.15\%.  
     In red, LPD of $n^{\rm LDA+{\rm noise},150}$ at the iteration  $i=150$, where the MaAPE is 0.54\%.
   In blue, LPD of $n^{\rm QMC,500}$  at the iteration  $i=500$, where the   MaAPE is 0.21\% and the MeAPE is 0.02\%, with respect to  $n_{\rm QMC}$. 
   Bottom panel: the LDA $v_{\rm xc}^{\rm LDA}$ potential (orange) is compared to the xc potentials obtained by inversion of the QMC density $v_{\rm xc}^{\rm QMC,500}$ (blue),  and by inversion of the noisy LDA density
$v_{\rm xc}^{\rm LDA+{\rm noise},24}$ (brown dashed line), $v_{\rm xc}^{\rm LDA+{\rm noise},39}$ (purple), and   $v_{\rm xc}^{\rm LDA+{\rm noise},150}$ (red) at $i=150$.
            }
     \label{fig:dens_xcpot_SILICON-QMC-noise}
 \end{figure}

To conclude on the reliability of the results, we also have to take into account the fact that the QMC results have statistical errors. In order to elucidate the influence of  the QMC stochastic noise on the results,
 we have taken the LDA density $n({\bf r})$ and added point-wise Gaussian noise, which is obtained from a normal distribution with mean zero and standard deviation given by the characteristic AFQMC statistical error bar scaled by $n({\bf r})$.
 To appreciate what this means, the yellow curve in the upper panel gives the relative difference of the noisy LDA density with respect to the clean one.
 Fig. \ref{fig:dens_xcpot_SILICON-QMC-noise} shows results for the inversion starting from this new reference density.  The error of the inversion of the noisy LDA data is displayed in Fig. \ref{fig:error_lda_pbd_qmc} {(upper panel,} purple curve). It behaves similarly to the QMC inversion error: the MaAPE decreases rapidly and reaches a minimum, from whereon a slight increase followed by a decrease is noted. The MeAPE, instead, reaches a plateau. The inversion error on the density is given by the purple curve in the upper panel of Fig. \ref{fig:dens_xcpot_SILICON-QMC-noise}, representing the LPD $100\times(n^{\rm LDA+{\rm noise,39}}({\bf r})/n^{\rm LDA+{\rm noise}}({\bf r})-1)$ at 39 iterations.
It is of similar magnitude as the noise itself, as in the QMC case. Iterating further to $i=150$ (red), only a slight smoothing of the error on the density is observed.
 The bottom panel { of Fig. \ref{fig:dens_xcpot_SILICON-QMC-noise}} shows xc potentials: the red curve is the xc potential resulting from inversion of the noisy LDA data at $i=150$. It has spikes that are of the same order as those of $v_{\rm xc}^{\rm QMC,500}$ (blue curve) and that are in percentage orders of magnitude larger than the noise of the density, again as in the case of $v_{\rm xc}^{\rm QMC,500}$. With such an error bar, one would not be able to distinguish the  LDA and QMC potentials. 
 By  way of contrast, the xc potential resulting from the noisy LDA data but at only $i=39$ iterations, where the MaAPE on the density has its minimum, shows only very small spikes (purple curve).
  The result is stable in the range of iterations preceding that minimum: the bottom panel also shows the result for $i=24$ (brown dashed curve), with a virtually indistinguishable potential. Moreover, this potential is close to the clean LDA potential, given by the orange curve. In this range of iterations, 
 we can consider the resulting potential to be reliable. The spikes that develop by iterating further, instead, may suffer from the fact that the noisy density and the KS LDA hamiltonian are not completely consistent, which means that a higher precision cannot be reached. 
 
 The observations concerning the behavior of the noisy LDA are strictly analogous to our QMC-based results, as we will also discuss in the next section. This gives strong evidence for the fact that the inversion problem of the QMC data after a certain number of iterations is indeed due to the stochastic noise of the QMC. Moreover, it suggests that a sufficiently reliable xc potential is obtained by taking the result in the range where a stable and relatively smooth potential is obtained, and before the MeAPE on the density stops to decrease. In the present case, this confirms the choice $i=20$, for which the QMC xc potential is given in Fig. \ref{fig:dens_xcpot_SILICON-QMC}.  
In other words, this xc potential is, to the best of our knowledge, today's most precise estimate for the true xc potential of bulk silicon.

 \begin{figure}[ht]
    \centering
    \includegraphics[width=1\linewidth]{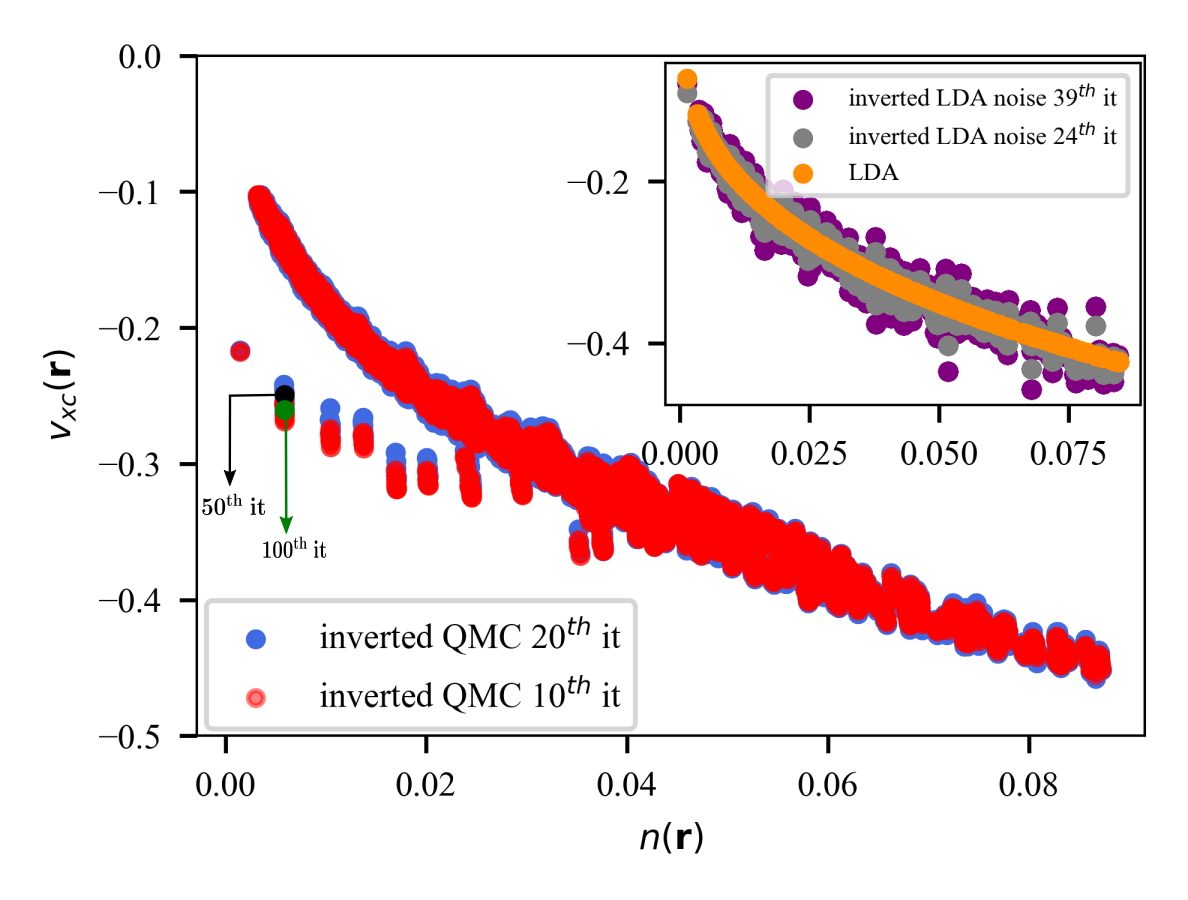}
    \caption{Map of the xc potential versus the local density in silicon. Main panel: Results of the inversion of QMC for different $i$.  Each $v_{\rm xc}$ is plotted against its own density. 
     QMC inversion results are shown at 10 (smooth potential) and 20 (minimum of MaAPE of the density) iterations. For one density below 0.01, additional results at $i=50$ and $i=100$ are given, to illustrate the stability of the extra branch.
     In the inset, KS LDA is compared to the inversion of noisy LDA at 24 (smooth potential) and 39  (minimum of MaAPE of the density) iterations. 
    }
    \label{fig:convergence_branch_si}
\end{figure}

Finally,  Fig. \ref{fig:convergence_branch_si} confirms that the noise is at least partially responsible for the blurring of the QMC result observed in the map [$v_{\rm xc}({\bf r}) \leftrightarrow n({\bf r})$] in Fig. \ref{fig:si_grad_tau}: 
inversion of the noisy LDA data leads to a more scattered potential, as shown in the inset of Fig. \ref{fig:convergence_branch_si}. The comparison of the result for $i=24$ and $i=39$ also shows that it remains essentially a scattered version of the clean LDA result, with a blurring according to the number of iterations, whereas no additional features are caused by the noise.
In the main panel, we compare different iterations of the QMC result. Also in this case higher $i$ leads to stronger blurring, but the extra branch is confirmed to be a stable feature. Note that the point at lowest density on the extra branch undergoes large oscillations with the number of iterations. All other points, instead, are well behaved, as it is illustrated on the figure for the point of second-lowest density.


\begin{figure}[h]
    \centering
    \includegraphics[width=\linewidth]{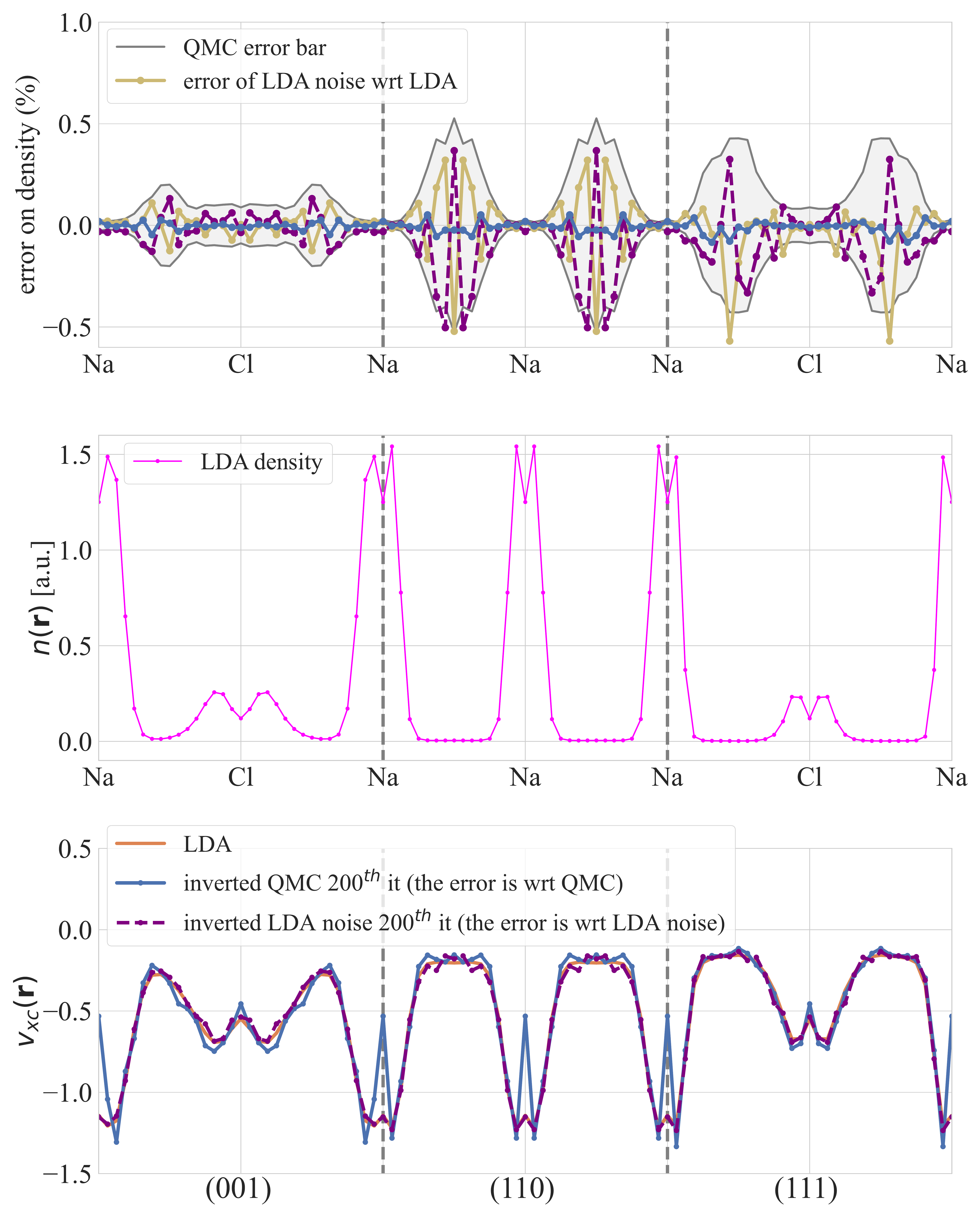}
    \caption{ Analog of Fig. \ref{fig:dens_xcpot_SILICON-QMC-noise}, but for NaCl. Here spikes in noisy LDA and QMC are less developed than in the case of silicon.  Note that the QMC error bar is smaller than in the case of silicon, in particular in regions of significant density.   
    }
    \label{fig:lda_GN_nacl}
\end{figure}

  For completeness, we investigate the issue of noise also in NaCl. 
 {Details are displayed in Fig. \ref{fig:lda_GN_nacl}.
As one can see, 
 the QMC error bar is smaller in this case, especially in the important region of high density. Therefore, although qualitatively the
 noise has similar effects as in the case of silicon, quantitatively the effect on the results is negligible, as one can see in the local potential-density relation displayed in Fig. \ref{fig:nacl-noise}. 
\begin{figure}[h]
    \centering
    \includegraphics[width=\linewidth]{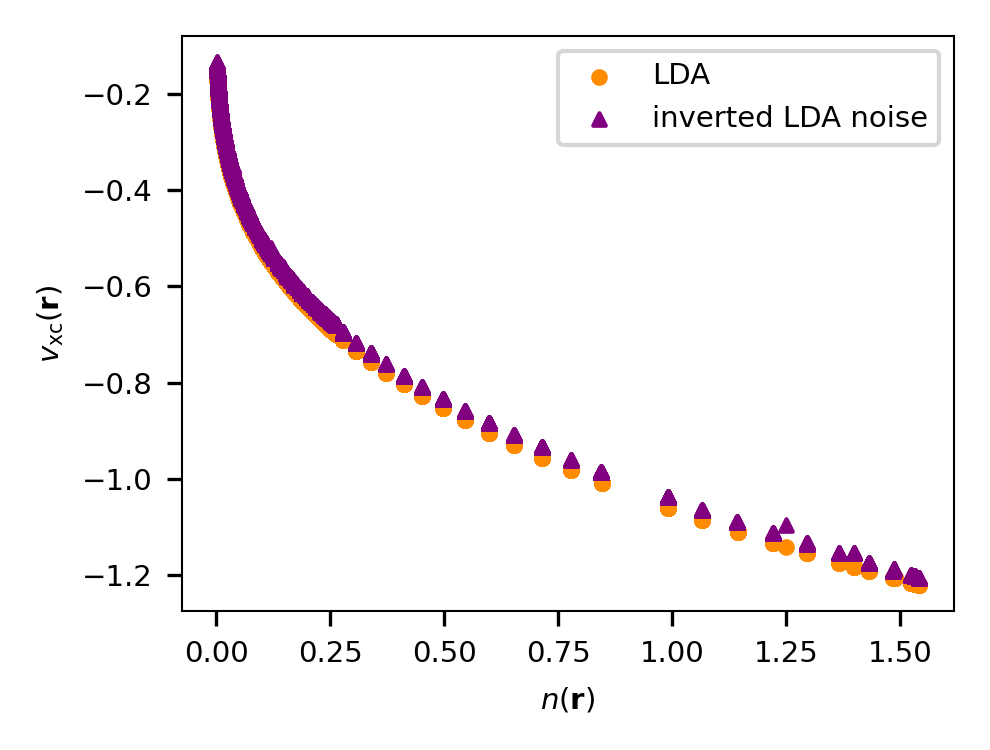}
    \caption{Map of the xc potential versus the local density in NaCl. The result of the inversion of LDA with gaussian noise for $i=200$ is compared to the clean LDA xc potential resulting from  the KS calculation. 
 }
    \label{fig:nacl-noise}
\end{figure}

\section{More about the inversion starting from the QMC density}
\label{sec:QMC-invert}

Here, we will examine the consequences of noise on the inversion of the QMC results of silicon, where the effect is more pronounced than in NaCl.

The first interesting fact is that the error of the QMC inversion behaves  similarly to that of the noisy LDA. This can be appreciated  in Fig. \ref{fig:error_lda_pbd_qmc} (upper panel), where the MaAPE of the density as a function of the number of iterations $i$ is given by the blue  curve.  It shows an overall decrease, but with a pronounced minimum at $i=20$. At this point, it has decreased to 0.38\%. The minimum is followed by a modest increase, after which the error decreases again monotonously. Instead, as the inset in Fig. \ref{fig:error_lda_pbd_qmc} shows, in correspondence to the minimum the MeAPE reaches a plateau of about 0.02\%
and a better precision cannot be reached. For this reason, 
the inversion error remains larger than what we obtained in the case of the inverted clean LDA, by about a factor 500 for the MaAPE and 1000 for the MeAPE.

Of course, one could think 
to continue the iterations, since Fig. \ref{fig:error_lda_pbd_qmc} shows that the MaAPE could be decreased further. However, the fact that a plateau is reached in the MeAPE anticipates that one might encounter problems when doing so. The blue curve in the upper panel of Fig. \ref{fig:dens_xcpot_SILICON-QMC-noise} shows the density from the QMC inversion at $i=500$, where the MaAPE has decreased from 0.38\% to 0.21 \%. Indeed, the error is now further away from the QMC error bar in the most critical points along the path, with respect to the $i=20$ result shown in Fig. \ref{fig:check-LDA}. Also the MeAPE has decreased from 0.04 \% at $i=20$ to 0.02 \% at $i=500$. However, the xc potential obtained from the QMC inversion, shown in the bottom panel of Fig. \ref{fig:dens_xcpot_SILICON-QMC-noise} (blue curve), is no longer smooth. It develops spikes that become even more pronounced when one iterates further, while still decreasing the MaAPE on the density, but with a quite constant MeAPE, which points to a mere redistribution of errors (see Fig. \ref{fig:si_QMC_spikes} for illustration).

 \begin{figure}[h]
    \centering
    \includegraphics[width=\linewidth]{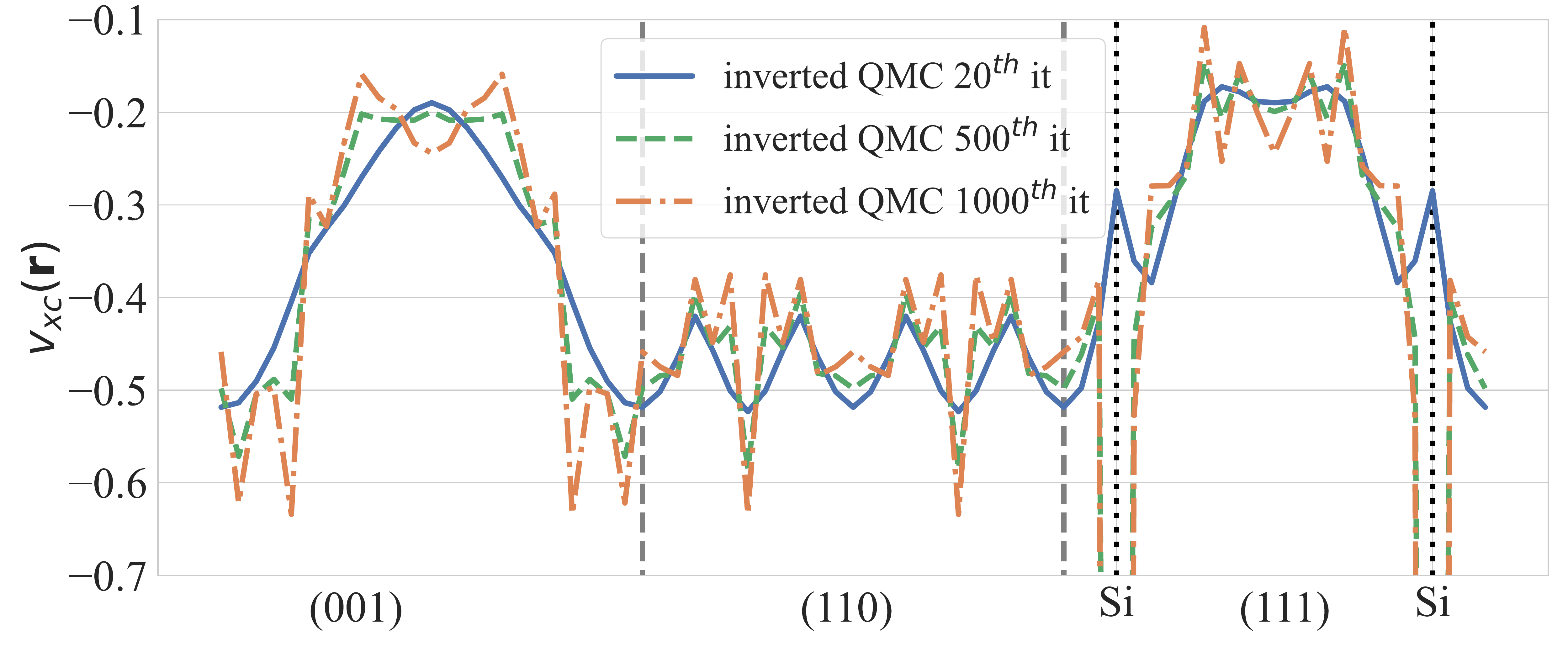}
    \caption{Exchange-correlation potentials obtained by inversion of the QMC density of bulk silicon, at three different iteration steps. All three potentials yield a very similar, accurate density. The MeAPE of the density is  0.04\% at $i=20$, 0.02 \% at $i=500$, and the same value at $i=1000$.   For the MaAPE, we get 0.38\% at $i=20$, 0.21\% at $i=500$, and 0.18\% at $i=1000$.            }
    \label{fig:si_QMC_spikes}
\end{figure}

 Visibly, the algorithm does not succeed in improving the result any further and introduces unexpected features when trying to do so. 
 Difficulties with the inversion procedure have also been reported for finite systems, and they have been attributed to the finite basis set which may introduce an  inconsistency between density and external potential \cite{Schipper1997,Mura1997,Heaton-Burgess2007,Jacob2011,Gaiduk2013}. 
 In the present work, as we have verified, the results are sufficiently well converged to exclude basis set problems.  Instead, the strict analogy to the behavior observed for the noisy LDA is strong evidence for the fact that here the QMC stochastic noise is the limiting factor. We stress again that noise leads to blurring, not to the appearance of spurious features, and its effect can therefore be detected and tested.

\begin{figure}
  \centering
    \includegraphics[width=\linewidth]{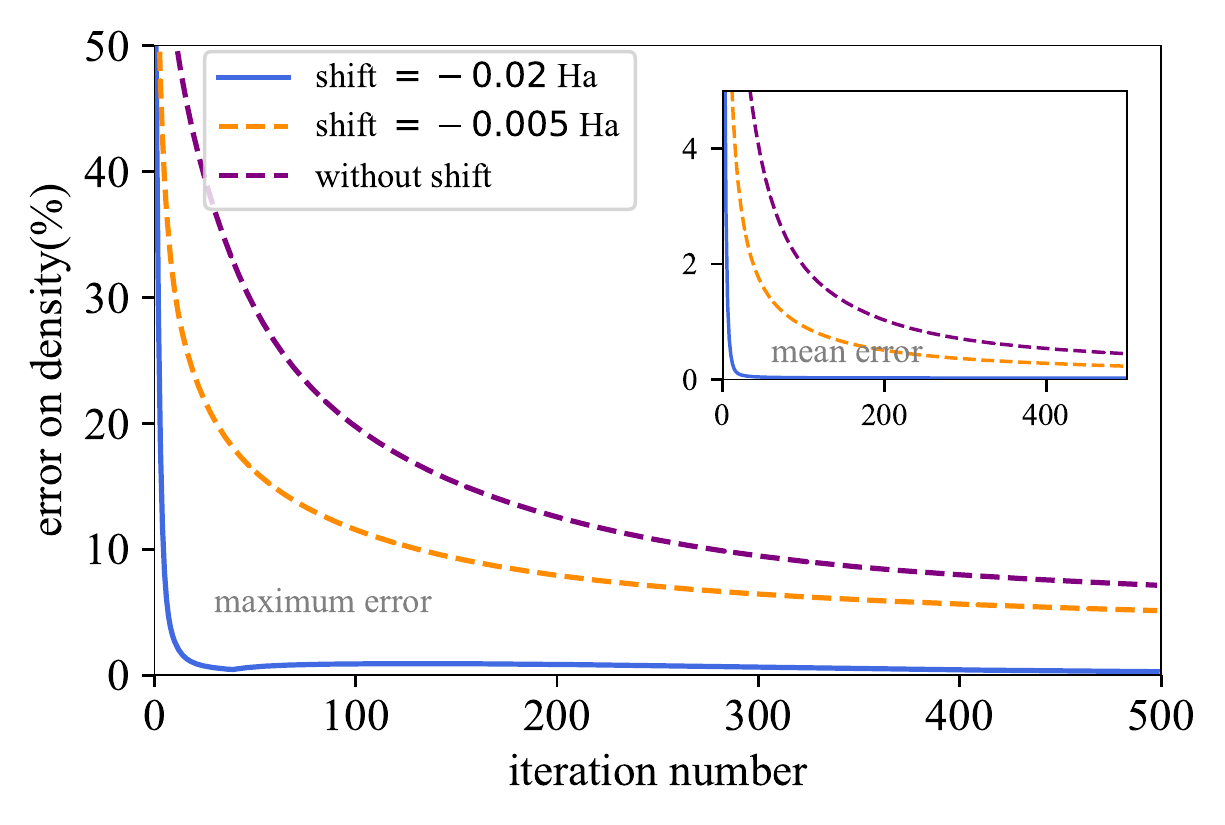}
    \caption{Silicon: error of the inverted QMC density as a function of the iteration number for three different choices of the shift of the starting point. The starting point is $0.1 v^{\rm LDA}_{\rm xc} + {\rm shift}$. 
    }
    \label{fig:error_qmc_shift}
\end{figure}

\begin{figure}
  \centering
    \includegraphics[width=\linewidth]{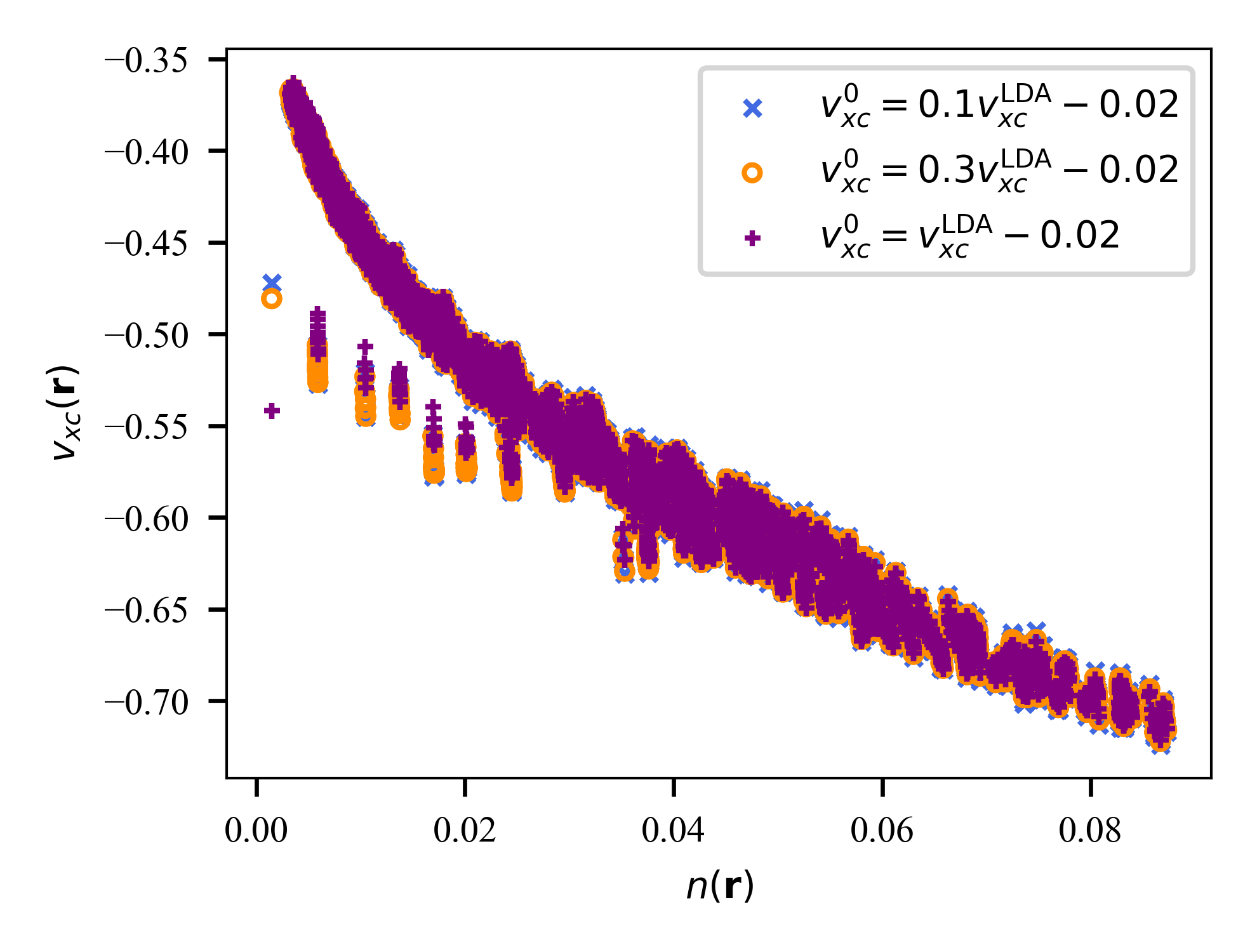}
    \caption{Map of the xc potential of silicon with respect to the local density for three different choices of the starting point. The three xc potentials have the same MeAPE of 0.02 \%. This precision is reached at $i=11, 20$ or $38$ when the starting point is a shifted $v_{\rm xc}^{\rm LDA}$, $0.3 v_{\rm xc}^{\rm LDA}$ or $0.1 v_{\rm xc}^{\rm LDA}$, respectively.   
    }
    \label{fig:vxc_qmc_stpoint}
\end{figure}

Finally, it is important to note that the results of the inversion do not depend on the starting point. This is true both for the shape and for the average value of the starting potential. The latter is important for the convergence behavior: Fig. \ref{fig:error_qmc_shift} shows that a value of the initial shift which guarantees that the potential does not change sign improves the convergence for silicon significantly. Fig. \ref{fig:vxc_qmc_stpoint} shows results obtained for a starting point corresponding to different fractions of  $v_{\rm xc}^{\rm LDA}$: the scatter plot of silicon shows stability both of the main and of the extra branch, except for the point of very low density.

\newpage

\bibliography{vxc-inversion,biblio2}

\begin{thebibliography}{98}%
\makeatletter
\providecommand \@ifxundefined [1]{%
 \@ifx{#1\undefined}
}%
\providecommand \@ifnum [1]{%
 \ifnum #1\expandafter \@firstoftwo
 \else \expandafter \@secondoftwo
 \fi
}%
\providecommand \@ifx [1]{%
 \ifx #1\expandafter \@firstoftwo
 \else \expandafter \@secondoftwo
 \fi
}%
\providecommand \natexlab [1]{#1}%
\providecommand \enquote  [1]{``#1''}%
\providecommand \bibnamefont  [1]{#1}%
\providecommand \bibfnamefont [1]{#1}%
\providecommand \citenamefont [1]{#1}%
\providecommand \href@noop [0]{\@secondoftwo}%
\providecommand \href [0]{\begingroup \@sanitize@url \@href}%
\providecommand \@href[1]{\@@startlink{#1}\@@href}%
\providecommand \@@href[1]{\endgroup#1\@@endlink}%
\providecommand \@sanitize@url [0]{\catcode `\\12\catcode `\$12\catcode
  `\&12\catcode `\#12\catcode `\^12\catcode `\_12\catcode `\%12\relax}%
\providecommand \@@startlink[1]{}%
\providecommand \@@endlink[0]{}%
\providecommand \url  [0]{\begingroup\@sanitize@url \@url }%
\providecommand \@url [1]{\endgroup\@href {#1}{\urlprefix }}%
\providecommand \urlprefix  [0]{URL }%
\providecommand \Eprint [0]{\href }%
\providecommand \doibase [0]{https://doi.org/}%
\providecommand \selectlanguage [0]{\@gobble}%
\providecommand \bibinfo  [0]{\@secondoftwo}%
\providecommand \bibfield  [0]{\@secondoftwo}%
\providecommand \translation [1]{[#1]}%
\providecommand \BibitemOpen [0]{}%
\providecommand \bibitemStop [0]{}%
\providecommand \bibitemNoStop [0]{.\EOS\space}%
\providecommand \EOS [0]{\spacefactor3000\relax}%
\providecommand \BibitemShut  [1]{\csname bibitem#1\endcsname}%
\let\auto@bib@innerbib\@empty
\bibitem [{\citenamefont {Hohenberg}\ and\ \citenamefont {Kohn}(1964)}]{DFT0}%
  \BibitemOpen
  \bibfield  {author} {\bibinfo {author} {\bibfnamefont {P.}~\bibnamefont
  {Hohenberg}}\ and\ \bibinfo {author} {\bibfnamefont {W.}~\bibnamefont
  {Kohn}},\ }\bibfield  {title} {\bibinfo {title} {Inhomogeneous electron
  gas},\ }\href {https://doi.org/10.1103/PhysRev.136.B864} {\bibfield
  {journal} {\bibinfo  {journal} {Phys. Rev.}\ }\textbf {\bibinfo {volume}
  {136}},\ \bibinfo {pages} {B864} (\bibinfo {year} {1964})}\BibitemShut
  {NoStop}%
\bibitem [{\citenamefont {Kohn}\ and\ \citenamefont {Sham}(1965)}]{DFT1}%
  \BibitemOpen
  \bibfield  {author} {\bibinfo {author} {\bibfnamefont {W.}~\bibnamefont
  {Kohn}}\ and\ \bibinfo {author} {\bibfnamefont {L.~J.}\ \bibnamefont
  {Sham}},\ }\bibfield  {title} {\bibinfo {title} {Self-consistent equations
  including exchange and correlation effects},\ }\href
  {https://doi.org/10.1103/PhysRev.140.A1133} {\bibfield  {journal} {\bibinfo
  {journal} {Phys. Rev.}\ }\textbf {\bibinfo {volume} {140}},\ \bibinfo {pages}
  {A1133} (\bibinfo {year} {1965})}\BibitemShut {NoStop}%
\bibitem [{\citenamefont {Kohn}(1999)}]{Kohn1999}%
  \BibitemOpen
  \bibfield  {author} {\bibinfo {author} {\bibfnamefont {W.}~\bibnamefont
  {Kohn}},\ }\bibfield  {title} {\bibinfo {title} {Nobel lecture: Electronic
  structure of matter---wave functions and density functionals},\ }\href
  {https://doi.org/10.1103/RevModPhys.71.1253} {\bibfield  {journal} {\bibinfo
  {journal} {Rev. Mod. Phys.}\ }\textbf {\bibinfo {volume} {71}},\ \bibinfo
  {pages} {1253} (\bibinfo {year} {1999})}\BibitemShut {NoStop}%
\bibitem [{\citenamefont {Martin}(2004)}]{martin2004}%
  \BibitemOpen
  \bibfield  {author} {\bibinfo {author} {\bibfnamefont {R.~M.}\ \bibnamefont
  {Martin}},\ }\href {https://doi.org/10.1017/CBO9780511805769} {\emph
  {\bibinfo {title} {Electronic Structure: Basic Theory and Practical
  Methods}}}\ (\bibinfo  {publisher} {Cambridge University Press},\ \bibinfo
  {year} {2004})\BibitemShut {NoStop}%
\bibitem [{\citenamefont {Ceperley}\ and\ \citenamefont
  {Alder}(1980)}]{Cepe1980}%
  \BibitemOpen
  \bibfield  {author} {\bibinfo {author} {\bibfnamefont {D.~M.}\ \bibnamefont
  {Ceperley}}\ and\ \bibinfo {author} {\bibfnamefont {B.~J.}\ \bibnamefont
  {Alder}},\ }\bibfield  {title} {\bibinfo {title} {Ground state of the
  electron gas by a stochastic method},\ }\href
  {https://doi.org/10.1103/PhysRevLett.45.566} {\bibfield  {journal} {\bibinfo
  {journal} {Phys. Rev. Lett.}\ }\textbf {\bibinfo {volume} {45}},\ \bibinfo
  {pages} {566} (\bibinfo {year} {1980})}\BibitemShut {NoStop}%
\bibitem [{\citenamefont {Medvedev}\ \emph {et~al.}(2017)\citenamefont
  {Medvedev}, \citenamefont {Bushmarinov}, \citenamefont {Sun}, \citenamefont
  {Perdew},\ and\ \citenamefont {Lyssenko}}]{Medvedev2017}%
  \BibitemOpen
  \bibfield  {author} {\bibinfo {author} {\bibfnamefont {M.~G.}\ \bibnamefont
  {Medvedev}}, \bibinfo {author} {\bibfnamefont {I.~S.}\ \bibnamefont
  {Bushmarinov}}, \bibinfo {author} {\bibfnamefont {J.}~\bibnamefont {Sun}},
  \bibinfo {author} {\bibfnamefont {J.~P.}\ \bibnamefont {Perdew}},\ and\
  \bibinfo {author} {\bibfnamefont {K.~A.}\ \bibnamefont {Lyssenko}},\
  }\bibfield  {title} {\bibinfo {title} {Density functional theory is straying
  from the path toward the exact functional},\ }\href
  {https://doi.org/10.1126/science.aah5975} {\bibfield  {journal} {\bibinfo
  {journal} {Science}\ }\textbf {\bibinfo {volume} {355}},\ \bibinfo {pages}
  {49} (\bibinfo {year} {2017})}\BibitemShut {NoStop}%
\bibitem [{\citenamefont {Becke}(2022)}]{Becke2022}%
  \BibitemOpen
  \bibfield  {author} {\bibinfo {author} {\bibfnamefont {A.~D.}\ \bibnamefont
  {Becke}},\ }\bibfield  {title} {\bibinfo {title} {Density-functional theory
  vs density-functional fits},\ }\href {https://doi.org/10.1063/5.0091198}
  {\bibfield  {journal} {\bibinfo  {journal} {The Journal of Chemical Physics}\
  }\textbf {\bibinfo {volume} {156}},\ \bibinfo {pages} {214101} (\bibinfo
  {year} {2022})}\BibitemShut {NoStop}%
\bibitem [{\citenamefont {Perdew}\ and\ \citenamefont
  {Yue}(1986)}]{PerdewWang1986}%
  \BibitemOpen
  \bibfield  {author} {\bibinfo {author} {\bibfnamefont {J.~P.}\ \bibnamefont
  {Perdew}}\ and\ \bibinfo {author} {\bibfnamefont {W.}~\bibnamefont {Yue}},\
  }\bibfield  {title} {\bibinfo {title} {Accurate and simple density functional
  for the electronic exchange energy: Generalized gradient approximation},\
  }\href {https://doi.org/10.1103/PhysRevB.33.8800} {\bibfield  {journal}
  {\bibinfo  {journal} {Phys. Rev. B}\ }\textbf {\bibinfo {volume} {33}},\
  \bibinfo {pages} {8800} (\bibinfo {year} {1986})}\BibitemShut {NoStop}%
\bibitem [{\citenamefont {Perdew}\ \emph {et~al.}(1996)\citenamefont {Perdew},
  \citenamefont {Burke},\ and\ \citenamefont {Ernzerhof}}]{PBE1996}%
  \BibitemOpen
  \bibfield  {author} {\bibinfo {author} {\bibfnamefont {J.~P.}\ \bibnamefont
  {Perdew}}, \bibinfo {author} {\bibfnamefont {K.}~\bibnamefont {Burke}},\ and\
  \bibinfo {author} {\bibfnamefont {M.}~\bibnamefont {Ernzerhof}},\ }\bibfield
  {title} {\bibinfo {title} {Generalized gradient approximation made simple},\
  }\href {https://doi.org/10.1103/PhysRevLett.77.3865} {\bibfield  {journal}
  {\bibinfo  {journal} {Phys. Rev. Lett.}\ }\textbf {\bibinfo {volume} {77}},\
  \bibinfo {pages} {3865} (\bibinfo {year} {1996})}\BibitemShut {NoStop}%
\bibitem [{\citenamefont {Perdew}\ \emph {et~al.}(2008)\citenamefont {Perdew},
  \citenamefont {Ruzsinszky}, \citenamefont {Csonka}, \citenamefont {Vydrov},
  \citenamefont {Scuseria}, \citenamefont {Constantin}, \citenamefont {Zhou},\
  and\ \citenamefont {Burke}}]{Perdew2008}%
  \BibitemOpen
  \bibfield  {author} {\bibinfo {author} {\bibfnamefont {J.~P.}\ \bibnamefont
  {Perdew}}, \bibinfo {author} {\bibfnamefont {A.}~\bibnamefont {Ruzsinszky}},
  \bibinfo {author} {\bibfnamefont {G.~I.}\ \bibnamefont {Csonka}}, \bibinfo
  {author} {\bibfnamefont {O.~A.}\ \bibnamefont {Vydrov}}, \bibinfo {author}
  {\bibfnamefont {G.~E.}\ \bibnamefont {Scuseria}}, \bibinfo {author}
  {\bibfnamefont {L.~A.}\ \bibnamefont {Constantin}}, \bibinfo {author}
  {\bibfnamefont {X.}~\bibnamefont {Zhou}},\ and\ \bibinfo {author}
  {\bibfnamefont {K.}~\bibnamefont {Burke}},\ }\bibfield  {title} {\bibinfo
  {title} {Restoring the density-gradient expansion for exchange in solids and
  surfaces},\ }\href {https://doi.org/10.1103/PhysRevLett.100.136406}
  {\bibfield  {journal} {\bibinfo  {journal} {Phys. Rev. Lett.}\ }\textbf
  {\bibinfo {volume} {100}},\ \bibinfo {pages} {136406} (\bibinfo {year}
  {2008})}\BibitemShut {NoStop}%
\bibitem [{\citenamefont {Sun}\ \emph {et~al.}(2015)\citenamefont {Sun},
  \citenamefont {Ruzsinszky},\ and\ \citenamefont {Perdew}}]{Sun2015}%
  \BibitemOpen
  \bibfield  {author} {\bibinfo {author} {\bibfnamefont {J.}~\bibnamefont
  {Sun}}, \bibinfo {author} {\bibfnamefont {A.}~\bibnamefont {Ruzsinszky}},\
  and\ \bibinfo {author} {\bibfnamefont {J.~P.}\ \bibnamefont {Perdew}},\
  }\bibfield  {title} {\bibinfo {title} {Strongly constrained and appropriately
  normed semilocal density functional},\ }\href
  {https://doi.org/10.1103/PhysRevLett.115.036402} {\bibfield  {journal}
  {\bibinfo  {journal} {Phys. Rev. Lett.}\ }\textbf {\bibinfo {volume} {115}},\
  \bibinfo {pages} {036402} (\bibinfo {year} {2015})}\BibitemShut {NoStop}%
\bibitem [{\citenamefont {Perdew}\ \emph {et~al.}(1982)\citenamefont {Perdew},
  \citenamefont {Parr}, \citenamefont {Levy},\ and\ \citenamefont
  {Balduz}}]{Perdew1982}%
  \BibitemOpen
  \bibfield  {author} {\bibinfo {author} {\bibfnamefont {J.~P.}\ \bibnamefont
  {Perdew}}, \bibinfo {author} {\bibfnamefont {R.~G.}\ \bibnamefont {Parr}},
  \bibinfo {author} {\bibfnamefont {M.}~\bibnamefont {Levy}},\ and\ \bibinfo
  {author} {\bibfnamefont {J.~L.}\ \bibnamefont {Balduz}},\ }\bibfield  {title}
  {\bibinfo {title} {Density-functional theory for fractional particle number:
  Derivative discontinuities of the energy},\ }\href
  {https://doi.org/10.1103/PhysRevLett.49.1691} {\bibfield  {journal} {\bibinfo
   {journal} {Phys. Rev. Lett.}\ }\textbf {\bibinfo {volume} {49}},\ \bibinfo
  {pages} {1691} (\bibinfo {year} {1982})}\BibitemShut {NoStop}%
\bibitem [{\citenamefont {Sham}\ and\ \citenamefont
  {Schl\"uter}(1983)}]{ShamSchlueter1983}%
  \BibitemOpen
  \bibfield  {author} {\bibinfo {author} {\bibfnamefont {L.~J.}\ \bibnamefont
  {Sham}}\ and\ \bibinfo {author} {\bibfnamefont {M.}~\bibnamefont
  {Schl\"uter}},\ }\bibfield  {title} {\bibinfo {title} {Density-functional
  theory of the energy gap},\ }\href
  {https://doi.org/10.1103/PhysRevLett.51.1888} {\bibfield  {journal} {\bibinfo
   {journal} {Phys. Rev. Lett.}\ }\textbf {\bibinfo {volume} {51}},\ \bibinfo
  {pages} {1888} (\bibinfo {year} {1983})}\BibitemShut {NoStop}%
\bibitem [{\citenamefont {Perdew}\ and\ \citenamefont
  {Levy}(1983)}]{Perdew1983}%
  \BibitemOpen
  \bibfield  {author} {\bibinfo {author} {\bibfnamefont {J.~P.}\ \bibnamefont
  {Perdew}}\ and\ \bibinfo {author} {\bibfnamefont {M.}~\bibnamefont {Levy}},\
  }\bibfield  {title} {\bibinfo {title} {Physical content of the exact
  kohn-sham orbital energies: Band gaps and derivative discontinuities},\
  }\href {https://doi.org/10.1103/PhysRevLett.51.1884} {\bibfield  {journal}
  {\bibinfo  {journal} {Phys. Rev. Lett.}\ }\textbf {\bibinfo {volume} {51}},\
  \bibinfo {pages} {1884} (\bibinfo {year} {1983})}\BibitemShut {NoStop}%
\bibitem [{\citenamefont {Perdew}\ and\ \citenamefont
  {Zunger}(1981)}]{Perdew1981}%
  \BibitemOpen
  \bibfield  {author} {\bibinfo {author} {\bibfnamefont {J.~P.}\ \bibnamefont
  {Perdew}}\ and\ \bibinfo {author} {\bibfnamefont {A.}~\bibnamefont
  {Zunger}},\ }\bibfield  {title} {\bibinfo {title} {Self-interaction
  correction to density-functional approximations for many-electron systems},\
  }\href {https://doi.org/10.1103/PhysRevB.23.5048} {\bibfield  {journal}
  {\bibinfo  {journal} {Phys. Rev. B}\ }\textbf {\bibinfo {volume} {23}},\
  \bibinfo {pages} {5048} (\bibinfo {year} {1981})}\BibitemShut {NoStop}%
\bibitem [{\citenamefont {van Schilfgaarde}\ \emph {et~al.}(2006)\citenamefont
  {van Schilfgaarde}, \citenamefont {Kotani},\ and\ \citenamefont
  {Faleev}}]{Schilfgaarde2006}%
  \BibitemOpen
  \bibfield  {author} {\bibinfo {author} {\bibfnamefont {M.}~\bibnamefont {van
  Schilfgaarde}}, \bibinfo {author} {\bibfnamefont {T.}~\bibnamefont
  {Kotani}},\ and\ \bibinfo {author} {\bibfnamefont {S.}~\bibnamefont
  {Faleev}},\ }\bibfield  {title} {\bibinfo {title} {Quasiparticle
  self-consistent $gw$ theory},\ }\href
  {https://doi.org/10.1103/PhysRevLett.96.226402} {\bibfield  {journal}
  {\bibinfo  {journal} {Phys. Rev. Lett.}\ }\textbf {\bibinfo {volume} {96}},\
  \bibinfo {pages} {226402} (\bibinfo {year} {2006})}\BibitemShut {NoStop}%
\bibitem [{\citenamefont {Sham}\ and\ \citenamefont {Kohn}(1966)}]{Sham1966}%
  \BibitemOpen
  \bibfield  {author} {\bibinfo {author} {\bibfnamefont {L.~J.}\ \bibnamefont
  {Sham}}\ and\ \bibinfo {author} {\bibfnamefont {W.}~\bibnamefont {Kohn}},\
  }\bibfield  {title} {\bibinfo {title} {One-particle properties of an
  inhomogeneous interacting electron gas},\ }\href
  {https://doi.org/10.1103/PhysRev.145.561} {\bibfield  {journal} {\bibinfo
  {journal} {Phys. Rev.}\ }\textbf {\bibinfo {volume} {145}},\ \bibinfo {pages}
  {561} (\bibinfo {year} {1966})}\BibitemShut {NoStop}%
\bibitem [{\citenamefont {Godby}\ \emph {et~al.}(1986)\citenamefont {Godby},
  \citenamefont {Schl\"uter},\ and\ \citenamefont {Sham}}]{Godby1986}%
  \BibitemOpen
  \bibfield  {author} {\bibinfo {author} {\bibfnamefont {R.~W.}\ \bibnamefont
  {Godby}}, \bibinfo {author} {\bibfnamefont {M.}~\bibnamefont {Schl\"uter}},\
  and\ \bibinfo {author} {\bibfnamefont {L.~J.}\ \bibnamefont {Sham}},\
  }\bibfield  {title} {\bibinfo {title} {Accurate exchange-correlation
  potential for silicon and its discontinuity on addition of an electron},\
  }\href {https://doi.org/10.1103/PhysRevLett.56.2415} {\bibfield  {journal}
  {\bibinfo  {journal} {Phys. Rev. Lett.}\ }\textbf {\bibinfo {volume} {56}},\
  \bibinfo {pages} {2415} (\bibinfo {year} {1986})}\BibitemShut {NoStop}%
\bibitem [{\citenamefont {Godby}\ \emph {et~al.}(1987)\citenamefont {Godby},
  \citenamefont {Schl\"uter},\ and\ \citenamefont {Sham}}]{Godby87}%
  \BibitemOpen
  \bibfield  {author} {\bibinfo {author} {\bibfnamefont {R.~W.}\ \bibnamefont
  {Godby}}, \bibinfo {author} {\bibfnamefont {M.}~\bibnamefont {Schl\"uter}},\
  and\ \bibinfo {author} {\bibfnamefont {L.~J.}\ \bibnamefont {Sham}},\
  }\bibfield  {title} {\bibinfo {title} {Trends in self-energy operators and
  their corresponding exchange-correlation potentials},\ }\href
  {https://doi.org/10.1103/PhysRevB.36.6497} {\bibfield  {journal} {\bibinfo
  {journal} {Phys. Rev. B}\ }\textbf {\bibinfo {volume} {36}},\ \bibinfo
  {pages} {6497} (\bibinfo {year} {1987})}\BibitemShut {NoStop}%
\bibitem [{\citenamefont {Godby}\ \emph {et~al.}(1988)\citenamefont {Godby},
  \citenamefont {Schl\"uter},\ and\ \citenamefont {Sham}}]{Godby88}%
  \BibitemOpen
  \bibfield  {author} {\bibinfo {author} {\bibfnamefont {R.~W.}\ \bibnamefont
  {Godby}}, \bibinfo {author} {\bibfnamefont {M.}~\bibnamefont {Schl\"uter}},\
  and\ \bibinfo {author} {\bibfnamefont {L.~J.}\ \bibnamefont {Sham}},\
  }\bibfield  {title} {\bibinfo {title} {Self-energy operators and
  exchange-correlation potentials in semiconductors},\ }\href
  {https://doi.org/10.1103/PhysRevB.37.10159} {\bibfield  {journal} {\bibinfo
  {journal} {Phys. Rev. B}\ }\textbf {\bibinfo {volume} {37}},\ \bibinfo
  {pages} {10159} (\bibinfo {year} {1988})}\BibitemShut {NoStop}%
\bibitem [{\citenamefont {Niquet}\ and\ \citenamefont
  {Gonze}(2004)}]{Niquet2004}%
  \BibitemOpen
  \bibfield  {author} {\bibinfo {author} {\bibfnamefont {Y.~M.}\ \bibnamefont
  {Niquet}}\ and\ \bibinfo {author} {\bibfnamefont {X.}~\bibnamefont {Gonze}},\
  }\bibfield  {title} {\bibinfo {title} {Band-gap energy in the random-phase
  approximation to density-functional theory},\ }\href
  {https://doi.org/10.1103/PhysRevB.70.245115} {\bibfield  {journal} {\bibinfo
  {journal} {Phys. Rev. B}\ }\textbf {\bibinfo {volume} {70}},\ \bibinfo
  {pages} {245115} (\bibinfo {year} {2004})}\BibitemShut {NoStop}%
\bibitem [{\citenamefont {Gr\"uning}\ \emph {et~al.}(2006)\citenamefont
  {Gr\"uning}, \citenamefont {Marini},\ and\ \citenamefont
  {Rubio}}]{Gruning2006}%
  \BibitemOpen
  \bibfield  {author} {\bibinfo {author} {\bibfnamefont {M.}~\bibnamefont
  {Gr\"uning}}, \bibinfo {author} {\bibfnamefont {A.}~\bibnamefont {Marini}},\
  and\ \bibinfo {author} {\bibfnamefont {A.}~\bibnamefont {Rubio}},\ }\bibfield
   {title} {\bibinfo {title} {Effect of spatial nonlocality on the density
  functional band gap},\ }\href {https://doi.org/10.1103/PhysRevB.74.161103}
  {\bibfield  {journal} {\bibinfo  {journal} {Phys. Rev. B}\ }\textbf {\bibinfo
  {volume} {74}},\ \bibinfo {pages} {161103} (\bibinfo {year}
  {2006})}\BibitemShut {NoStop}%
\bibitem [{\citenamefont {Grüning}\ \emph {et~al.}(2006)\citenamefont
  {Grüning}, \citenamefont {Marini},\ and\ \citenamefont
  {Rubio}}]{GruningJCP2006}%
  \BibitemOpen
  \bibfield  {author} {\bibinfo {author} {\bibfnamefont {M.}~\bibnamefont
  {Grüning}}, \bibinfo {author} {\bibfnamefont {A.}~\bibnamefont {Marini}},\
  and\ \bibinfo {author} {\bibfnamefont {A.}~\bibnamefont {Rubio}},\ }\bibfield
   {title} {\bibinfo {title} {Density functionals from many-body perturbation
  theory: The band gap for semiconductors and insulators},\ }\href
  {https://doi.org/10.1063/1.2189226} {\bibfield  {journal} {\bibinfo
  {journal} {The Journal of Chemical Physics}\ }\textbf {\bibinfo {volume}
  {124}},\ \bibinfo {pages} {154108} (\bibinfo {year} {2006})}\BibitemShut
  {NoStop}%
\bibitem [{\citenamefont {Kotani}(1998)}]{Kotani1998}%
  \BibitemOpen
  \bibfield  {author} {\bibinfo {author} {\bibfnamefont {T.}~\bibnamefont
  {Kotani}},\ }\bibfield  {title} {\bibinfo {title} {An
  optimized-effective-potential method for solids with exact exchange and
  random-phase approximation correlation},\ }\href@noop {} {\bibfield
  {journal} {\bibinfo  {journal} {J. Phys.: Condens. Matter}\ }\textbf
  {\bibinfo {volume} {10}},\ \bibinfo {pages} {9241} (\bibinfo {year}
  {1998})}\BibitemShut {NoStop}%
\bibitem [{\citenamefont {Klimeš}\ and\ \citenamefont
  {Kresse}(2014)}]{Klimes2014}%
  \BibitemOpen
  \bibfield  {author} {\bibinfo {author} {\bibfnamefont {J.}~\bibnamefont
  {Klimeš}}\ and\ \bibinfo {author} {\bibfnamefont {G.}~\bibnamefont
  {Kresse}},\ }\bibfield  {title} {\bibinfo {title} {Kohn-sham band gaps and
  potentials of solids from the optimised effective potential method within the
  random phase approximation},\ }\href {https://doi.org/10.1063/1.4863502}
  {\bibfield  {journal} {\bibinfo  {journal} {The Journal of Chemical Physics}\
  }\textbf {\bibinfo {volume} {140}},\ \bibinfo {pages} {054516} (\bibinfo
  {year} {2014})}\BibitemShut {NoStop}%
\bibitem [{\citenamefont {Riemelmoser}\ \emph {et~al.}(2021)\citenamefont
  {Riemelmoser}, \citenamefont {Kaltak},\ and\ \citenamefont
  {Kresse}}]{Riemelmoser2021}%
  \BibitemOpen
  \bibfield  {author} {\bibinfo {author} {\bibfnamefont {S.}~\bibnamefont
  {Riemelmoser}}, \bibinfo {author} {\bibfnamefont {M.}~\bibnamefont
  {Kaltak}},\ and\ \bibinfo {author} {\bibfnamefont {G.}~\bibnamefont
  {Kresse}},\ }\bibfield  {title} {\bibinfo {title} {Optimized effective
  potentials from the random-phase approximation: Accuracy of the quasiparticle
  approximation},\ }\href {https://doi.org/10.1063/5.0045400} {\bibfield
  {journal} {\bibinfo  {journal} {The Journal of Chemical Physics}\ }\textbf
  {\bibinfo {volume} {154}},\ \bibinfo {pages} {154103} (\bibinfo {year}
  {2021})}\BibitemShut {NoStop}%
\bibitem [{\citenamefont {van Leeuwen}\ and\ \citenamefont
  {Baerends}(1994)}]{PhysRevA.49.2421}%
  \BibitemOpen
  \bibfield  {author} {\bibinfo {author} {\bibfnamefont {R.}~\bibnamefont {van
  Leeuwen}}\ and\ \bibinfo {author} {\bibfnamefont {E.~J.}\ \bibnamefont
  {Baerends}},\ }\bibfield  {title} {\bibinfo {title} {Exchange-correlation
  potential with correct asymptotic behavior},\ }\href
  {https://doi.org/10.1103/PhysRevA.49.2421} {\bibfield  {journal} {\bibinfo
  {journal} {Phys. Rev. A}\ }\textbf {\bibinfo {volume} {49}},\ \bibinfo
  {pages} {2421} (\bibinfo {year} {1994})}\BibitemShut {NoStop}%
\bibitem [{\citenamefont {Zhao}\ \emph {et~al.}(1994)\citenamefont {Zhao},
  \citenamefont {Morrison},\ and\ \citenamefont {Parr}}]{Zhao1994}%
  \BibitemOpen
  \bibfield  {author} {\bibinfo {author} {\bibfnamefont {Q.}~\bibnamefont
  {Zhao}}, \bibinfo {author} {\bibfnamefont {R.~C.}\ \bibnamefont {Morrison}},\
  and\ \bibinfo {author} {\bibfnamefont {R.~G.}\ \bibnamefont {Parr}},\
  }\bibfield  {title} {\bibinfo {title} {From electron densities to kohn-sham
  kinetic energies, orbital energies, exchange-correlation potentials, and
  exchange-correlation energies},\ }\href
  {https://doi.org/10.1103/PhysRevA.50.2138} {\bibfield  {journal} {\bibinfo
  {journal} {Phys. Rev. A}\ }\textbf {\bibinfo {volume} {50}},\ \bibinfo
  {pages} {2138} (\bibinfo {year} {1994})}\BibitemShut {NoStop}%
\bibitem [{\citenamefont {Wu}\ and\ \citenamefont {Yang}(2003)}]{Wu2003}%
  \BibitemOpen
  \bibfield  {author} {\bibinfo {author} {\bibfnamefont {Q.}~\bibnamefont
  {Wu}}\ and\ \bibinfo {author} {\bibfnamefont {W.}~\bibnamefont {Yang}},\
  }\bibfield  {title} {\bibinfo {title} {A direct optimization method for
  calculating density functionals and exchange–correlation potentials from
  electron densities},\ }\href {https://doi.org/10.1063/1.1535422} {\bibfield
  {journal} {\bibinfo  {journal} {The Journal of Chemical Physics}\ }\textbf
  {\bibinfo {volume} {118}},\ \bibinfo {pages} {2498} (\bibinfo {year}
  {2003})}\BibitemShut {NoStop}%
\bibitem [{\citenamefont {Peirs}\ \emph {et~al.}(2003)\citenamefont {Peirs},
  \citenamefont {Van~Neck},\ and\ \citenamefont {Waroquier}}]{Peirs2003}%
  \BibitemOpen
  \bibfield  {author} {\bibinfo {author} {\bibfnamefont {K.}~\bibnamefont
  {Peirs}}, \bibinfo {author} {\bibfnamefont {D.}~\bibnamefont {Van~Neck}},\
  and\ \bibinfo {author} {\bibfnamefont {M.}~\bibnamefont {Waroquier}},\
  }\bibfield  {title} {\bibinfo {title} {Algorithm to derive exact
  exchange-correlation potentials from correlated densities in atoms},\ }\href
  {https://doi.org/10.1103/PhysRevA.67.012505} {\bibfield  {journal} {\bibinfo
  {journal} {Phys. Rev. A}\ }\textbf {\bibinfo {volume} {67}},\ \bibinfo
  {pages} {012505} (\bibinfo {year} {2003})}\BibitemShut {NoStop}%
\bibitem [{\citenamefont {Almbladh}\ and\ \citenamefont
  {Pedroza}(1984)}]{Almbladh1984}%
  \BibitemOpen
  \bibfield  {author} {\bibinfo {author} {\bibfnamefont {C.~O.}\ \bibnamefont
  {Almbladh}}\ and\ \bibinfo {author} {\bibfnamefont {A.~C.}\ \bibnamefont
  {Pedroza}},\ }\bibfield  {title} {\bibinfo {title} {Density-functional
  exchange-correlation potentials and orbital eigenvalues for light atoms},\
  }\href {https://doi.org/10.1103/PhysRevA.29.2322} {\bibfield  {journal}
  {\bibinfo  {journal} {Phys. Rev. A}\ }\textbf {\bibinfo {volume} {29}},\
  \bibinfo {pages} {2322} (\bibinfo {year} {1984})}\BibitemShut {NoStop}%
\bibitem [{\citenamefont {Aryasetiawan}\ and\ \citenamefont
  {Stott}(1988)}]{Aryasetiawan1988}%
  \BibitemOpen
  \bibfield  {author} {\bibinfo {author} {\bibfnamefont {F.}~\bibnamefont
  {Aryasetiawan}}\ and\ \bibinfo {author} {\bibfnamefont {M.~J.}\ \bibnamefont
  {Stott}},\ }\bibfield  {title} {\bibinfo {title} {Effective potentials in
  density-functional theory},\ }\href
  {https://doi.org/10.1103/PhysRevB.38.2974} {\bibfield  {journal} {\bibinfo
  {journal} {Phys. Rev. B}\ }\textbf {\bibinfo {volume} {38}},\ \bibinfo
  {pages} {2974} (\bibinfo {year} {1988})}\BibitemShut {NoStop}%
\bibitem [{\citenamefont {G\"orling}(1992)}]{Goerling1992}%
  \BibitemOpen
  \bibfield  {author} {\bibinfo {author} {\bibfnamefont {A.}~\bibnamefont
  {G\"orling}},\ }\bibfield  {title} {\bibinfo {title} {Kohn-sham potentials
  and wave functions from electron densities},\ }\href
  {https://doi.org/10.1103/PhysRevA.46.3753} {\bibfield  {journal} {\bibinfo
  {journal} {Phys. Rev. A}\ }\textbf {\bibinfo {volume} {46}},\ \bibinfo
  {pages} {3753} (\bibinfo {year} {1992})}\BibitemShut {NoStop}%
\bibitem [{\citenamefont {Knorr}\ and\ \citenamefont
  {Godby}(1992)}]{Knorr1992}%
  \BibitemOpen
  \bibfield  {author} {\bibinfo {author} {\bibfnamefont {W.}~\bibnamefont
  {Knorr}}\ and\ \bibinfo {author} {\bibfnamefont {R.~W.}\ \bibnamefont
  {Godby}},\ }\bibfield  {title} {\bibinfo {title} {Investigating exact
  density-functional theory of a model semiconductor},\ }\href
  {https://doi.org/10.1103/PhysRevLett.68.639} {\bibfield  {journal} {\bibinfo
  {journal} {Phys. Rev. Lett.}\ }\textbf {\bibinfo {volume} {68}},\ \bibinfo
  {pages} {639} (\bibinfo {year} {1992})}\BibitemShut {NoStop}%
\bibitem [{\citenamefont {Knorr}\ and\ \citenamefont
  {Godby}(1994)}]{Knorr1994}%
  \BibitemOpen
  \bibfield  {author} {\bibinfo {author} {\bibfnamefont {W.}~\bibnamefont
  {Knorr}}\ and\ \bibinfo {author} {\bibfnamefont {R.~W.}\ \bibnamefont
  {Godby}},\ }\bibfield  {title} {\bibinfo {title} {Quantum monte carlo study
  of density-functional theory for a semiconducting wire},\ }\href
  {https://doi.org/10.1103/PhysRevB.50.1779} {\bibfield  {journal} {\bibinfo
  {journal} {Phys. Rev. B}\ }\textbf {\bibinfo {volume} {50}},\ \bibinfo
  {pages} {1779} (\bibinfo {year} {1994})}\BibitemShut {NoStop}%
\bibitem [{\citenamefont {Umrigar}\ and\ \citenamefont
  {Gonze}(1994)}]{Umrigar1994}%
  \BibitemOpen
  \bibfield  {author} {\bibinfo {author} {\bibfnamefont {C.~J.}\ \bibnamefont
  {Umrigar}}\ and\ \bibinfo {author} {\bibfnamefont {X.}~\bibnamefont
  {Gonze}},\ }\bibfield  {title} {\bibinfo {title} {Accurate
  exchange-correlation potentials and total-energy components for the helium
  isoelectronic series},\ }\href {https://doi.org/10.1103/PhysRevA.50.3827}
  {\bibfield  {journal} {\bibinfo  {journal} {Phys. Rev. A}\ }\textbf {\bibinfo
  {volume} {50}},\ \bibinfo {pages} {3827} (\bibinfo {year}
  {1994})}\BibitemShut {NoStop}%
\bibitem [{\citenamefont {Gritsenko}\ \emph {et~al.}(1995)\citenamefont
  {Gritsenko}, \citenamefont {van Leeuwen},\ and\ \citenamefont
  {Baerends}}]{Gritsenko1995}%
  \BibitemOpen
  \bibfield  {author} {\bibinfo {author} {\bibfnamefont {O.~V.}\ \bibnamefont
  {Gritsenko}}, \bibinfo {author} {\bibfnamefont {R.}~\bibnamefont {van
  Leeuwen}},\ and\ \bibinfo {author} {\bibfnamefont {E.~J.}\ \bibnamefont
  {Baerends}},\ }\bibfield  {title} {\bibinfo {title} {Molecular kohn-sham
  exchange-correlation potential from the correlated ab initio electron
  density},\ }\href {https://doi.org/10.1103/PhysRevA.52.1870} {\bibfield
  {journal} {\bibinfo  {journal} {Phys. Rev. A}\ }\textbf {\bibinfo {volume}
  {52}},\ \bibinfo {pages} {1870} (\bibinfo {year} {1995})}\BibitemShut
  {NoStop}%
\bibitem [{\citenamefont {Tozer}\ \emph {et~al.}(1996)\citenamefont {Tozer},
  \citenamefont {Ingamells},\ and\ \citenamefont {Handy}}]{Tozer1996}%
  \BibitemOpen
  \bibfield  {author} {\bibinfo {author} {\bibfnamefont {D.~J.}\ \bibnamefont
  {Tozer}}, \bibinfo {author} {\bibfnamefont {V.~E.}\ \bibnamefont
  {Ingamells}},\ and\ \bibinfo {author} {\bibfnamefont {N.~C.}\ \bibnamefont
  {Handy}},\ }\bibfield  {title} {\bibinfo {title} {Exchange‐correlation
  potentials},\ }\href {https://doi.org/10.1063/1.472753} {\bibfield  {journal}
  {\bibinfo  {journal} {The Journal of Chemical Physics}\ }\textbf {\bibinfo
  {volume} {105}},\ \bibinfo {pages} {9200} (\bibinfo {year}
  {1996})}\BibitemShut {NoStop}%
\bibitem [{\citenamefont {Helbig}\ \emph {et~al.}(2009)\citenamefont {Helbig},
  \citenamefont {Tokatly},\ and\ \citenamefont {Rubio}}]{Helbig2009}%
  \BibitemOpen
  \bibfield  {author} {\bibinfo {author} {\bibfnamefont {N.}~\bibnamefont
  {Helbig}}, \bibinfo {author} {\bibfnamefont {I.~V.}\ \bibnamefont
  {Tokatly}},\ and\ \bibinfo {author} {\bibfnamefont {A.}~\bibnamefont
  {Rubio}},\ }\bibfield  {title} {\bibinfo {title} {Exact kohn–sham potential
  of strongly correlated finite systems},\ }\href
  {https://doi.org/10.1063/1.3271392} {\bibfield  {journal} {\bibinfo
  {journal} {The Journal of Chemical Physics}\ }\textbf {\bibinfo {volume}
  {131}},\ \bibinfo {pages} {224105} (\bibinfo {year} {2009})}\BibitemShut
  {NoStop}%
\bibitem [{\citenamefont {Stoudenmire}\ \emph {et~al.}(2012)\citenamefont
  {Stoudenmire}, \citenamefont {Wagner}, \citenamefont {White},\ and\
  \citenamefont {Burke}}]{Burke1D2012}%
  \BibitemOpen
  \bibfield  {author} {\bibinfo {author} {\bibfnamefont {E.~M.}\ \bibnamefont
  {Stoudenmire}}, \bibinfo {author} {\bibfnamefont {L.~O.}\ \bibnamefont
  {Wagner}}, \bibinfo {author} {\bibfnamefont {S.~R.}\ \bibnamefont {White}},\
  and\ \bibinfo {author} {\bibfnamefont {K.}~\bibnamefont {Burke}},\ }\bibfield
   {title} {\bibinfo {title} {One-dimensional continuum electronic structure
  with the density-matrix renormalization group and its implications for
  density-functional theory},\ }\href
  {https://doi.org/10.1103/PhysRevLett.109.056402} {\bibfield  {journal}
  {\bibinfo  {journal} {Phys. Rev. Lett.}\ }\textbf {\bibinfo {volume} {109}},\
  \bibinfo {pages} {056402} (\bibinfo {year} {2012})}\BibitemShut {NoStop}%
\bibitem [{\citenamefont {Varsano}\ \emph {et~al.}(2014)\citenamefont
  {Varsano}, \citenamefont {Barborini},\ and\ \citenamefont
  {Guidoni}}]{Varsano2014}%
  \BibitemOpen
  \bibfield  {author} {\bibinfo {author} {\bibfnamefont {D.}~\bibnamefont
  {Varsano}}, \bibinfo {author} {\bibfnamefont {M.}~\bibnamefont {Barborini}},\
  and\ \bibinfo {author} {\bibfnamefont {L.}~\bibnamefont {Guidoni}},\
  }\bibfield  {title} {\bibinfo {title} {Kohn-sham orbitals and potentials from
  quantum monte carlo molecular densities},\ }\href
  {https://doi.org/10.1063/1.4863213} {\bibfield  {journal} {\bibinfo
  {journal} {The Journal of Chemical Physics}\ }\textbf {\bibinfo {volume}
  {140}},\ \bibinfo {pages} {054102} (\bibinfo {year} {2014})}\BibitemShut
  {NoStop}%
\bibitem [{\citenamefont {Hollins}\ \emph {et~al.}(2016)\citenamefont
  {Hollins}, \citenamefont {Clark}, \citenamefont {Refson},\ and\ \citenamefont
  {Gidopoulos}}]{Hollins2016}%
  \BibitemOpen
  \bibfield  {author} {\bibinfo {author} {\bibfnamefont {T.~W.}\ \bibnamefont
  {Hollins}}, \bibinfo {author} {\bibfnamefont {S.~J.}\ \bibnamefont {Clark}},
  \bibinfo {author} {\bibfnamefont {K.}~\bibnamefont {Refson}},\ and\ \bibinfo
  {author} {\bibfnamefont {N.~I.}\ \bibnamefont {Gidopoulos}},\ }\bibfield
  {title} {\bibinfo {title} {A local fock-exchange potential in
  kohn{\textendash}sham equations},\ }\href
  {https://doi.org/10.1088/1361-648x/29/4/04lt01} {\bibfield  {journal}
  {\bibinfo  {journal} {Journal of Physics: Condensed Matter}\ }\textbf
  {\bibinfo {volume} {29}},\ \bibinfo {pages} {04LT01} (\bibinfo {year}
  {2016})}\BibitemShut {NoStop}%
\bibitem [{\citenamefont {Hodgson}\ \emph {et~al.}(2016)\citenamefont
  {Hodgson}, \citenamefont {Ramsden},\ and\ \citenamefont
  {Godby}}]{Hodgson2016}%
  \BibitemOpen
  \bibfield  {author} {\bibinfo {author} {\bibfnamefont {M.~J.~P.}\
  \bibnamefont {Hodgson}}, \bibinfo {author} {\bibfnamefont {J.~D.}\
  \bibnamefont {Ramsden}},\ and\ \bibinfo {author} {\bibfnamefont {R.~W.}\
  \bibnamefont {Godby}},\ }\bibfield  {title} {\bibinfo {title} {Origin of
  static and dynamic steps in exact kohn-sham potentials},\ }\href
  {https://doi.org/10.1103/PhysRevB.93.155146} {\bibfield  {journal} {\bibinfo
  {journal} {Phys. Rev. B}\ }\textbf {\bibinfo {volume} {93}},\ \bibinfo
  {pages} {155146} (\bibinfo {year} {2016})}\BibitemShut {NoStop}%
\bibitem [{\citenamefont {Hodgson}\ \emph {et~al.}(2017)\citenamefont
  {Hodgson}, \citenamefont {Kraisler}, \citenamefont {Schild},\ and\
  \citenamefont {Gross}}]{Hodgson2017}%
  \BibitemOpen
  \bibfield  {author} {\bibinfo {author} {\bibfnamefont {M.~J.~P.}\
  \bibnamefont {Hodgson}}, \bibinfo {author} {\bibfnamefont {E.}~\bibnamefont
  {Kraisler}}, \bibinfo {author} {\bibfnamefont {A.}~\bibnamefont {Schild}},\
  and\ \bibinfo {author} {\bibfnamefont {E.~K.~U.}\ \bibnamefont {Gross}},\
  }\bibfield  {title} {\bibinfo {title} {How interatomic steps in the exact
  kohn–sham potential relate to derivative discontinuities of the energy},\
  }\href {https://doi.org/10.1021/acs.jpclett.7b02615} {\bibfield  {journal}
  {\bibinfo  {journal} {The Journal of Physical Chemistry Letters}\ }\textbf
  {\bibinfo {volume} {8}},\ \bibinfo {pages} {5974} (\bibinfo {year}
  {2017})}\BibitemShut {NoStop}%
\bibitem [{\citenamefont {Wetherell}\ \emph {et~al.}(2019)\citenamefont
  {Wetherell}, \citenamefont {Hodgson}, \citenamefont {Talirz},\ and\
  \citenamefont {Godby}}]{Wetherell2019}%
  \BibitemOpen
  \bibfield  {author} {\bibinfo {author} {\bibfnamefont {J.}~\bibnamefont
  {Wetherell}}, \bibinfo {author} {\bibfnamefont {M.~J.~P.}\ \bibnamefont
  {Hodgson}}, \bibinfo {author} {\bibfnamefont {L.}~\bibnamefont {Talirz}},\
  and\ \bibinfo {author} {\bibfnamefont {R.~W.}\ \bibnamefont {Godby}},\
  }\bibfield  {title} {\bibinfo {title} {Advantageous nearsightedness of
  many-body perturbation theory contrasted with kohn-sham density functional
  theory},\ }\href {https://doi.org/10.1103/PhysRevB.99.045129} {\bibfield
  {journal} {\bibinfo  {journal} {Phys. Rev. B}\ }\textbf {\bibinfo {volume}
  {99}},\ \bibinfo {pages} {045129} (\bibinfo {year} {2019})}\BibitemShut
  {NoStop}%
\bibitem [{\citenamefont {Nam}\ \emph {et~al.}(2020)\citenamefont {Nam},
  \citenamefont {Song}, \citenamefont {Sim},\ and\ \citenamefont
  {Burke}}]{Nam2020}%
  \BibitemOpen
  \bibfield  {author} {\bibinfo {author} {\bibfnamefont {S.}~\bibnamefont
  {Nam}}, \bibinfo {author} {\bibfnamefont {S.}~\bibnamefont {Song}}, \bibinfo
  {author} {\bibfnamefont {E.}~\bibnamefont {Sim}},\ and\ \bibinfo {author}
  {\bibfnamefont {K.}~\bibnamefont {Burke}},\ }\bibfield  {title} {\bibinfo
  {title} {Measuring density-driven errors using kohn–sham inversion},\
  }\href {https://doi.org/10.1021/acs.jctc.0c00391} {\bibfield  {journal}
  {\bibinfo  {journal} {Journal of Chemical Theory and Computation}\ }\textbf
  {\bibinfo {volume} {16}},\ \bibinfo {pages} {5014} (\bibinfo {year}
  {2020})}\BibitemShut {NoStop}%
\bibitem [{\citenamefont {Savin}\ \emph {et~al.}(1998)\citenamefont {Savin},
  \citenamefont {Umrigar},\ and\ \citenamefont {Gonze}}]{Savin1998}%
  \BibitemOpen
  \bibfield  {author} {\bibinfo {author} {\bibfnamefont {A.}~\bibnamefont
  {Savin}}, \bibinfo {author} {\bibfnamefont {C.}~\bibnamefont {Umrigar}},\
  and\ \bibinfo {author} {\bibfnamefont {X.}~\bibnamefont {Gonze}},\ }\bibfield
   {title} {\bibinfo {title} {Relationship of kohn–sham eigenvalues to
  excitation energies},\ }\href
  {https://doi.org/https://doi.org/10.1016/S0009-2614(98)00316-9} {\bibfield
  {journal} {\bibinfo  {journal} {Chemical Physics Letters}\ }\textbf {\bibinfo
  {volume} {288}},\ \bibinfo {pages} {391} (\bibinfo {year}
  {1998})}\BibitemShut {NoStop}%
\bibitem [{\citenamefont {Li}\ \emph {et~al.}(2019)\citenamefont {Li},
  \citenamefont {Drummond}, \citenamefont {Schuck},\ and\ \citenamefont
  {Olevano}}]{Li2019}%
  \BibitemOpen
  \bibfield  {author} {\bibinfo {author} {\bibfnamefont {J.}~\bibnamefont
  {Li}}, \bibinfo {author} {\bibfnamefont {N.~D.}\ \bibnamefont {Drummond}},
  \bibinfo {author} {\bibfnamefont {P.}~\bibnamefont {Schuck}},\ and\ \bibinfo
  {author} {\bibfnamefont {V.}~\bibnamefont {Olevano}},\ }\bibfield  {title}
  {\bibinfo {title} {{Comparing many-body approaches against the helium atom
  exact solution}},\ }\href {https://doi.org/10.21468/SciPostPhys.6.4.040}
  {\bibfield  {journal} {\bibinfo  {journal} {SciPost Phys.}\ }\textbf
  {\bibinfo {volume} {6}},\ \bibinfo {pages} {40} (\bibinfo {year}
  {2019})}\BibitemShut {NoStop}%
\bibitem [{\citenamefont {Schipper}\ \emph {et~al.}(1997)\citenamefont
  {Schipper}, \citenamefont {Gritsenko},\ and\ \citenamefont
  {Baerends}}]{Schipper1997}%
  \BibitemOpen
  \bibfield  {author} {\bibinfo {author} {\bibfnamefont {P.~R.~T.}\
  \bibnamefont {Schipper}}, \bibinfo {author} {\bibfnamefont {O.~V.}\
  \bibnamefont {Gritsenko}},\ and\ \bibinfo {author} {\bibfnamefont {E.~J.}\
  \bibnamefont {Baerends}},\ }\bibfield  {title} {\bibinfo {title} {Kohn-sham
  potentials corresponding to slater and gaussian basis set densities},\ }\href
  {https://doi.org/10.1007/s002140050273} {\bibfield  {journal} {\bibinfo
  {journal} {Theoretical Chemistry Accounts}\ }\textbf {\bibinfo {volume}
  {98}},\ \bibinfo {pages} {16} (\bibinfo {year} {1997})}\BibitemShut {NoStop}%
\bibitem [{\citenamefont {Mura}\ \emph {et~al.}(1997)\citenamefont {Mura},
  \citenamefont {Knowles},\ and\ \citenamefont {Reynolds}}]{Mura1997}%
  \BibitemOpen
  \bibfield  {author} {\bibinfo {author} {\bibfnamefont {M.~E.}\ \bibnamefont
  {Mura}}, \bibinfo {author} {\bibfnamefont {P.~J.}\ \bibnamefont {Knowles}},\
  and\ \bibinfo {author} {\bibfnamefont {C.~A.}\ \bibnamefont {Reynolds}},\
  }\bibfield  {title} {\bibinfo {title} {Accurate numerical determination of
  kohn-sham potentials from electronic densities: I. two-electron systems},\
  }\href {https://doi.org/10.1063/1.473838} {\bibfield  {journal} {\bibinfo
  {journal} {The Journal of Chemical Physics}\ }\textbf {\bibinfo {volume}
  {106}},\ \bibinfo {pages} {9659} (\bibinfo {year} {1997})}\BibitemShut
  {NoStop}%
\bibitem [{\citenamefont {Heaton-Burgess}\ \emph {et~al.}(2007)\citenamefont
  {Heaton-Burgess}, \citenamefont {Bulat},\ and\ \citenamefont
  {Yang}}]{Heaton-Burgess2007}%
  \BibitemOpen
  \bibfield  {author} {\bibinfo {author} {\bibfnamefont {T.}~\bibnamefont
  {Heaton-Burgess}}, \bibinfo {author} {\bibfnamefont {F.~A.}\ \bibnamefont
  {Bulat}},\ and\ \bibinfo {author} {\bibfnamefont {W.}~\bibnamefont {Yang}},\
  }\bibfield  {title} {\bibinfo {title} {Optimized effective potentials in
  finite basis sets},\ }\href {https://doi.org/10.1103/PhysRevLett.98.256401}
  {\bibfield  {journal} {\bibinfo  {journal} {Phys. Rev. Lett.}\ }\textbf
  {\bibinfo {volume} {98}},\ \bibinfo {pages} {256401} (\bibinfo {year}
  {2007})}\BibitemShut {NoStop}%
\bibitem [{\citenamefont {Jacob}(2011)}]{Jacob2011}%
  \BibitemOpen
  \bibfield  {author} {\bibinfo {author} {\bibfnamefont {C.~R.}\ \bibnamefont
  {Jacob}},\ }\bibfield  {title} {\bibinfo {title} {Unambiguous optimization of
  effective potentials in finite basis sets},\ }\href
  {https://doi.org/10.1063/1.3670414} {\bibfield  {journal} {\bibinfo
  {journal} {The Journal of Chemical Physics}\ }\textbf {\bibinfo {volume}
  {135}},\ \bibinfo {pages} {244102} (\bibinfo {year} {2011})}\BibitemShut
  {NoStop}%
\bibitem [{\citenamefont {Gaiduk}\ \emph {et~al.}(2013)\citenamefont {Gaiduk},
  \citenamefont {Ryabinkin},\ and\ \citenamefont {Staroverov}}]{Gaiduk2013}%
  \BibitemOpen
  \bibfield  {author} {\bibinfo {author} {\bibfnamefont {A.~P.}\ \bibnamefont
  {Gaiduk}}, \bibinfo {author} {\bibfnamefont {I.~G.}\ \bibnamefont
  {Ryabinkin}},\ and\ \bibinfo {author} {\bibfnamefont {V.~N.}\ \bibnamefont
  {Staroverov}},\ }\bibfield  {title} {\bibinfo {title} {Removal of basis-set
  artifacts in kohn–sham potentials recovered from electron densities},\
  }\href {https://doi.org/10.1021/ct4004146} {\bibfield  {journal} {\bibinfo
  {journal} {Journal of Chemical Theory and Computation}\ }\textbf {\bibinfo
  {volume} {9}},\ \bibinfo {pages} {3959} (\bibinfo {year} {2013})}\BibitemShut
  {NoStop}%
\bibitem [{\citenamefont {Jensen}\ and\ \citenamefont
  {Wasserman}(2018)}]{Jensen2018}%
  \BibitemOpen
  \bibfield  {author} {\bibinfo {author} {\bibfnamefont {D.~S.}\ \bibnamefont
  {Jensen}}\ and\ \bibinfo {author} {\bibfnamefont {A.}~\bibnamefont
  {Wasserman}},\ }\bibfield  {title} {\bibinfo {title} {Numerical methods for
  the inverse problem of density functional theory},\ }\href
  {https://doi.org/https://doi.org/10.1002/qua.25425} {\bibfield  {journal}
  {\bibinfo  {journal} {International Journal of Quantum Chemistry}\ }\textbf
  {\bibinfo {volume} {118}},\ \bibinfo {pages} {e25425} (\bibinfo {year}
  {2018})}\BibitemShut {NoStop}%
\bibitem [{\citenamefont {Shi}\ and\ \citenamefont
  {Wasserman}(2021)}]{Shi2021}%
  \BibitemOpen
  \bibfield  {author} {\bibinfo {author} {\bibfnamefont {Y.}~\bibnamefont
  {Shi}}\ and\ \bibinfo {author} {\bibfnamefont {A.}~\bibnamefont
  {Wasserman}},\ }\bibfield  {title} {\bibinfo {title} {Inverse kohn–sham
  density functional theory: Progress and challenges},\ }\href
  {https://doi.org/10.1021/acs.jpclett.1c00752} {\bibfield  {journal} {\bibinfo
   {journal} {The Journal of Physical Chemistry Letters}\ }\textbf {\bibinfo
  {volume} {12}},\ \bibinfo {pages} {5308} (\bibinfo {year}
  {2021})}\BibitemShut {NoStop}%
\bibitem [{\citenamefont {Ryabinkin}\ and\ \citenamefont
  {Staroverov}(2012)}]{Ryabinkin2012}%
  \BibitemOpen
  \bibfield  {author} {\bibinfo {author} {\bibfnamefont {I.~G.}\ \bibnamefont
  {Ryabinkin}}\ and\ \bibinfo {author} {\bibfnamefont {V.~N.}\ \bibnamefont
  {Staroverov}},\ }\bibfield  {title} {\bibinfo {title} {Determination of
  kohn–sham effective potentials from electron densities using the
  differential virial theorem},\ }\href {https://doi.org/10.1063/1.4763481}
  {\bibfield  {journal} {\bibinfo  {journal} {The Journal of Chemical Physics}\
  }\textbf {\bibinfo {volume} {137}},\ \bibinfo {pages} {164113} (\bibinfo
  {year} {2012})}\BibitemShut {NoStop}%
\bibitem [{\citenamefont {Ou}\ and\ \citenamefont {Carter}(2018)}]{Ou2018}%
  \BibitemOpen
  \bibfield  {author} {\bibinfo {author} {\bibfnamefont {Q.}~\bibnamefont
  {Ou}}\ and\ \bibinfo {author} {\bibfnamefont {E.~A.}\ \bibnamefont
  {Carter}},\ }\bibfield  {title} {\bibinfo {title} {Potential functional
  embedding theory with an improved kohn–sham inversion algorithm},\ }\href
  {https://doi.org/10.1021/acs.jctc.8b00717} {\bibfield  {journal} {\bibinfo
  {journal} {Journal of Chemical Theory and Computation}\ }\textbf {\bibinfo
  {volume} {14}},\ \bibinfo {pages} {5680} (\bibinfo {year} {2018})},\ \bibinfo
  {note} {pMID: 30216062}\BibitemShut {NoStop}%
\bibitem [{\citenamefont {Kanungo}\ \emph {et~al.}(2019)\citenamefont
  {Kanungo}, \citenamefont {Zimmerman},\ and\ \citenamefont
  {Gavini}}]{Kanungo2019}%
  \BibitemOpen
  \bibfield  {author} {\bibinfo {author} {\bibfnamefont {B.}~\bibnamefont
  {Kanungo}}, \bibinfo {author} {\bibfnamefont {P.~M.}\ \bibnamefont
  {Zimmerman}},\ and\ \bibinfo {author} {\bibfnamefont {V.}~\bibnamefont
  {Gavini}},\ }\bibfield  {title} {\bibinfo {title} {Exact exchange-correlation
  potentials from ground-state electron densities},\ }\href
  {https://doi.org/10.1038/s41467-019-12467-0} {\bibfield  {journal} {\bibinfo
  {journal} {Nature Communications}\ }\textbf {\bibinfo {volume} {10}},\
  \bibinfo {pages} {4497} (\bibinfo {year} {2019})}\BibitemShut {NoStop}%
\bibitem [{\citenamefont {Kumar}\ \emph {et~al.}(2019)\citenamefont {Kumar},
  \citenamefont {Singh},\ and\ \citenamefont {Harbola}}]{Kumar2019}%
  \BibitemOpen
  \bibfield  {author} {\bibinfo {author} {\bibfnamefont {A.}~\bibnamefont
  {Kumar}}, \bibinfo {author} {\bibfnamefont {R.}~\bibnamefont {Singh}},\ and\
  \bibinfo {author} {\bibfnamefont {M.~K.}\ \bibnamefont {Harbola}},\
  }\bibfield  {title} {\bibinfo {title} {Universal nature of different methods
  of obtaining the exact kohn{\textendash}sham exchange-correlation potential
  for a given density},\ }\href {https://doi.org/10.1088/1361-6455/ab04e8}
  {\bibfield  {journal} {\bibinfo  {journal} {Journal of Physics B: Atomic,
  Molecular and Optical Physics}\ }\textbf {\bibinfo {volume} {52}},\ \bibinfo
  {pages} {075007} (\bibinfo {year} {2019})}\BibitemShut {NoStop}%
\bibitem [{\citenamefont {Kumar}\ and\ \citenamefont
  {Harbola}(2020)}]{Kumar2020}%
  \BibitemOpen
  \bibfield  {author} {\bibinfo {author} {\bibfnamefont {A.}~\bibnamefont
  {Kumar}}\ and\ \bibinfo {author} {\bibfnamefont {M.~K.}\ \bibnamefont
  {Harbola}},\ }\bibfield  {title} {\bibinfo {title} {A general penalty method
  for density-to-potential inversion},\ }\href
  {https://doi.org/https://doi.org/10.1002/qua.26400} {\bibfield  {journal}
  {\bibinfo  {journal} {International Journal of Quantum Chemistry}\ }\textbf
  {\bibinfo {volume} {120}},\ \bibinfo {pages} {e26400} (\bibinfo {year}
  {2020})}\BibitemShut {NoStop}%
\bibitem [{\citenamefont {Kumar}\ \emph {et~al.}(2020)\citenamefont {Kumar},
  \citenamefont {Singh},\ and\ \citenamefont {Harbola}}]{Kumar2020a}%
  \BibitemOpen
  \bibfield  {author} {\bibinfo {author} {\bibfnamefont {A.}~\bibnamefont
  {Kumar}}, \bibinfo {author} {\bibfnamefont {R.}~\bibnamefont {Singh}},\ and\
  \bibinfo {author} {\bibfnamefont {M.~K.}\ \bibnamefont {Harbola}},\
  }\bibfield  {title} {\bibinfo {title} {Accurate effective potential for
  density amplitude and the corresponding kohn{\textendash}sham
  exchange{\textendash}correlation potential calculated from approximate
  wavefunctions},\ }\href {https://doi.org/10.1088/1361-6455/ab9768} {\bibfield
   {journal} {\bibinfo  {journal} {Journal of Physics B: Atomic, Molecular and
  Optical Physics}\ }\textbf {\bibinfo {volume} {53}},\ \bibinfo {pages}
  {165002} (\bibinfo {year} {2020})}\BibitemShut {NoStop}%
\bibitem [{\citenamefont {Callow}\ \emph {et~al.}(2020)\citenamefont {Callow},
  \citenamefont {Lathiotakis},\ and\ \citenamefont {Gidopoulos}}]{Callow2020}%
  \BibitemOpen
  \bibfield  {author} {\bibinfo {author} {\bibfnamefont {T.~J.}\ \bibnamefont
  {Callow}}, \bibinfo {author} {\bibfnamefont {N.~N.}\ \bibnamefont
  {Lathiotakis}},\ and\ \bibinfo {author} {\bibfnamefont {N.~I.}\ \bibnamefont
  {Gidopoulos}},\ }\bibfield  {title} {\bibinfo {title} {Density-inversion
  method for the kohn–sham potential: Role of the screening density},\ }\href
  {https://doi.org/10.1063/5.0005781} {\bibfield  {journal} {\bibinfo
  {journal} {The Journal of Chemical Physics}\ }\textbf {\bibinfo {volume}
  {152}},\ \bibinfo {pages} {164114} (\bibinfo {year} {2020})}\BibitemShut
  {NoStop}%
\bibitem [{\citenamefont {Nam}\ \emph {et~al.}(2021)\citenamefont {Nam},
  \citenamefont {McCarty}, \citenamefont {Park},\ and\ \citenamefont
  {Sim}}]{Nam2021}%
  \BibitemOpen
  \bibfield  {author} {\bibinfo {author} {\bibfnamefont {S.}~\bibnamefont
  {Nam}}, \bibinfo {author} {\bibfnamefont {R.~J.}\ \bibnamefont {McCarty}},
  \bibinfo {author} {\bibfnamefont {H.}~\bibnamefont {Park}},\ and\ \bibinfo
  {author} {\bibfnamefont {E.}~\bibnamefont {Sim}},\ }\bibfield  {title}
  {\bibinfo {title} {Ks-pies: Kohn–sham inversion toolkit},\ }\href
  {https://doi.org/10.1063/5.0040941} {\bibfield  {journal} {\bibinfo
  {journal} {The Journal of Chemical Physics}\ }\textbf {\bibinfo {volume}
  {154}},\ \bibinfo {pages} {124122} (\bibinfo {year} {2021})}\BibitemShut
  {NoStop}%
\bibitem [{\citenamefont {Erhard}\ \emph {et~al.}(0)\citenamefont {Erhard},
  \citenamefont {Trushin},\ and\ \citenamefont {Görling}}]{erhard2022}%
  \BibitemOpen
  \bibfield  {author} {\bibinfo {author} {\bibfnamefont {J.}~\bibnamefont
  {Erhard}}, \bibinfo {author} {\bibfnamefont {E.}~\bibnamefont {Trushin}},\
  and\ \bibinfo {author} {\bibfnamefont {A.}~\bibnamefont {Görling}},\
  }\bibfield  {title} {\bibinfo {title} {Numerically stable inversion approach
  to construct kohn-sham potentials for given electron densities within a
  gaussian basis set framework},\ }\href {https://doi.org/10.1063/5.0087356}
  {\bibfield  {journal} {\bibinfo  {journal} {The Journal of Chemical Physics}\
  }\textbf {\bibinfo {volume} {0}},\ \bibinfo {pages} {null} (\bibinfo {year}
  {0})}\BibitemShut {NoStop}%
\bibitem [{\citenamefont {Shi}\ \emph {et~al.}()\citenamefont {Shi},
  \citenamefont {Chávez},\ and\ \citenamefont {Wasserman}}]{Shi2022}%
  \BibitemOpen
  \bibfield  {author} {\bibinfo {author} {\bibfnamefont {Y.}~\bibnamefont
  {Shi}}, \bibinfo {author} {\bibfnamefont {V.~H.}\ \bibnamefont {Chávez}},\
  and\ \bibinfo {author} {\bibfnamefont {A.}~\bibnamefont {Wasserman}},\
  }\bibfield  {title} {\bibinfo {title} {n2v: A density-to-potential inversion
  suite. a sandbox for creating, testing, and benchmarking density functional
  theory inversion methods},\ }\href
  {https://doi.org/https://doi.org/10.1002/wcms.1617} {\bibfield  {journal}
  {\bibinfo  {journal} {WIREs Computational Molecular Science}\ }\textbf
  {\bibinfo {volume} {n/a}},\ \bibinfo {pages} {e1617}}\BibitemShut {NoStop}%
\bibitem [{\citenamefont {Jayatilaka}(1998)}]{Jayatilaka1998}%
  \BibitemOpen
  \bibfield  {author} {\bibinfo {author} {\bibfnamefont {D.}~\bibnamefont
  {Jayatilaka}},\ }\bibfield  {title} {\bibinfo {title} {Wave function for
  beryllium from x-ray diffraction data},\ }\href
  {https://doi.org/10.1103/PhysRevLett.80.798} {\bibfield  {journal} {\bibinfo
  {journal} {Phys. Rev. Lett.}\ }\textbf {\bibinfo {volume} {80}},\ \bibinfo
  {pages} {798} (\bibinfo {year} {1998})}\BibitemShut {NoStop}%
\bibitem [{Note1()}]{Note1}%
  \BibitemOpen
  \bibinfo {note} {Given the pseudopotential. In other words, all discussions
  are valid because we have used the same hamiltonian, with a fixed LDA
  pseudopotential, for the valence electron problem throughout, including for
  the QMC calculations. A full all-electron $v_{\protect \rm xc}$ would of
  course look different.}\BibitemShut {Stop}%
\bibitem [{\citenamefont {Zhang}\ and\ \citenamefont
  {Krakauer}(2003)}]{Zhang2003}%
  \BibitemOpen
  \bibfield  {author} {\bibinfo {author} {\bibfnamefont {S.}~\bibnamefont
  {Zhang}}\ and\ \bibinfo {author} {\bibfnamefont {H.}~\bibnamefont
  {Krakauer}},\ }\bibfield  {title} {\bibinfo {title} {Quantum monte carlo
  method using phase-free random walks with slater determinants},\ }\href
  {https://doi.org/10.1103/PhysRevLett.90.136401} {\bibfield  {journal}
  {\bibinfo  {journal} {Phys. Rev. Lett.}\ }\textbf {\bibinfo {volume} {90}},\
  \bibinfo {pages} {136401} (\bibinfo {year} {2003})}\BibitemShut {NoStop}%
\bibitem [{\citenamefont {Motta}\ and\ \citenamefont
  {Zhang}(2018)}]{Motta2018}%
  \BibitemOpen
  \bibfield  {author} {\bibinfo {author} {\bibfnamefont {M.}~\bibnamefont
  {Motta}}\ and\ \bibinfo {author} {\bibfnamefont {S.}~\bibnamefont {Zhang}},\
  }\bibfield  {title} {\bibinfo {title} {Ab initio computations of molecular
  systems by the auxiliary-field quantum monte carlo method},\ }\href@noop {}
  {\bibfield  {journal} {\bibinfo  {journal} {WIREs Computational Molecular
  Science}\ }\textbf {\bibinfo {volume} {8}},\ \bibinfo {pages} {e1364}
  (\bibinfo {year} {2018})}\BibitemShut {NoStop}%
\bibitem [{\citenamefont {Chen}\ \emph {et~al.}(2021)\citenamefont {Chen},
  \citenamefont {Motta}, \citenamefont {Ma},\ and\ \citenamefont
  {Zhang}}]{SiyuanChen}%
  \BibitemOpen
  \bibfield  {author} {\bibinfo {author} {\bibfnamefont {S.}~\bibnamefont
  {Chen}}, \bibinfo {author} {\bibfnamefont {M.}~\bibnamefont {Motta}},
  \bibinfo {author} {\bibfnamefont {F.}~\bibnamefont {Ma}},\ and\ \bibinfo
  {author} {\bibfnamefont {S.}~\bibnamefont {Zhang}},\ }\bibfield  {title}
  {\bibinfo {title} {Ab initio electronic density in solids by many-body
  plane-wave auxiliary-field quantum monte carlo calculations},\ }\href
  {https://doi.org/10.1103/PhysRevB.103.075138} {\bibfield  {journal} {\bibinfo
   {journal} {Phys. Rev. B}\ }\textbf {\bibinfo {volume} {103}},\ \bibinfo
  {pages} {075138} (\bibinfo {year} {2021})}\BibitemShut {NoStop}%
\bibitem [{\citenamefont {Nazarov}(2021)}]{Nazarov2021}%
  \BibitemOpen
  \bibfield  {author} {\bibinfo {author} {\bibfnamefont {V.~U.}\ \bibnamefont
  {Nazarov}},\ }\bibfield  {title} {\bibinfo {title} {Breakdown of the
  ionization potential theorem of density functional theory in mesoscopic
  systems},\ }\href {https://doi.org/10.1063/5.0070429} {\bibfield  {journal}
  {\bibinfo  {journal} {The Journal of Chemical Physics}\ }\textbf {\bibinfo
  {volume} {155}},\ \bibinfo {pages} {194105} (\bibinfo {year}
  {2021})}\BibitemShut {NoStop}%
\bibitem [{Note2()}]{Note2}%
  \BibitemOpen
  \bibinfo {note} {Note that we have defined the error with opposite sign with
  respect to Ref. \cite {SiyuanChen}}\BibitemShut {NoStop}%
\bibitem [{\citenamefont {Tran}\ and\ \citenamefont
  {Blaha}(2009)}]{TranBlaha2009}%
  \BibitemOpen
  \bibfield  {author} {\bibinfo {author} {\bibfnamefont {F.}~\bibnamefont
  {Tran}}\ and\ \bibinfo {author} {\bibfnamefont {P.}~\bibnamefont {Blaha}},\
  }\bibfield  {title} {\bibinfo {title} {Accurate band gaps of semiconductors
  and insulators with a semilocal exchange-correlation potential},\ }\href
  {https://doi.org/10.1103/PhysRevLett.102.226401} {\bibfield  {journal}
  {\bibinfo  {journal} {Phys. Rev. Lett.}\ }\textbf {\bibinfo {volume} {102}},\
  \bibinfo {pages} {226401} (\bibinfo {year} {2009})}\BibitemShut {NoStop}%
\bibitem [{\citenamefont {Perdew}\ \emph {et~al.}(1992)\citenamefont {Perdew},
  \citenamefont {Chevary}, \citenamefont {Vosko}, \citenamefont {Jackson},
  \citenamefont {Pederson}, \citenamefont {Singh},\ and\ \citenamefont
  {Fiolhais}}]{PerdewChevary1992}%
  \BibitemOpen
  \bibfield  {author} {\bibinfo {author} {\bibfnamefont {J.~P.}\ \bibnamefont
  {Perdew}}, \bibinfo {author} {\bibfnamefont {J.~A.}\ \bibnamefont {Chevary}},
  \bibinfo {author} {\bibfnamefont {S.~H.}\ \bibnamefont {Vosko}}, \bibinfo
  {author} {\bibfnamefont {K.~A.}\ \bibnamefont {Jackson}}, \bibinfo {author}
  {\bibfnamefont {M.~R.}\ \bibnamefont {Pederson}}, \bibinfo {author}
  {\bibfnamefont {D.~J.}\ \bibnamefont {Singh}},\ and\ \bibinfo {author}
  {\bibfnamefont {C.}~\bibnamefont {Fiolhais}},\ }\bibfield  {title} {\bibinfo
  {title} {Atoms, molecules, solids, and surfaces: Applications of the
  generalized gradient approximation for exchange and correlation},\ }\href
  {https://doi.org/10.1103/PhysRevB.46.6671} {\bibfield  {journal} {\bibinfo
  {journal} {Phys. Rev. B}\ }\textbf {\bibinfo {volume} {46}},\ \bibinfo
  {pages} {6671} (\bibinfo {year} {1992})}\BibitemShut {NoStop}%
\bibitem [{\citenamefont {Kalita}\ \emph {et~al.}(2021)\citenamefont {Kalita},
  \citenamefont {Li}, \citenamefont {McCarty},\ and\ \citenamefont
  {Burke}}]{Kalita2021}%
  \BibitemOpen
  \bibfield  {author} {\bibinfo {author} {\bibfnamefont {B.}~\bibnamefont
  {Kalita}}, \bibinfo {author} {\bibfnamefont {L.}~\bibnamefont {Li}}, \bibinfo
  {author} {\bibfnamefont {R.~J.}\ \bibnamefont {McCarty}},\ and\ \bibinfo
  {author} {\bibfnamefont {K.}~\bibnamefont {Burke}},\ }\bibfield  {title}
  {\bibinfo {title} {Learning to approximate density functionals},\ }\href
  {https://doi.org/10.1021/acs.accounts.0c00742} {\bibfield  {journal}
  {\bibinfo  {journal} {Accounts of Chemical Research}\ }\textbf {\bibinfo
  {volume} {54}},\ \bibinfo {pages} {818} (\bibinfo {year} {2021})}\BibitemShut
  {NoStop}%
\bibitem [{\citenamefont {Chong}\ \emph {et~al.}(2002)\citenamefont {Chong},
  \citenamefont {Gritsenko},\ and\ \citenamefont {Baerends}}]{Chong2002}%
  \BibitemOpen
  \bibfield  {author} {\bibinfo {author} {\bibfnamefont {D.~P.}\ \bibnamefont
  {Chong}}, \bibinfo {author} {\bibfnamefont {O.~V.}\ \bibnamefont
  {Gritsenko}},\ and\ \bibinfo {author} {\bibfnamefont {E.~J.}\ \bibnamefont
  {Baerends}},\ }\bibfield  {title} {\bibinfo {title} {Interpretation of the
  kohn–sham orbital energies as approximate vertical ionization potentials},\
  }\href {https://doi.org/10.1063/1.1430255} {\bibfield  {journal} {\bibinfo
  {journal} {The Journal of Chemical Physics}\ }\textbf {\bibinfo {volume}
  {116}},\ \bibinfo {pages} {1760} (\bibinfo {year} {2002})}\BibitemShut
  {NoStop}%
\bibitem [{\citenamefont {Filippi}\ \emph {et~al.}(1997)\citenamefont
  {Filippi}, \citenamefont {Umrigar},\ and\ \citenamefont
  {Gonze}}]{Filippi1997}%
  \BibitemOpen
  \bibfield  {author} {\bibinfo {author} {\bibfnamefont {C.}~\bibnamefont
  {Filippi}}, \bibinfo {author} {\bibfnamefont {C.~J.}\ \bibnamefont
  {Umrigar}},\ and\ \bibinfo {author} {\bibfnamefont {X.}~\bibnamefont
  {Gonze}},\ }\bibfield  {title} {\bibinfo {title} {Excitation energies from
  density functional perturbation theory},\ }\href
  {https://doi.org/10.1063/1.475304} {\bibfield  {journal} {\bibinfo  {journal}
  {The Journal of Chemical Physics}\ }\textbf {\bibinfo {volume} {107}},\
  \bibinfo {pages} {9994} (\bibinfo {year} {1997})}\BibitemShut {NoStop}%
\bibitem [{\citenamefont {Martin}\ \emph {et~al.}(2016)\citenamefont {Martin},
  \citenamefont {Reining},\ and\ \citenamefont {Ceperley}}]{Martin2016}%
  \BibitemOpen
  \bibfield  {author} {\bibinfo {author} {\bibfnamefont {R.}~\bibnamefont
  {Martin}}, \bibinfo {author} {\bibfnamefont {L.}~\bibnamefont {Reining}},\
  and\ \bibinfo {author} {\bibfnamefont {D.}~\bibnamefont {Ceperley}},\
  }\href@noop {} {\emph {\bibinfo {title} {Interacting Electrons: Theory and
  Computational Approaches}}}\ (\bibinfo  {publisher} {Cambridge University
  Press},\ \bibinfo {year} {2016})\BibitemShut {NoStop}%
\bibitem [{\citenamefont {Lannoo}\ \emph {et~al.}(1985)\citenamefont {Lannoo},
  \citenamefont {Schl\"uter},\ and\ \citenamefont {Sham}}]{Lannoo1985}%
  \BibitemOpen
  \bibfield  {author} {\bibinfo {author} {\bibfnamefont {M.}~\bibnamefont
  {Lannoo}}, \bibinfo {author} {\bibfnamefont {M.}~\bibnamefont {Schl\"uter}},\
  and\ \bibinfo {author} {\bibfnamefont {L.~J.}\ \bibnamefont {Sham}},\
  }\bibfield  {title} {\bibinfo {title} {Calculation of the kohn-sham potential
  and its discontinuity for a model-semiconductor},\ }\href
  {https://doi.org/10.1103/PhysRevB.32.3890} {\bibfield  {journal} {\bibinfo
  {journal} {Phys. Rev. B}\ }\textbf {\bibinfo {volume} {32}},\ \bibinfo
  {pages} {3890} (\bibinfo {year} {1985})}\BibitemShut {NoStop}%
\bibitem [{\citenamefont {Eguiluz}\ \emph {et~al.}(1992)\citenamefont
  {Eguiluz}, \citenamefont {Heinrichsmeier}, \citenamefont {Fleszar},\ and\
  \citenamefont {Hanke}}]{Eguiluz1992}%
  \BibitemOpen
  \bibfield  {author} {\bibinfo {author} {\bibfnamefont {A.~G.}\ \bibnamefont
  {Eguiluz}}, \bibinfo {author} {\bibfnamefont {M.}~\bibnamefont
  {Heinrichsmeier}}, \bibinfo {author} {\bibfnamefont {A.}~\bibnamefont
  {Fleszar}},\ and\ \bibinfo {author} {\bibfnamefont {W.}~\bibnamefont
  {Hanke}},\ }\bibfield  {title} {\bibinfo {title} {First-principles evaluation
  of the surface barrier for a kohn-sham electron at a metal surface},\ }\href
  {https://doi.org/10.1103/PhysRevLett.68.1359} {\bibfield  {journal} {\bibinfo
   {journal} {Phys. Rev. Lett.}\ }\textbf {\bibinfo {volume} {68}},\ \bibinfo
  {pages} {1359} (\bibinfo {year} {1992})}\BibitemShut {NoStop}%
\bibitem [{\citenamefont {Hellgren}\ and\ \citenamefont {von
  Barth}(2007)}]{Hellgren2007}%
  \BibitemOpen
  \bibfield  {author} {\bibinfo {author} {\bibfnamefont {M.}~\bibnamefont
  {Hellgren}}\ and\ \bibinfo {author} {\bibfnamefont {U.}~\bibnamefont {von
  Barth}},\ }\bibfield  {title} {\bibinfo {title} {Correlation potential in
  density functional theory at the gwa level: Spherical atoms},\ }\href
  {https://doi.org/10.1103/PhysRevB.76.075107} {\bibfield  {journal} {\bibinfo
  {journal} {Phys. Rev. B}\ }\textbf {\bibinfo {volume} {76}},\ \bibinfo
  {pages} {075107} (\bibinfo {year} {2007})}\BibitemShut {NoStop}%
\bibitem [{\citenamefont {Hellgren}\ and\ \citenamefont {von
  Barth}(2010)}]{Hellgren2010}%
  \BibitemOpen
  \bibfield  {author} {\bibinfo {author} {\bibfnamefont {M.}~\bibnamefont
  {Hellgren}}\ and\ \bibinfo {author} {\bibfnamefont {U.}~\bibnamefont {von
  Barth}},\ }\bibfield  {title} {\bibinfo {title} {Correlation energy
  functional and potential from time-dependent exact-exchange theory},\ }\href
  {https://doi.org/10.1063/1.3290947} {\bibfield  {journal} {\bibinfo
  {journal} {The Journal of Chemical Physics}\ }\textbf {\bibinfo {volume}
  {132}},\ \bibinfo {pages} {044101} (\bibinfo {year} {2010})}\BibitemShut
  {NoStop}%
\bibitem [{\citenamefont {Bleiziffer}\ \emph {et~al.}(2013)\citenamefont
  {Bleiziffer}, \citenamefont {Heßelmann},\ and\ \citenamefont
  {Görling}}]{Bleiziffer2013}%
  \BibitemOpen
  \bibfield  {author} {\bibinfo {author} {\bibfnamefont {P.}~\bibnamefont
  {Bleiziffer}}, \bibinfo {author} {\bibfnamefont {A.}~\bibnamefont
  {Heßelmann}},\ and\ \bibinfo {author} {\bibfnamefont {A.}~\bibnamefont
  {Görling}},\ }\bibfield  {title} {\bibinfo {title} {Efficient
  self-consistent treatment of electron correlation within the random phase
  approximation},\ }\href {https://doi.org/10.1063/1.4818984} {\bibfield
  {journal} {\bibinfo  {journal} {The Journal of Chemical Physics}\ }\textbf
  {\bibinfo {volume} {139}},\ \bibinfo {pages} {084113} (\bibinfo {year}
  {2013})}\BibitemShut {NoStop}%
\bibitem [{\citenamefont {Hedin}(1965)}]{Hedin-GW1965}%
  \BibitemOpen
  \bibfield  {author} {\bibinfo {author} {\bibfnamefont {L.}~\bibnamefont
  {Hedin}},\ }\bibfield  {title} {\bibinfo {title} {New method for calculating
  the one-particle green's function with application to the electron-gas
  problem},\ }\href {https://doi.org/10.1103/PhysRev.139.A796} {\bibfield
  {journal} {\bibinfo  {journal} {Phys. Rev.}\ }\textbf {\bibinfo {volume}
  {139}},\ \bibinfo {pages} {A796} (\bibinfo {year} {1965})}\BibitemShut
  {NoStop}%
\bibitem [{\citenamefont {Landolt-B\"ornstein}(1982)}]{silicon-gap}%
  \BibitemOpen
  \bibfield  {author} {\bibinfo {author} {\bibnamefont {Landolt-B\"ornstein}},\
  }\href@noop {} {\emph {\bibinfo {title} {Numerical Data and Functional
  Relationships in Science and Technology}}},\ \bibinfo {series} {New Series
  Group III, Subvolume B: II-VII and I-VII Compounds}, Vol.\ \bibinfo {volume}
  {17 Pt A}\ (\bibinfo  {publisher} {Springer-Verlag},\ \bibinfo {address}
  {Berlin-Heidelberg, New York},\ \bibinfo {year} {1982})\BibitemShut {NoStop}%
\bibitem [{\citenamefont {Poole}\ \emph {et~al.}(1975)\citenamefont {Poole},
  \citenamefont {Jenkin}, \citenamefont {Liesegang},\ and\ \citenamefont
  {Leckey}}]{nacl-gap}%
  \BibitemOpen
  \bibfield  {author} {\bibinfo {author} {\bibfnamefont {R.~T.}\ \bibnamefont
  {Poole}}, \bibinfo {author} {\bibfnamefont {J.~G.}\ \bibnamefont {Jenkin}},
  \bibinfo {author} {\bibfnamefont {J.}~\bibnamefont {Liesegang}},\ and\
  \bibinfo {author} {\bibfnamefont {R.~C.~G.}\ \bibnamefont {Leckey}},\
  }\bibfield  {title} {\bibinfo {title} {Electronic band structure of the
  alkali halides. i. experimental parameters},\ }\href
  {https://doi.org/10.1103/PhysRevB.11.5179} {\bibfield  {journal} {\bibinfo
  {journal} {Phys. Rev. B}\ }\textbf {\bibinfo {volume} {11}},\ \bibinfo
  {pages} {5179} (\bibinfo {year} {1975})}\BibitemShut {NoStop}%
\bibitem [{Note3()}]{Note3}%
  \BibitemOpen
  \bibinfo {note} {Note that the difference between the exact KS band structure
  and the true quasiparticle one is not limited to a rigid shift of conduction
  states, since the KS potential is only supposed to correctly describe the
  highest occupied state of the N-electron and (when the derivative
  discontinuity is taken into account) N+1-electron systems.}\BibitemShut
  {Stop}%
\bibitem [{\citenamefont {Ortega}\ and\ \citenamefont
  {Himpsel}(1993)}]{Ortega1993}%
  \BibitemOpen
  \bibfield  {author} {\bibinfo {author} {\bibfnamefont {J.~E.}\ \bibnamefont
  {Ortega}}\ and\ \bibinfo {author} {\bibfnamefont {F.~J.}\ \bibnamefont
  {Himpsel}},\ }\bibfield  {title} {\bibinfo {title} {Inverse-photoemission
  study of ge(100), si(100), and gaas(100): Bulk bands and surface states},\
  }\href {https://doi.org/10.1103/PhysRevB.47.2130} {\bibfield  {journal}
  {\bibinfo  {journal} {Phys. Rev. B}\ }\textbf {\bibinfo {volume} {47}},\
  \bibinfo {pages} {2130} (\bibinfo {year} {1993})}\BibitemShut {NoStop}%
\bibitem [{\citenamefont {Savin}\ \emph {et~al.}(2003)\citenamefont {Savin},
  \citenamefont {Colonna},\ and\ \citenamefont {Pollet}}]{Savin2003}%
  \BibitemOpen
  \bibfield  {author} {\bibinfo {author} {\bibfnamefont {A.}~\bibnamefont
  {Savin}}, \bibinfo {author} {\bibfnamefont {F.}~\bibnamefont {Colonna}},\
  and\ \bibinfo {author} {\bibfnamefont {R.}~\bibnamefont {Pollet}},\
  }\bibfield  {title} {\bibinfo {title} {Adiabatic connection approach to
  density functional theory of electronic systems},\ }\href
  {https://doi.org/https://doi.org/10.1002/qua.10551} {\bibfield  {journal}
  {\bibinfo  {journal} {International Journal of Quantum Chemistry}\ }\textbf
  {\bibinfo {volume} {93}},\ \bibinfo {pages} {166} (\bibinfo {year}
  {2003})}\BibitemShut {NoStop}%
\bibitem [{\citenamefont {Kim}\ \emph {et~al.}(2013)\citenamefont {Kim},
  \citenamefont {Sim},\ and\ \citenamefont {Burke}}]{Kim2013}%
  \BibitemOpen
  \bibfield  {author} {\bibinfo {author} {\bibfnamefont {M.-C.}\ \bibnamefont
  {Kim}}, \bibinfo {author} {\bibfnamefont {E.}~\bibnamefont {Sim}},\ and\
  \bibinfo {author} {\bibfnamefont {K.}~\bibnamefont {Burke}},\ }\bibfield
  {title} {\bibinfo {title} {Understanding and reducing errors in density
  functional calculations},\ }\href
  {https://doi.org/10.1103/PhysRevLett.111.073003} {\bibfield  {journal}
  {\bibinfo  {journal} {Phys. Rev. Lett.}\ }\textbf {\bibinfo {volume} {111}},\
  \bibinfo {pages} {073003} (\bibinfo {year} {2013})}\BibitemShut {NoStop}%
\bibitem [{\citenamefont {Wasserman}\ \emph {et~al.}(2017)\citenamefont
  {Wasserman}, \citenamefont {Nafziger}, \citenamefont {Jiang}, \citenamefont
  {Kim}, \citenamefont {Sim},\ and\ \citenamefont {Burke}}]{Wasserman2017}%
  \BibitemOpen
  \bibfield  {author} {\bibinfo {author} {\bibfnamefont {A.}~\bibnamefont
  {Wasserman}}, \bibinfo {author} {\bibfnamefont {J.}~\bibnamefont {Nafziger}},
  \bibinfo {author} {\bibfnamefont {K.}~\bibnamefont {Jiang}}, \bibinfo
  {author} {\bibfnamefont {M.-C.}\ \bibnamefont {Kim}}, \bibinfo {author}
  {\bibfnamefont {E.}~\bibnamefont {Sim}},\ and\ \bibinfo {author}
  {\bibfnamefont {K.}~\bibnamefont {Burke}},\ }\bibfield  {title} {\bibinfo
  {title} {The importance of being inconsistent},\ }\href
  {https://doi.org/10.1146/annurev-physchem-052516-044957} {\bibfield
  {journal} {\bibinfo  {journal} {Annual Review of Physical Chemistry}\
  }\textbf {\bibinfo {volume} {68}},\ \bibinfo {pages} {555} (\bibinfo {year}
  {2017})}\BibitemShut {NoStop}%
\bibitem [{\citenamefont {Hamann}(2013)}]{Hamann2013}%
  \BibitemOpen
  \bibfield  {author} {\bibinfo {author} {\bibfnamefont {D.~R.}\ \bibnamefont
  {Hamann}},\ }\bibfield  {title} {\bibinfo {title} {Optimized norm-conserving
  vanderbilt pseudopotentials},\ }\href
  {https://doi.org/10.1103/PhysRevB.88.085117} {\bibfield  {journal} {\bibinfo
  {journal} {Phys. Rev. B}\ }\textbf {\bibinfo {volume} {88}},\ \bibinfo
  {pages} {085117} (\bibinfo {year} {2013})}\BibitemShut {NoStop}%
\bibitem [{\citenamefont {Gonze}\ \emph {et~al.}(2020)\citenamefont {Gonze},
  \citenamefont {Amadon}, \citenamefont {Antonius}, \citenamefont {Arnardi},
  \citenamefont {Baguet}, \citenamefont {Beuken}, \citenamefont {Bieder},
  \citenamefont {Bottin}, \citenamefont {Bouchet}, \citenamefont {Bousquet},
  \citenamefont {Brouwer}, \citenamefont {Bruneval}, \citenamefont {Brunin},
  \citenamefont {Cavignac}, \citenamefont {Charraud}, \citenamefont {Chen},
  \citenamefont {Côté}, \citenamefont {Cottenier}, \citenamefont {Denier},
  \citenamefont {Geneste}, \citenamefont {Ghosez}, \citenamefont {Giantomassi},
  \citenamefont {Gillet}, \citenamefont {Gingras}, \citenamefont {Hamann},
  \citenamefont {Hautier}, \citenamefont {He}, \citenamefont {Helbig},
  \citenamefont {Holzwarth}, \citenamefont {Jia}, \citenamefont {Jollet},
  \citenamefont {Lafargue-Dit-Hauret}, \citenamefont {Lejaeghere},
  \citenamefont {Marques}, \citenamefont {Martin}, \citenamefont {Martins},
  \citenamefont {Miranda}, \citenamefont {Naccarato}, \citenamefont {Persson},
  \citenamefont {Petretto}, \citenamefont {Planes}, \citenamefont {Pouillon},
  \citenamefont {Prokhorenko}, \citenamefont {Ricci}, \citenamefont
  {Rignanese}, \citenamefont {Romero}, \citenamefont {Schmitt}, \citenamefont
  {Torrent}, \citenamefont {{van Setten}}, \citenamefont {{Van Troeye}},
  \citenamefont {Verstraete}, \citenamefont {Zérah},\ and\ \citenamefont
  {Zwanziger}}]{abinit2020}%
  \BibitemOpen
  \bibfield  {author} {\bibinfo {author} {\bibfnamefont {X.}~\bibnamefont
  {Gonze}}, \bibinfo {author} {\bibfnamefont {B.}~\bibnamefont {Amadon}},
  \bibinfo {author} {\bibfnamefont {G.}~\bibnamefont {Antonius}}, \bibinfo
  {author} {\bibfnamefont {F.}~\bibnamefont {Arnardi}}, \bibinfo {author}
  {\bibfnamefont {L.}~\bibnamefont {Baguet}}, \bibinfo {author} {\bibfnamefont
  {J.-M.}\ \bibnamefont {Beuken}}, \bibinfo {author} {\bibfnamefont
  {J.}~\bibnamefont {Bieder}}, \bibinfo {author} {\bibfnamefont
  {F.}~\bibnamefont {Bottin}}, \bibinfo {author} {\bibfnamefont
  {J.}~\bibnamefont {Bouchet}}, \bibinfo {author} {\bibfnamefont
  {E.}~\bibnamefont {Bousquet}}, \bibinfo {author} {\bibfnamefont
  {N.}~\bibnamefont {Brouwer}}, \bibinfo {author} {\bibfnamefont
  {F.}~\bibnamefont {Bruneval}}, \bibinfo {author} {\bibfnamefont
  {G.}~\bibnamefont {Brunin}}, \bibinfo {author} {\bibfnamefont
  {T.}~\bibnamefont {Cavignac}}, \bibinfo {author} {\bibfnamefont {J.-B.}\
  \bibnamefont {Charraud}}, \bibinfo {author} {\bibfnamefont {W.}~\bibnamefont
  {Chen}}, \bibinfo {author} {\bibfnamefont {M.}~\bibnamefont {Côté}},
  \bibinfo {author} {\bibfnamefont {S.}~\bibnamefont {Cottenier}}, \bibinfo
  {author} {\bibfnamefont {J.}~\bibnamefont {Denier}}, \bibinfo {author}
  {\bibfnamefont {G.}~\bibnamefont {Geneste}}, \bibinfo {author} {\bibfnamefont
  {P.}~\bibnamefont {Ghosez}}, \bibinfo {author} {\bibfnamefont
  {M.}~\bibnamefont {Giantomassi}}, \bibinfo {author} {\bibfnamefont
  {Y.}~\bibnamefont {Gillet}}, \bibinfo {author} {\bibfnamefont
  {O.}~\bibnamefont {Gingras}}, \bibinfo {author} {\bibfnamefont {D.~R.}\
  \bibnamefont {Hamann}}, \bibinfo {author} {\bibfnamefont {G.}~\bibnamefont
  {Hautier}}, \bibinfo {author} {\bibfnamefont {X.}~\bibnamefont {He}},
  \bibinfo {author} {\bibfnamefont {N.}~\bibnamefont {Helbig}}, \bibinfo
  {author} {\bibfnamefont {N.}~\bibnamefont {Holzwarth}}, \bibinfo {author}
  {\bibfnamefont {Y.}~\bibnamefont {Jia}}, \bibinfo {author} {\bibfnamefont
  {F.}~\bibnamefont {Jollet}}, \bibinfo {author} {\bibfnamefont
  {W.}~\bibnamefont {Lafargue-Dit-Hauret}}, \bibinfo {author} {\bibfnamefont
  {K.}~\bibnamefont {Lejaeghere}}, \bibinfo {author} {\bibfnamefont {M.~A.}\
  \bibnamefont {Marques}}, \bibinfo {author} {\bibfnamefont {A.}~\bibnamefont
  {Martin}}, \bibinfo {author} {\bibfnamefont {C.}~\bibnamefont {Martins}},
  \bibinfo {author} {\bibfnamefont {H.~P.}\ \bibnamefont {Miranda}}, \bibinfo
  {author} {\bibfnamefont {F.}~\bibnamefont {Naccarato}}, \bibinfo {author}
  {\bibfnamefont {K.}~\bibnamefont {Persson}}, \bibinfo {author} {\bibfnamefont
  {G.}~\bibnamefont {Petretto}}, \bibinfo {author} {\bibfnamefont
  {V.}~\bibnamefont {Planes}}, \bibinfo {author} {\bibfnamefont
  {Y.}~\bibnamefont {Pouillon}}, \bibinfo {author} {\bibfnamefont
  {S.}~\bibnamefont {Prokhorenko}}, \bibinfo {author} {\bibfnamefont
  {F.}~\bibnamefont {Ricci}}, \bibinfo {author} {\bibfnamefont {G.-M.}\
  \bibnamefont {Rignanese}}, \bibinfo {author} {\bibfnamefont {A.~H.}\
  \bibnamefont {Romero}}, \bibinfo {author} {\bibfnamefont {M.~M.}\
  \bibnamefont {Schmitt}}, \bibinfo {author} {\bibfnamefont {M.}~\bibnamefont
  {Torrent}}, \bibinfo {author} {\bibfnamefont {M.~J.}\ \bibnamefont {{van
  Setten}}}, \bibinfo {author} {\bibfnamefont {B.}~\bibnamefont {{Van
  Troeye}}}, \bibinfo {author} {\bibfnamefont {M.~J.}\ \bibnamefont
  {Verstraete}}, \bibinfo {author} {\bibfnamefont {G.}~\bibnamefont {Zérah}},\
  and\ \bibinfo {author} {\bibfnamefont {J.~W.}\ \bibnamefont {Zwanziger}},\
  }\bibfield  {title} {\bibinfo {title} {The abinitproject: Impact, environment
  and recent developments},\ }\href
  {https://doi.org/https://doi.org/10.1016/j.cpc.2019.107042} {\bibfield
  {journal} {\bibinfo  {journal} {Computer Physics Communications}\ }\textbf
  {\bibinfo {volume} {248}},\ \bibinfo {pages} {107042} (\bibinfo {year}
  {2020})}\BibitemShut {NoStop}%
\bibitem [{\citenamefont {Giannozzi}\ \emph {et~al.}(2017)\citenamefont
  {Giannozzi}, \citenamefont {Andreussi}, \citenamefont {Brumme}, \citenamefont
  {Bunau}, \citenamefont {Nardelli}, \citenamefont {Calandra}, \citenamefont
  {Car}, \citenamefont {Cavazzoni}, \citenamefont {Ceresoli}, \citenamefont
  {Cococcioni}, \citenamefont {Colonna}, \citenamefont {Carnimeo},
  \citenamefont {Corso}, \citenamefont {de~Gironcoli}, \citenamefont {Delugas},
  \citenamefont {DiStasio}, \citenamefont {Ferretti}, \citenamefont {Floris},
  \citenamefont {Fratesi}, \citenamefont {Fugallo}, \citenamefont {Gebauer},
  \citenamefont {Gerstmann}, \citenamefont {Giustino}, \citenamefont {Gorni},
  \citenamefont {Jia}, \citenamefont {Kawamura}, \citenamefont {Ko},
  \citenamefont {Kokalj}, \citenamefont {Kü{\c{c}}ükbenli}, \citenamefont
  {Lazzeri}, \citenamefont {Marsili}, \citenamefont {Marzari}, \citenamefont
  {Mauri}, \citenamefont {Nguyen}, \citenamefont {Nguyen}, \citenamefont {de-la
  Roza}, \citenamefont {Paulatto}, \citenamefont {Ponc{\'{e}}}, \citenamefont
  {Rocca}, \citenamefont {Sabatini}, \citenamefont {Santra}, \citenamefont
  {Schlipf}, \citenamefont {Seitsonen}, \citenamefont {Smogunov}, \citenamefont
  {Timrov}, \citenamefont {Thonhauser}, \citenamefont {Umari}, \citenamefont
  {Vast}, \citenamefont {Wu},\ and\ \citenamefont {Baroni}}]{quantum-espresso}%
  \BibitemOpen
  \bibfield  {author} {\bibinfo {author} {\bibfnamefont {P.}~\bibnamefont
  {Giannozzi}}, \bibinfo {author} {\bibfnamefont {O.}~\bibnamefont
  {Andreussi}}, \bibinfo {author} {\bibfnamefont {T.}~\bibnamefont {Brumme}},
  \bibinfo {author} {\bibfnamefont {O.}~\bibnamefont {Bunau}}, \bibinfo
  {author} {\bibfnamefont {M.~B.}\ \bibnamefont {Nardelli}}, \bibinfo {author}
  {\bibfnamefont {M.}~\bibnamefont {Calandra}}, \bibinfo {author}
  {\bibfnamefont {R.}~\bibnamefont {Car}}, \bibinfo {author} {\bibfnamefont
  {C.}~\bibnamefont {Cavazzoni}}, \bibinfo {author} {\bibfnamefont
  {D.}~\bibnamefont {Ceresoli}}, \bibinfo {author} {\bibfnamefont
  {M.}~\bibnamefont {Cococcioni}}, \bibinfo {author} {\bibfnamefont
  {N.}~\bibnamefont {Colonna}}, \bibinfo {author} {\bibfnamefont
  {I.}~\bibnamefont {Carnimeo}}, \bibinfo {author} {\bibfnamefont {A.~D.}\
  \bibnamefont {Corso}}, \bibinfo {author} {\bibfnamefont {S.}~\bibnamefont
  {de~Gironcoli}}, \bibinfo {author} {\bibfnamefont {P.}~\bibnamefont
  {Delugas}}, \bibinfo {author} {\bibfnamefont {R.~A.}\ \bibnamefont
  {DiStasio}}, \bibinfo {author} {\bibfnamefont {A.}~\bibnamefont {Ferretti}},
  \bibinfo {author} {\bibfnamefont {A.}~\bibnamefont {Floris}}, \bibinfo
  {author} {\bibfnamefont {G.}~\bibnamefont {Fratesi}}, \bibinfo {author}
  {\bibfnamefont {G.}~\bibnamefont {Fugallo}}, \bibinfo {author} {\bibfnamefont
  {R.}~\bibnamefont {Gebauer}}, \bibinfo {author} {\bibfnamefont
  {U.}~\bibnamefont {Gerstmann}}, \bibinfo {author} {\bibfnamefont
  {F.}~\bibnamefont {Giustino}}, \bibinfo {author} {\bibfnamefont
  {T.}~\bibnamefont {Gorni}}, \bibinfo {author} {\bibfnamefont
  {J.}~\bibnamefont {Jia}}, \bibinfo {author} {\bibfnamefont {M.}~\bibnamefont
  {Kawamura}}, \bibinfo {author} {\bibfnamefont {H.-Y.}\ \bibnamefont {Ko}},
  \bibinfo {author} {\bibfnamefont {A.}~\bibnamefont {Kokalj}}, \bibinfo
  {author} {\bibfnamefont {E.}~\bibnamefont {Kü{\c{c}}ükbenli}}, \bibinfo
  {author} {\bibfnamefont {M.}~\bibnamefont {Lazzeri}}, \bibinfo {author}
  {\bibfnamefont {M.}~\bibnamefont {Marsili}}, \bibinfo {author} {\bibfnamefont
  {N.}~\bibnamefont {Marzari}}, \bibinfo {author} {\bibfnamefont
  {F.}~\bibnamefont {Mauri}}, \bibinfo {author} {\bibfnamefont {N.~L.}\
  \bibnamefont {Nguyen}}, \bibinfo {author} {\bibfnamefont {H.-V.}\
  \bibnamefont {Nguyen}}, \bibinfo {author} {\bibfnamefont {A.~O.}\
  \bibnamefont {de-la Roza}}, \bibinfo {author} {\bibfnamefont
  {L.}~\bibnamefont {Paulatto}}, \bibinfo {author} {\bibfnamefont
  {S.}~\bibnamefont {Ponc{\'{e}}}}, \bibinfo {author} {\bibfnamefont
  {D.}~\bibnamefont {Rocca}}, \bibinfo {author} {\bibfnamefont
  {R.}~\bibnamefont {Sabatini}}, \bibinfo {author} {\bibfnamefont
  {B.}~\bibnamefont {Santra}}, \bibinfo {author} {\bibfnamefont
  {M.}~\bibnamefont {Schlipf}}, \bibinfo {author} {\bibfnamefont {A.~P.}\
  \bibnamefont {Seitsonen}}, \bibinfo {author} {\bibfnamefont {A.}~\bibnamefont
  {Smogunov}}, \bibinfo {author} {\bibfnamefont {I.}~\bibnamefont {Timrov}},
  \bibinfo {author} {\bibfnamefont {T.}~\bibnamefont {Thonhauser}}, \bibinfo
  {author} {\bibfnamefont {P.}~\bibnamefont {Umari}}, \bibinfo {author}
  {\bibfnamefont {N.}~\bibnamefont {Vast}}, \bibinfo {author} {\bibfnamefont
  {X.}~\bibnamefont {Wu}},\ and\ \bibinfo {author} {\bibfnamefont
  {S.}~\bibnamefont {Baroni}},\ }\bibfield  {title} {\bibinfo {title} {Advanced
  capabilities for materials modelling with quantum {ESPRESSO}},\ }\href
  {https://doi.org/10.1088/1361-648x/aa8f79} {\bibfield  {journal} {\bibinfo
  {journal} {Journal of Physics: Condensed Matter}\ }\textbf {\bibinfo {volume}
  {29}},\ \bibinfo {pages} {465901} (\bibinfo {year} {2017})}\BibitemShut
  {NoStop}%
\bibitem [{\citenamefont {Becke}\ and\ \citenamefont
  {Johnson}(2006)}]{Becke2006}%
  \BibitemOpen
  \bibfield  {author} {\bibinfo {author} {\bibfnamefont {A.~D.}\ \bibnamefont
  {Becke}}\ and\ \bibinfo {author} {\bibfnamefont {E.~R.}\ \bibnamefont
  {Johnson}},\ }\bibfield  {title} {\bibinfo {title} {A simple effective
  potential for exchange},\ }\href {https://doi.org/10.1063/1.2213970}
  {\bibfield  {journal} {\bibinfo  {journal} {The Journal of Chemical Physics}\
  }\textbf {\bibinfo {volume} {124}},\ \bibinfo {pages} {221101} (\bibinfo
  {year} {2006})}\BibitemShut {NoStop}%
\bibitem [{\citenamefont {Waroquiers}\ \emph {et~al.}(2013)\citenamefont
  {Waroquiers}, \citenamefont {Lherbier}, \citenamefont {Miglio}, \citenamefont
  {Stankovski}, \citenamefont {Ponc\'e}, \citenamefont {Oliveira},
  \citenamefont {Giantomassi}, \citenamefont {Rignanese},\ and\ \citenamefont
  {Gonze}}]{Waroquiers2013}%
  \BibitemOpen
  \bibfield  {author} {\bibinfo {author} {\bibfnamefont {D.}~\bibnamefont
  {Waroquiers}}, \bibinfo {author} {\bibfnamefont {A.}~\bibnamefont
  {Lherbier}}, \bibinfo {author} {\bibfnamefont {A.}~\bibnamefont {Miglio}},
  \bibinfo {author} {\bibfnamefont {M.}~\bibnamefont {Stankovski}}, \bibinfo
  {author} {\bibfnamefont {S.}~\bibnamefont {Ponc\'e}}, \bibinfo {author}
  {\bibfnamefont {M.~J.~T.}\ \bibnamefont {Oliveira}}, \bibinfo {author}
  {\bibfnamefont {M.}~\bibnamefont {Giantomassi}}, \bibinfo {author}
  {\bibfnamefont {G.-M.}\ \bibnamefont {Rignanese}},\ and\ \bibinfo {author}
  {\bibfnamefont {X.}~\bibnamefont {Gonze}},\ }\bibfield  {title} {\bibinfo
  {title} {Band widths and gaps from the tran-blaha functional: Comparison with
  many-body perturbation theory},\ }\href
  {https://doi.org/10.1103/PhysRevB.87.075121} {\bibfield  {journal} {\bibinfo
  {journal} {Phys. Rev. B}\ }\textbf {\bibinfo {volume} {87}},\ \bibinfo
  {pages} {075121} (\bibinfo {year} {2013})}\BibitemShut {NoStop}%
\bibitem [{\citenamefont {Marques}\ \emph {et~al.}(2012)\citenamefont
  {Marques}, \citenamefont {Oliveira},\ and\ \citenamefont
  {Burnus}}]{Marques2012}%
  \BibitemOpen
  \bibfield  {author} {\bibinfo {author} {\bibfnamefont {M.~A.}\ \bibnamefont
  {Marques}}, \bibinfo {author} {\bibfnamefont {M.~J.}\ \bibnamefont
  {Oliveira}},\ and\ \bibinfo {author} {\bibfnamefont {T.}~\bibnamefont
  {Burnus}},\ }\bibfield  {title} {\bibinfo {title} {Libxc: A library of
  exchange and correlation functionals for density functional theory},\ }\href
  {https://doi.org/https://doi.org/10.1016/j.cpc.2012.05.007} {\bibfield
  {journal} {\bibinfo  {journal} {Computer Physics Communications}\ }\textbf
  {\bibinfo {volume} {183}},\ \bibinfo {pages} {2272} (\bibinfo {year}
  {2012})}\BibitemShut {NoStop}%
\bibitem [{Note4()}]{Note4}%
  \BibitemOpen
  \bibinfo {note} {One might think that inverting the LDA is trivial, since the
  LDA xc potential is a monotonic function of the density, and the density is
  therefore a function (and not a non-local functional) of the LDA xc
  potential. This might suggest that our algorithm should obviously work well
  for the LDA. However, one has to consider that at the start of the iteration
  procedure the xc potential is not the LDA one, so the density-potential
  relation is not local. Therefore, the algorithm could in principle be
  non-convergent or lead to a wrong solution, even for the LDA.}\BibitemShut
  {Stop}%
\end{thebibliography}%
\end{document}